\pdfoutput=1
\documentclass[a4paper,12pt]{article}
\usepackage[text={453pt,686pt},centering]{geometry}
\usepackage[noconfig]{refstyle}
\usepackage{lmodern}
\usepackage[T1]{fontenc}
\usepackage{microtype,amsmath,amsfonts,amssymb}
\usepackage[sort&compress,square,comma,numbers]{natbib}
\usepackage[usenames,dvipsnames,svgnames,table]{xcolor}
\usepackage{graphicx}
\usepackage[bf]{caption}
\usepackage[colorlinks, citecolor=blue, linkcolor=blue, urlcolor=Maroon, filecolor=Maroon, linktocpage=true]{hyperref}

\allowdisplaybreaks[4]

\let\eqref=\relax
\newref{eq}{name={eq.~},Name={Eq.~},names={eqs.~},Names={Eqs.~},rngtxt={-},refcmd=(\ref{#1})}
\newref{f}{name={footnote~},Name={Footnote~},names={footnotes~},Names={Footnotes~}}
\newref{app}{name={appendix~},Name={Appendix~},names={appendixes~},Names={Appendixes~}}
\newref{tab}{name={table~},Name={Table~},names={tables~},Names={Tables~}}
\newref{ch}{name={chapter~},Name={Chapter~},names={chapters~},Names={Chapters~}}
\newref{sec}{name={section~},Name={Section~},names={sections~},Names={Sections~}}
\newref{fig}{name={figure~},Name={Figure~},names={figures~},Names={Figures~}}
\numberwithin{equation}{section}

\newcommand{\sfrac}[2]{{\textstyle\frac{#1}{#2}}}
\newcommand{\sfr}[2]{{\textstyle\frac{#1}{#2}}}

\newcommand{\dd}{\mathrm{d}}
\newcommand{\im}{\mathrm{i}}
\newcommand{\RR}{\mathbb{R}}
\newcommand{\M}{\mathcal{M}}

\DeclareMathOperator\diag{diag}

\DeclareMathOperator\tr{tr}
\let\Re=\relax\DeclareMathOperator\Re{Re}
\let\Im=\relax\DeclareMathOperator\Im{Im}

\newcommand{\at}{\tilde{a}}
\newcommand{\bt}{\tilde{b}}
\newcommand{\ct}{\tilde{c}}
\newcommand{\ah}{\hat{a}}
\newcommand{\bh}{\hat{b}}
\newcommand{\ch}{\hat{c}}
\def\a{\alpha}
\def\b{\beta}

\def\s{\zeta}
\def\ab{\bar\alpha}
\def\bb{\bar\beta}
\def\vp{\varphi}
\def\De{\Delta}
\def\1{{\bar 1}}
\def\2{{\bar 2}}
\def\3{{\bar 3}}
\def\4{{\bar 4}}
\def\Ct{C}
\def\et{e}
\def\It{I}
\def\J{J}
\def\and{\quad\textrm{and}\quad}

\def\doublerowheight{\vphantom{\parbox[c]{1cm}{\begin{tabular}{c} \vspace{-2pt} A \\ B \end{tabular}}}}
\def\triplerowheight{\vphantom{\parbox[c]{1cm}{\begin{tabular}{c} \vspace{-2pt} A \\ \vspace{-2pt} B \\ C \end{tabular}}}}
\newcommand\varpm{\mathbin{\vcenter{\hbox{%
  \oalign{\hfil$\scriptstyle+$\hfil\cr
          \noalign{\kern-.5ex}
          $\scriptscriptstyle({-})$\cr}%
}}}}
\newcommand{\stoptocwriting}{%
  \addtocontents{toc}{\protect\setcounter{tocdepth}{-5}}}
\newcommand{\resumetocwriting}{%
  \addtocontents{toc}{\protect\setcounter{tocdepth}{\arabic{tocdepth}}}}
\begin{document}

\begin{titlepage}

\vfill
\begin{flushright}
\normalsize{ZMP--HH/15--28}\\
\normalsize{MPP--2015--309}\\
\end{flushright}

\vfill

\begin{center}
   \baselineskip=16pt
   {\Large \bf Yang-Mills solutions and Spin$(7)$-instantons on\\\vspace{1mm} cylinders over coset spaces with $G_2$-structure}
   \vskip 2cm
   Alexander~S.~Haupt
   \vskip .6cm
   \begin{small}
    {\it Department of Mathematics \\ and Center for Mathematical Physics,\\
         University of Hamburg,\\ Bundesstrasse 55, 20146 Hamburg, Germany} \\[0.5cm]
    {\it II. Institute for Theoretical Physics,\\ University of Hamburg,\\
         Luruper Chaussee 149, 22761 Hamburg, Germany} \\[0.5cm]
    {\it Max-Planck-Institut f\"ur Physik,\\
         F\"ohringer Ring 6, 80805 Munich, Germany} \\[0.5cm]
    \texttt{alexander.haupt@uni-hamburg.de}
   \end{small}
\end{center}

\vfill 
\begin{abstract}
\noindent
We study $\mathfrak{g}$-valued Yang-Mills fields on cylinders $Z(G/H)=\RR \times G/H$, where $G/H$ is a compact seven-dimensional coset space with $G_2$-structure, $\mathfrak{g}$ is the Lie algebra of $G$, and $Z(G/H)$ inherits a Spin$(7)$-structure. After imposing a general $G$-invariance condition, Yang-Mills theory with torsion on $Z(G/H)$ reduces to Newtonian mechanics of a point particle moving in $\RR^n$ under the influence of some quartic potential and possibly additional constraints. The kinematics and dynamics depends on the chosen coset space. We consider in detail three coset spaces with nearly parallel $G_2$-structure and four coset spaces with $SU(3)$-structure. For each case, we analyze the critical points of the potential and present a range of finite-energy solutions. We also study a higher-dimensional analog of the instanton equation. Its solutions yield $G$-invariant Spin$(7)$-instanton configurations on $Z(G/H)$, which are special cases of Yang-Mills configurations with torsion.
\end{abstract}

\begin{quote}

\end{quote} 
\vfill
\setcounter{footnote}{0}
\setcounter{page}{0}

\end{titlepage}

\tableofcontents

\section{Introduction}\seclabel{intro}

Higher-dimensional Yang-Mills theory naturally appears in superstring theories~\cite{Green:1987, Blumenhagen:2013fgp}, for example as part of the low-energy limit in the presence of D-branes and as a sub-sector of heterotic supergravity. Moreover, the massless spectrum of closed strings contains an NS-NS 2-form, called $B$-field, whose coupling to the gauge field leads to Yang-Mills theory with torsion. The torsion 3-form $H$ appearing in the Yang-Mills equations is given by the field strength of the $B$-field plus $\a'$-corrections.

In string/M-theory compactifications (for reviews we refer for example to~\cite{Green:1987, Blumenhagen:2013fgp, Grana:2005jc, Douglas:2006es, Blumenhagen:2006ci} and references therein), the higher-dimensional space-time is assumed to be a (possibly warped) product of a lower-dimensional external part and a compact internal Riemannian manifold, denoted $\M$. The most interesting choices for phenomenology seem to be $d=6$, $7$ or $8$, where $d=\dim\M$. Of particular interest are furthermore solutions that preserve some amount of supersymmetry. The condition of supersymmetry preservation leads to first-order differential equations in the fields, known as BPS equations. In the gauge sector, the condition is a higher-dimensional generalization of four-dimensional Yang-Mills anti-self-duality and the corresponding solutions are called higher-dimensional Yang-Mills instantons. In the gravity sector, the BPS equations place stringent restrictions on the geometry of $\M$. In the simplest examples, the metric on $\M$ has special holonomy, or equivalently one says that $\M$ carries an integrable $G$-structure. However, in the presence of flux, $\M$ must be a $G$-structure manifold, where $G$ is a subgroup of $SO(d)$ and the $G$-structure is typically non-integrable. Prominent examples are $SU(3)$-structures in $d=6$, $G_2$-structures in $d=7$ and Spin$(7)$-structures in $d=8$.

Henceforth, we denote by $(\M, g)$ a Riemannian manifold of dimension $d > 4$, with $G$-structure, $G \subset SO(d)$. Such a manifold always possesses a $G$-invariant globally well-defined 4-form $Q_\M$. In addition, let $E\to\M$ be a vector bundle over $\M$ with connection $A$. The curvature of this connection is denoted by $F = \dd A + A \wedge A \in\Gamma(\Lambda^2 T^\ast \M \otimes \mathrm{End}(E))$. 

There are several \emph{a priori} different ways to define an instanton condition for $F$ (see for example~\cite{Harland:2011zs}). In this paper, an instanton is taken to be a solution of the generalized anti-self-duality equation~\cite{Corrigan:1982th,Ward:1983zm},
\begin{equation}\eqlabel{insteq}
 \ast F = - F \wedge \ast Q_\M \; ,
\end{equation}
where $\ast$ is the Hodge star operator with respect to the metric $g$. The Yang-Mills equation with torsion,
\begin{equation}\eqlabel{YM}
 \dd \ast F + [A, \ast F]_\wedge + F \wedge \ast H = 0 \; ,
\end{equation}
follows from the instanton equation~\eqref*{insteq} by applying the gauge covariant derivative $D := \dd + [A, \cdot]_\wedge$ and using the Bianchi identity. The torsion 3-form $H$ is related to $Q_\M$ via $\ast H := \dd\ast Q_\M$. \Eqref{YM} reduces to the standard Yang-Mills equation \emph{without} torsion, if $Q_\M$ is co-closed. It is important to note that \eqref{YM} is the equation of motion for the action
\begin{equation}\eqlabel{YMaction}
 S = \int_\M \tr\left(F \wedge\ast F + (-1)^d F \wedge F \wedge \ast Q_\M \right) \; .
\end{equation}

Numerous solutions of~\eqref{insteq,YM} have been constructed, for example, in~\cite{Ivanova:2009yi, Harland:2009yu, Bauer:2010fia, Haupt:2011mg, Gemmer:2011cp, Ivanova:2012vz, Bunk:2014kva, Tormahlen:2014zia}. These constructions share some common features. Namely, $\M$ is taken to be a (possibly) warped product of $\RR$ with $G/H$, which is a $(d-1)$-dimensional compact coset space. Together with some simplifying assumptions regarding the gauge field $A$, \eqref{insteq,YM} reduce to a system of gradient flow equations for a set of scalar fields that only depend on the $\RR$-direction. In some cases, the flow equations are supplemented by constraints, which can be algebraic or first-order differential equations. For \eqref{YM}, the flow is second-order and determined by the gradient of a quartic potential, whereas for \eqref{insteq}, the flow is first-order and determined by the gradient of a cubic superpotential. For this set-up, both numerical and analytical solutions have been found in various dimensions and for different kinds of $G$-structures~\cite{Ivanova:2009yi, Harland:2009yu, Bauer:2010fia, Haupt:2011mg, Gemmer:2011cp, Ivanova:2012vz, Bunk:2014kva, Tormahlen:2014zia}. In this paper, we consider the instanton equation and the Yang-Mills equation with torsion on cylinders $\RR\times G/H$, where $G/H$ is a compact seven-dimensional coset space with $G_2$-structure and the cylinder has an induced Spin$(7)$-structure. 

The paper is organized as follows. In the next section, we set up our conventions on cylinders of the general form $\RR\times G/H$. The specialization to $G/H$ being seven-dimensional with $G_2$-structure is the subject of \secref{G_over_H_G2}. After reviewing the key properties of $G_2$-structures in seven dimensions and Spin$(7)$-structures in eight dimensions, we compute the instanton and Yang-Mills equations. We discuss some simple solutions common to all such coset spaces. However, exploring the full space of solutions requires explicit knowledge of the coset space in question. We thus study the instanton and Yang-Mills equations on a case-by-case basis for three explicit $G_2$-structure coset spaces, namely the Berger space $SO(5)/SO(3)_{\text{max}}$, the squashed 7-sphere $Sp(2)\times Sp(1)/Sp(1)^2$, and the Aloff-Wallach spaces $SU(3)/U(1)_{k,l}$. In \secref{G_over_H_SU3}, we consider a related set-up, where now $G/H$ is a compact seven-dimensional coset space with $SU(3)$-structure. This section begins with a brief review of $SU(3)$-structures in seven dimensions. We then construct solutions on four explicit examples of such coset spaces, namely $SO(5)/SO(3)_{A+B}$, $N^{pqr} = (SU(3)\times U(1)) / (U(1) \times U(1))$, $M^{pqr} = (SU(3)\times SU(2)\times U(1)) / (SU(2)\times U(1) \times U(1))$ and $Q^{pqr} = (SU(2)\times SU(2)\times SU(2)) / (U(1) \times U(1))$. We conclude in \secref{conclusions} with a summary of the main results and a discussion of possible future directions. Throughout the main text, we frequently need the structure constants of the coset spaces. For reference and to make our presentation self-contained, we thus include in \appref{struct_consts} a list of the structure constants of all the coset spaces appearing in this paper.

\section{Instantons and Yang-Mills equation with torsion on \texorpdfstring{$\RR\times G/H$}{real line times coset space}}\seclabel{M_equal_to_R_times_G_over_H}

We shall now briefly review the specialization to the case $\M=\RR\times G/H$, following~\cite{Ivanova:2009yi}. Throughout this paper, $G/H$ is assumed to be a compact coset space. That is, $G$ denotes a compact connected Lie group and $H$ a closed subgroup, such that $G/H$ is a compact reductive homogeneous space.

The metric $g$ on the cylinder $Z(G/H):=\RR\times G/H$ with a coordinate $\tau$ on $\RR$ can be written as
\begin{equation}\eqlabel{metric}
 g = \delta_{\mu\nu} e^\mu \otimes e^\nu = \dd\tau\otimes\dd\tau + g_{G/H} = \dd\tau\otimes\dd\tau + \delta_{ab} e^a \otimes e^b \; ,
\end{equation}
where $\{e^\mu \} = \{e^0 = \dd\tau, e^a \}$ is a local orthonormal basis of $T^\ast (\RR\times G/H)$. Note that Greek and lower case Latin indices label the coordinates of $\RR\times G/H$ and $G/H$, respectively. That is, $\mu=(0,a)=0,1,\ldots,(d-1)$. The connection $A$ and curvature $F$ can also be expanded in the basis of 1-forms according to
\begin{equation}
 A = A_0 e^0 + A_a e^a \; , \qquad\text{and}\qquad F=F_{0a} e^{0a} + \sfrac12 F_{ab} e^{ab} \; ,
\end{equation}
where $e^{\mu_1\ldots\mu_n} := e^{\mu_1} \wedge\cdots\wedge e^{\mu_n}$ and we will choose \emph{temporal gauge}, that is $A_0 = 0$, throughout this paper.

Since $G/H$ is assumed to be a coset space, the Lie algebra $\mathfrak{g}$ of $G$ can be decomposed as $\mathfrak{g}=\mathfrak{h}\oplus\mathfrak{m}$. Here, $\mathfrak{m}$ is the orthogonal complement of the Lie algebra $\mathfrak{h}$ of $H$ in $\mathfrak{g}$. The generators $\{I_A\}_{A=1,\ldots,\dim\mathfrak{g}}$ spanning $\mathfrak{g}$ can be divided into two sets $\{I_i\}_{i=\dim\mathfrak{m}+1,\ldots,\dim\mathfrak{m}+\dim\mathfrak{h}}$ and $\{I_a\}_{a=1,\ldots,\dim\mathfrak{m}}$. The latter are generators of $\mathfrak{h}$ and $\mathfrak{m}$, respectively. They satisfy the following commutation relations,
\begin{equation}
 [I_i, I_j] = f^k_{ij} I_k \; , \qquad
 [I_i, I_a] = f^b_{ia} I_b \; , \qquad\text{and}\qquad
 [I_a, I_b] = f^i_{ab} I_i + f^c_{ab} I_c \; .
\end{equation}
The generators are normalized such that the usual Killing-Cartan metric on $\mathfrak{g}$ can be written as
\begin{equation}\eqlabel{g_CK}
 (g_\mathfrak{g})_{ab} = 2 f^i_{ad} f^d_{ib} + f^c_{ad} f^d_{cb} = \delta_{ab} \; , \;\;\;
 (g_\mathfrak{g})_{ij} = f^k_{il} f^l_{kj} + f^b_{ia} f^a_{bj} = \delta_{ij} \; , \;\;\;\text{and}\;\;\; (g_\mathfrak{g})_{ia} = 0 \; .
\end{equation}
Throughout this paper, we use the adjoint representation, $(I_A)^C_B = f^C_{AB}$, for the generators of $\mathfrak{g}$. The Killing-Cartan metric can also be succinctly expressed as $(g_\mathfrak{g})_{AB} = - \tr(I_A I_B) = - f^C_{AD} f^D_{BC} = \delta_{AB}$ by virtue of~\eqref*{g_CK}.

Making use of the above coset space construction, we choose a $G$-invariant ansatz for the gauge connection $A$~\cite{Bauer:2010fia},
\begin{equation}\eqlabel{Aansatz}
 A = e^i I_i + e^a X_a \; ,
\end{equation}
with $X_a = X_a^b I_b$ and $\{e^i\}$ being left-invariant 1-forms on $G/H$ dual to $\{I_i\}$. The $e^i$ can be expressed as linear combinations of the 1-forms $\{e^a\}$ via $e^i = e^i_a e^a$, with real functions $e^i_a$. For the precise definitions of $e^a$ and $e^i$ we refer to~\cite{Ivanova:2009yi,Harland:2009yu,Bauer:2010fia}, and for a more pedagogical introduction to~\cite{Tormahlen:2015}. For the calculations carried out in this paper, it suffices to keep in mind that the 1-forms $\{ e^A \} = \{ e^a, e^i \}$ are always well-defined on a contractible open subset $U$ of $G/H$, and moreover that they obey the Maurer-Cartan equations,
\begin{equation}
 \dd e^a = - f^a_{ib} e^{ib} - \sfrac12 f^a_{bc} e^{bc} \; , \qquad\qquad\qquad 
 \dd e^i = - \sfrac12 f^i_{bc} e^{bc} - \sfrac12 f^i_{jk} e^{jk} \; .
\end{equation}
In order to ensure the $G$-invariance of the ansatz~\eqref*{Aansatz}, the matrices $X_a$ need to satisfy
\begin{equation}\eqlabel{Ginvcond_gaugefield}
 [I_i , X_a] = f^b_{ia} X_b \; .
\end{equation}
In component notation, this condition reads $X_a^b f^c_{ib} = f^b_{ia} X_b^c$. The curvature $F$ of the connection~\eqref*{Aansatz} becomes
\begin{equation}\eqlabel{Fansatz}
  F = \dd A + A \wedge A = \dot{X}_a e^{0a} - \sfr12 \left( f^i_{bc} I_i + f^a_{bc} X_a - [X_b,X_c] \right) e^{bc} \; ,
\end{equation}
where the dot denotes the derivative with respect to $\tau$.

\section{Seven-dimensional coset spaces with \texorpdfstring{$G_2$}{G2}-structure}\seclabel{G_over_H_G2}

In this section, we study the instanton equation and the Yang-Mills equation with torsion on cylinders over compact seven-dimensional coset spaces $G/H$ admitting a $G_2$-structure. Before considering Yang-Mills theory in this set-up, we begin by reviewing some important facts about $G_2$-structures in seven dimensions and Spin$(7)$-structures in eight dimensions.

\subsection{\texorpdfstring{$G_2$}{G2}-structures in seven dimensions}\seclabel{G2_structures}

The $G_2$-structure on a compact seven-dimensional coset space $G/H$ is defined by a 3-form $P$. The 4-form dual to $P$ is denoted $Q:=\ast_7 P$. In a suitably chosen basis, the canonical forms $(P,Q)$ are given by the standard expressions~\cite{Bryant:2003, Agricola:2006},
\begin{equation}\eqlabel{PQdef}
\begin{aligned}
 P &= e^{123} + e^{145} - e^{167} - e^{246} - e^{257} - e^{347} + e^{356} \; , \\
 Q &= e^{1247} - e^{2345} + e^{4567} + e^{2367} - e^{1357} - e^{1346} - e^{1256} \; .
\end{aligned}
\end{equation}
These forms and their respective components satisfy the following useful relations
\begin{align}
 P_{acd} P_{bcd} &= 6\, \delta_{ab} \; , \eqlabel{PQrelns1} \\
 Q_{acde} Q_{bcde} &= 24\, \delta_{ab} \; , \eqlabel{PQrelns2} \\
 P_{abe} P_{cde} &= - Q_{abcd} + \delta_{ac} \delta_{bd} - \delta_{ad} \delta_{bc} \; , \eqlabel{PQrelns3} \\
 P \wedge Q &= 7 \, e^{1\ldots 7} = 7 \, \text{Vol}(G/H) \; . \eqlabel{PQrelns4}
\end{align}
Note that in our conventions, the orientations of the seven-dimensional coset space $G/H$ and of the cylinder $\RR\times G/H$ are chosen such that $\varepsilon^{(7)}_{1234567} = +1$ and $\varepsilon^{(8)}_{01234567} = +1$, respectively. Consequently, the Hodge star operator $\ast$ on the cylinder is in our conventions related to the Hodge star operator $\ast_7$ on the coset space via
\begin{equation}\eqlabel{Hodge_star_8_7_conversion}
 \ast \omega^{(7)}_p = - (\ast_7 \omega^{(7)}_p) \wedge \dd\tau\; , \qquad
 \ast (\dd\tau \wedge \omega^{(7)}_p) = \ast_7 \omega^{(7)}_p \; .
\end{equation}
Here, $\omega^{(7)}_p$ is a $p$-form with legs only in the coset space directions.

$G_2$-structures can be distinguished and classified by means of the intrinsic torsion. This is analogous to the case of six-dimensional $SU(3)$-structures, reviewed for example in~\cite{Grana:2005jc}. In the present case, it holds that
\begin{equation}
  T_{ab} {}^c \in \Lambda^1 \otimes (\mathfrak{g}_2 \oplus \mathfrak{g}_2^\perp ) \; ,
\end{equation}
where $T_{ab} {}^c$ are the components of the torsion tensor, $\Lambda^1$ is the space of 1-forms and $\mathfrak{so}(7) = \mathfrak{g}_2 \oplus \mathfrak{g}_2^\perp$. The $\mathfrak{g}_2$ piece drops when acting on $G_2$-invariant forms. The corresponding torsion is called \emph{intrinsic torsion}, denoted $T^0$. It can be decomposed into irreducible $G_2$ representations~\cite{Fernandez:1982, Chiossi:2002tw, Cleyton:2006, Agricola:2006} according to
\begin{equation}
\begin{aligned}
 T^0_{ab} {}^c \in \Lambda^1 \otimes \mathfrak{g}_2^\perp = \bf{7}\otimes \bf{7}
 = &\bf{1} \oplus \bf{7} \oplus \bf{14} \oplus \bf{27} \; .\\
    &\tau_0 \hspace{2ex} \tau_1 \hspace{2.7ex} \tau_2 \hspace{3.5ex} \tau_3
\end{aligned}
\end{equation}
The tensors $\tau_p$ are $p$-forms ($p\in\{0,1,2,3\}$), and they correspond to the four torsion classes appearing in the exterior derivatives of the structure forms,
\begin{equation}\eqlabel{dPdQ}
\begin{aligned}
 \dd P & = \tau_0 \, Q + 3\, \tau_1 \wedge P + \ast_7 \tau_3 \; ,\\
 \dd Q & = 4\, \tau_1 \wedge Q + \tau_2 \wedge P \; .
\end{aligned}
\end{equation}
Important examples of types of $G_2$-structures, some of which will occur later, are listed in the following table (where TCs $=$ torsion classes):
\begin{center}
\begin{tabular}{c|c|l}
  Type & TCs & Properties \\\hline
  parallel & $\emptyset$ & $\dd P = 0$, $\dd Q = 0$ \\
  nearly parallel & $\tau_0$ & $\dd P = \tau_0\, Q$, $\dd Q = 0$ \\
  almost parallel & $\tau_2$ & $\dd P = 0$, $\dd Q = \tau_2 \wedge P$ \\
  balanced & $\tau_3$ & $\dd P = \ast_7 \tau_3$, $\dd Q = 0$ \\
  locally conformal & $\tau_0 \oplus \tau_1$ & $\dd P = \tau_0 \, Q + 3\, \tau_1 \wedge P$, $\dd Q = 4\, \tau_1 \wedge Q$ \\
  cocalibrated (or semi-parallel) & $\tau_0 \oplus \tau_3$ & $\dd P = \tau_0 \, Q + \ast_7 \tau_3$, $\dd Q = 0$
\end{tabular}
\end{center}

For a given $G_2$-structure on $G/H$, one may ask whether continuous $G$-invariant deformations exist. This is possible provided $g$ and $P$ obey $G$-invariance conditions given by (see for example~\cite{Cassani:2011fu} section 6.1)
\begin{align}
 f_{i\left(a\right.}^c g_{\left.b\right)c} &= 0 \; , \qquad\text{and} \eqlabel{Ginvcondg} \\
 f_{i\left[a\right.}^d P_{\left.bc\right]d} &= 0 \; , \eqlabel{GinvcondP}
\end{align}
respectively.

\subsection{Spin\texorpdfstring{$(7)$}{(7)}-structures in eight dimensions}\seclabel{Spin7_structures}

The $G_2$-structure on $G/H$ can be lifted to a Spin$(7)$-structure on the cylinder $Z(G/H)$. The 4-form $\Psi$ defining the Spin$(7)$-structure is constructed from $P$ and $Q$ as
\begin{equation}\eqlabel{PsiPQ}
 \Psi = P\wedge\dd\tau - Q \; .
\end{equation}
It is self-dual, $\ast\Psi = \Psi$, and satisfies 
\begin{equation}\eqlabel{dPsiPQ}
 \dd\Psi = P \wedge \dd\tau \wedge (3\, \tau_1) - Q \wedge (4\, \tau_1 - \tau_0 \, \dd\tau) - \tau_2 \wedge P + (\ast_7 \tau_3) \wedge \dd\tau \; .
\end{equation}

In general, Spin$(7)$-structures in eight dimensions can be classified analogously to seven-dimensional $G_2$-structures described above. Here, the torsion tensor satisfies
\begin{equation}
  T_{ab} {}^c \in \Lambda^1 \otimes (\mathfrak{so}(7) \oplus \mathfrak{so}(7)^\perp ) \; ,
\end{equation}
where $\mathfrak{so}(8) = \mathfrak{spin}(7) \oplus \mathfrak{spin}(7)^\perp$ and $\mathfrak{spin}(7) \simeq \mathfrak{so}(7)$. The $\mathfrak{so}(7)$ piece drops when acting on Spin$(7)$-invariant forms. The corresponding intrinsic torsion can be decomposed into irreducible $SO(7)$ representations~\cite{Fernandez:1986} according to
\begin{equation}
\begin{aligned}
 T^0_{ab} {}^c \in \Lambda^1 \otimes \mathfrak{so}(7)^\perp = \bf{8}\otimes \bf{7}
 = &\bf{8} \oplus \bf{48}\; .\\
    &W_8 \hspace{2ex} W_{48}
\end{aligned}
\end{equation}
Thus, there are two torsion classes $W_8$, $W_{48}$ and four different types of Spin$(7)$-structures in eight dimensions. They are listed in the following table (where TCs $=$ torsion classes):
\begin{center}
\begin{tabular}{c|c|l}
  Type & TCs & Properties \\\hline
  parallel & $\emptyset$ & $\dd \Psi = 0$ \\
  locally conformal & $W_8$ & $\dd\Psi + \sfrac17 \Theta\wedge\Psi = 0$ \\
  balanced & $W_{48}$ & $\Theta = 0$ \\
  --- & $W_8 \oplus W_{48}$ & no relation
\end{tabular}
\end{center}
In the above table, we have used the Lee-form $\Theta$, generally defined as~\cite{Cabrera:1995}
\begin{equation}
 \Theta = \ast\left((\ast\dd\Psi)\wedge\Psi \right) \; .
\end{equation}

It is interesting to ask how the properties of the $G_2$-structure on $G/H$ lift to the Spin$(7)$-structure on the cylinder $Z(G/H)$. First, we compute $\Theta$,
\begin{equation}
 \Theta = 2 \theta + 7\, \tau_0 \, \dd\tau \; , \qquad\quad \text{with}\;\; \theta = -12\,\tau_1 \; ,
\end{equation}
where $\theta$ is the seven-dimensional Lee-form, generally defined as $\theta := \ast_7 \left((\ast_7 \dd P)\wedge P \right)$, and we have used that $\tau_3\wedge P = 0$, $\tau_3\wedge Q = 0$ (see for example~\cite{Karigiannis:2003} eq. (2.21)). Thus, the Spin$(7)$-structure is balanced, if and only if $\tau_0 = 0$ and $\tau_1 = 0$. In other words, the intrinsic torsion of the $G_2$-structure has components\footnote{To our knowledge, there is no name assigned in the literature to this particular combination of $G_2$-structure torsion classes.} $\tau_2 \oplus \tau_3$, which includes the special cases of almost parallel (that is, $\tau_2$ only) and balanced (that is, $\tau_3$ only) $G_2$-structures. 

On the other hand, the Spin$(7)$-structure on $Z(G/H)$ is locally conformal, if and only if $\tau_2 = 0$ and $\tau_3 = 0$, which corresponds to a locally conformal $G_2$-structure on $G/H$. This includes the special case of a nearly parallel $G_2$-structure, which plays an important role in the present work. In this case (and after setting $\tau_0 = 4$, without loss of generality), the Spin$(7)$-structure equation becomes $\dd\Psi = 4\, \dd\tau\wedge Q$, which can be re-written as
\begin{equation}
 \dd\Psi + \sfrac17 \Theta\wedge\Psi = 0 \; , \qquad \text{with} \qquad \Theta = 28\, \dd\tau \; ,
\end{equation}
implying that $\Psi$ then indeed defines a \emph{locally conformal} Spin$(7)$-structure~\cite{Ivanov:2005}. 

Before closing this subsection, we mention that the cone over a nearly parallel $G_2$-structure manifold and the cylinder over a $G_2$ manifold (that is, a manifold with parallel $G_2$-structure) are examples of Spin$(7)$ manifolds (that is, manifolds with parallel Spin$(7)$-structure).

\subsection{Spin\texorpdfstring{$(7)$}{(7)}-instantons}\seclabel{Spin7inst}

For the special case of a Spin$(7)$-structure in eight dimensions, the instanton equation~\eqref*{insteq} with $Q_\M = \Psi$ becomes
\begin{equation}\eqlabel{insteqSpin7}
 F_{\mu\nu} = - \sfrac12 \Psi_{\mu\nu\rho\sigma} F^{\rho\sigma} \; .
\end{equation}
This is equivalent to the statement that $F\in\mathfrak{spin}(7)$. Further specializing to the case where the eight-dimensional manifold is a cylinder over a $G_2$-structure manifold and using~\eqref{PsiPQ}, leads to 
\begin{equation}\eqlabel{insteqSpin7coset}
 F_{0a} = \sfrac12 P_{abc} F^{bc} \; , \qquad\text{and}\qquad
 F_{ab} = \sfrac12 \left( P_{abe} P^e {}_{cd} + Q_{abcd} \right) F^{cd} \; .
\end{equation}
In our case, the second equation is identically satisfied by means of the identity~\eqref*{PQrelns3}. The remaining equation, $F_{0a} = \sfrac12 P_{abc} F^{bc}$, can be rewritten on $Z(G/H)$ as
\begin{equation}\eqlabel{insteqSpin7cosetXvec}
 \dot{X}_a + \sfr12 P_a {}^{bc} \left( f^i_{bc} I_i + f^d_{bc} X_d - [ X_b , X_c ] \right) = 0 \; ,
\end{equation}
after inserting~\eqref*{Fansatz}. 

Expressed in components, $X_a = X_a^b I_b$, \eqref{insteqSpin7cosetXvec} reads as follows,
\begin{equation}\eqlabel{insteqSpin7cosetX}
 \dot{X}_a^b + \sfr12 P_a {}^{cd} \left( f^e_{cd} X_e^b - X_c^e X_d^f f_{ef}^b \right) = 0 \; , \qquad\text{and}\qquad
 P_a {}^{bc} \left( f^i_{bc} - X_b^d X_c^e f_{de}^i \right) = 0 \; .
\end{equation}
The decomposition into two equations is due to the fact that $F$ has a part valued in $\mathfrak{m}$, a part valued in $\mathfrak{h}$ and both need to satisfy~\eqref*{insteqSpin7cosetXvec} separately. The first equation in~\eqref*{insteqSpin7cosetX} comes from the $\mathfrak{m}$-part of $F$ and is a first-order ordinary differential equation. The second equation in~\eqref*{insteqSpin7cosetX}, coming from the $\mathfrak{h}$-part of $F$, yields a set of purely algebraic conditions known as quiver relations~\cite{Lechtenfeld:2008nh}. The origin of the latter can be traced back to the reduced ansatz for $X_a$ which was taken to be $\mathfrak{m}$-valued only, $X_a = X_a^b I_b$. For the full $\mathfrak{g}$-valued $X_a = X_a^b I_b + X_a^i I_i$, the algebraic conditions instead become first-order differential equations in the extra components $X_a^i$.

In general, to proceed further, we require explicit knowledge of the groups $G$ and $H$. This can be seen for example from \eqref{Ginvcond_gaugefield}, which involves the structure constants. We will thus study the instanton equation in \secref{G2structure_cosets} on a case-by-case basis on cylinders over explicit seven-dimensional coset spaces $G/H$ that admit a nearly parallel $G_2$-structure.

On the other hand, there exists a specialized ansatz, which has already been considered in~\cite{Ivanova:2009yi} and which can be solved explicitly without the need to specify $G/H$, provided some additional assumptions are imposed. The idea is to trivially solve \eqref{Ginvcond_gaugefield} by reducing the number of degrees of freedom to just a single real scalar field $\phi(\tau)$. The $G$-invariance condition~\eqref*{Ginvcond_gaugefield} is obviously satisfied upon setting
\begin{equation}\eqlabel{single_field_X}
 X_a = \phi I_a \; .
\end{equation}
Consequently, the gauge connection $A$ and the corresponding curvature $F$ are given by
\begin{equation}\eqlabel{single_field_A_F}
 A = e^i I_i + \phi \, e^a I_a \; , \qquad
 F = \dot{\phi}\, e^{0a} I_a - \sfrac12 \left( [ 1-\phi^2 ] f^i_{ab} I_i + \phi [ 1-\phi ] f^c_{ab} I_c \right) e^{ab} \; .
\end{equation}
Inserting~\eqref*{single_field_X} into~\eqref*{insteqSpin7cosetX} yields
\begin{equation}\eqlabel{insteq_1field_1}
 \dot{\phi}\, \delta_a^b = \sfr12 \phi (\phi-1) P_a {}^{cd} f^b_{cd} \; , \qquad\text{and}\qquad
 0 = (\phi^2 - 1) P_a {}^{bc} f^i_{bc} \, .
\end{equation}
Now, we assume\footnote{Cases meeting these assumptions are presented in \secref{SO5_SO3_max, SquashedS7}. On the other hand, counterexamples where some of these assumptions do not hold will be encountered in \secref{AW, G_over_H_SU3}.} that the structure constants with all indices lowered, $f_{ABC} := \delta_{CD} f^D_{AB}$, are totally anti-symmetric, $f_{[ABC]}=f_{ABC}$, and that they can be used to construct a well-defined $G_2$-structure on $G/H$. That is, we set $P_{abc} = \sigma f_{abc}$ for some $\sigma\in\RR\setminus\{0\}$. The second equation in~\eqref*{insteq_1field_1} is then identically satisfied by virtue of~\eqref*{g_CK}.

In addition, we assume that the structure constants satisfy the following two equivalent relations
\begin{equation}\eqlabel{YM_struct_const_formula}
 \sum_{c,i} f_{aci} f_{bci} = \sfrac12 (1-\alpha) \delta_{ab} \qquad\Leftrightarrow\qquad \sum_{c,d} f_{acd} f_{bcd} = \alpha \delta_{ab} \; ,
\end{equation}
for some $\alpha\in\RR$, where~\eqref*{g_CK} has been taken into account. This assumption is valid for the coset spaces considered in \secref{SO5_SO3_max, SquashedS7}, for example. Compatibility with~\eqref*{PQrelns1} implies $\alpha=6/\sigma^2$. Note that the Killing-Cartan metric is assumed to be $\delta_{AB}$ in this subsection. Otherwise it must be reinstated in all expressions where indices of structure constants are manipulated.

The instanton equation~\eqref*{insteqSpin7cosetX} reduces in the above set-up to a first-order ordinary differential equation for the single function $\phi(\tau)$,
\begin{equation}\eqlabel{kink_ODE}
 \dot\phi = \sfrac{\alpha\sigma}{2} \phi (\phi-1) \; .
\end{equation}
It has two static solutions, $\phi = 0$, $1$, and the well-known interpolating kink solution~\cite{Ivanova:2009yi}
\begin{equation}\eqlabel{kink_sol}
 \phi(\tau) = \sfrac12 \left( 1 - \tanh\left[\sfrac{\alpha\sigma}{4} (\tau - \tau_0) \right] \right) \; .
\end{equation}
Here, $\tau_0 \in \RR$ is an arbitrary integration constant fixing the position of the instanton in the $\tau$ direction.

\subsection{Yang-Mills equation with torsion}\seclabel{YMsol}

We now turn to the Yang-Mills equation with torsion,~\eqref*{YM}, on a cylinder over a compact seven-dimensional coset space $G/H$ admitting a $G_2$-structure. It is clear that the instanton solutions discussed in the previous section also solve~\eqref*{YM}. Our aim in this section is to analyze whether there are additional analytical solutions.

To start, it is useful to write out~\eqref*{YM} in components (for a more detailed derivation see~\cite{Tormahlen:2014zia,Tormahlen:2015}),
\begin{multline}\eqlabel{YMComp}
	\partial_\mu F^{\mu\nu} - F^{\rho\sigma}\left(\sfrac{1}{2}T_{\rho\sigma}^{\nu}-\omega_{\rho\sigma}^{\nu}\right) + F^{\rho\nu}\left(\sfrac{1}{2}T_{\rho\sigma}^\sigma - \omega_{\rho\sigma}^\sigma \right)
		\\ - F^{\rho\nu}\left(\sfrac{1}{2}T_{\sigma\rho}^\sigma-\omega_{\sigma\rho}^\sigma \right) + \left[A_\mu, F^{\mu\nu}\right] - \sfrac{1}{2} H^\nu {}_{\rho\sigma} F^{\rho\sigma} = 0 \; .
\end{multline}
Here, $\omega_{\nu\rho}^{\mu}$ are the components of the affine spin connection $\omega^{\mu} {}_\rho = \omega_{\nu\rho}^{\mu} e^\nu$ on $Z(G/H)$ with torsion $T^\mu = \sfr12 T_{\nu\rho}^{\mu} e^{\nu\rho} = \dd e^\mu + \omega^\mu {}_\nu \wedge e^\nu$, and $H_{\mu\nu\rho}$ are the components of the 3-form $H = \sfrac{1}{3!} H_{\mu\nu\rho} e^{\mu\nu\rho}$ defined in \secref{intro}. For the cylinder (product) metric~\eqref*{metric}, one finds
\begin{equation}\eqlabel{cylinder_torsion_spin_conn}
 \omega^0_{0b} = \omega^a_{0b} = \omega^0_{cb} = 0 \; , \qquad\quad\text{and}\qquad\quad
 T^0_{0b} = T^a_{0b} = T^0_{cb} = 0 \; .
\end{equation}
Given the coset space construction, it is natural to relate the components of the torsion tensor of the affine spin connection on $G/H$ to the structure constants~\cite{Ivanova:2009yi},
\begin{equation}\eqlabel{G_over_H_torsion_kappa}
 T^a_{bc} = \kappa f^a_{bc} \; ,
\end{equation}
parameterized by $\kappa\in\RR$. The affine spin connection on $G/H$ becomes
\begin{equation}\eqlabel{G_over_H_spin_connection}
 \omega^a {}_b = f^a_{ib} e^i + \sfr12 (\kappa+1) f^a_{cb} e^c \; .
\end{equation}
Moreover, in this section we set\footnote{This set-up agrees with the definition for $H$ given in \secref{intro}, where we demanded compatibility between the instanton equation~\eqref*{insteq} and the Yang-Mills equation with torsion~\eqref*{YM}. Indeed, inserting $Q_\M = \Psi = P\wedge\dd\tau - Q$ into $H=-\ast\dd\ast Q_\M$, using~\eqref*{Hodge_star_8_7_conversion},~\eqref*{dPdQ} and $\tau_1=\tau_2=\tau_3=0$, one finds $H=-\tau_0 P$ in the nearly parallel $G_2$ case. Comparing with~\eqref*{YM_T_ansatz}, we thus see that in this case, $\kappa$ is related to the $G_2$ torsion class $\tau_0$ via $\kappa=\sigma\,\tau_0$.\flabel{H_propto_P}}
\begin{equation}\eqlabel{YM_T_ansatz}
 T_{bc}^a = \kappa f_{bc}^a = \sfrac{\kappa}{\sigma} P^a {}_{bc} \; , \qquad\qquad H_{bc}^a = - T_{bc}^a \; .
\end{equation}
Note that $\kappa$ drops out of the combination $\left(\sfrac{1}{2}T_{\rho\sigma}^{\nu}-\omega_{\rho\sigma}^{\nu}\right)$ and contractions thereof. Instead, it enters in the Yang-Mills equation~\eqref*{YMComp} only via the choice~\eqref*{YM_T_ansatz}. In \secref{AW,G_over_H_SU3}, we will encounter examples where $P_{abc} \not\propto f_{abc}$. In such a situation we still use~\eqrangeref*{G_over_H_torsion_kappa}{G_over_H_spin_connection} but $H_{bc}^a \not\propto T_{bc}^a$ and hence $\kappa$ does not enter in the Yang-Mills equation.

Throughout this paper, $\omega$, $T$ and $H$ never acquire components along the $\tau$-direction. Under this assumption, we find that the $\nu=0$ component of~\eqref*{YMComp} reduces to the Gauss-law constraint on the matrices $X_a$~\cite{Bauer:2010fia}, 
\begin{equation}\eqlabel{gauss_law}
 \sum_a [X_a , \dot{X}_a] = 0 \; ,
\end{equation}
after inserting~\eqref*{Aansatz}. This is a consequence of the gauge fixing $A_0 = 0$ and becomes non-trivial when $X_a^b$ is non-diagonal. Carrying over the assumptions on the structure constants from \secref{Spin7inst} and using~\eqrangeref*{cylinder_torsion_spin_conn}{G_over_H_spin_connection}, we find that the remaining $\nu=a$ component of~\eqref*{YMComp} reduces to a second-order equation given by
\begin{multline}\eqlabel{ddotX_eqnH}
 \ddot{X}_a = \left( \sfrac{1}{2}(f_{acd}-H_{acd}) f_{bcd} - f_{aci} f_{bci} \right) X_b \\ - \sfrac{1}{2} (3 f_{abc} - H_{abc})  [X_b, X_c] - \left[ X_b, [X_b, X_a] \right] - \sfrac{1}{2} H_{abc} f_{ibc} I_i \; ,
\end{multline}
where repeated indices are to be summed over. If $P_{abc} \propto f_{abc}$ holds, we may assume, in addition,~\eqref*{YM_T_ansatz} and obtain~\cite{Bauer:2010fia}
\begin{equation}\eqlabel{ddotX_eqn}
 \ddot{X}_a = \left( \sfrac{1}{2}(\kappa+1)f_{acd} f_{bcd} - f_{aci} f_{bci} \right) X_b - \sfrac{1}{2} (\kappa+3) f_{abc} [X_b, X_c] - \left[ X_b, [X_b, X_a] \right] \; .
\end{equation}
Note that in obtaining~\eqref*{ddotX_eqnH,ddotX_eqn} from~\eqref*{YMComp}, we used the fact that the terms involving the functions $e_c^i$ vanish by virtue of the Jacobi identity (for details on this calculation, we refer to~\cite{Ivanova:2009yi,Tormahlen:2015}). We also remark that the choice $\kappa=1$ corresponds to the canonical connection introduced in~\cite{Harland:2011zs}. With~\eqref*{YM_struct_const_formula}, $\kappa=6$ and $\alpha=1/5$ (i.e. $\sigma^2 = 30$), one finds that~\eqref*{ddotX_eqn} is the $\tau$-derivative of the instanton equation~\eqref*{insteqSpin7cosetXvec}.

Beyond this point, we again need explicit knowledge of the groups $G$ and $H$, as in the case of the instanton equation. We will study the Yang-Mills equation with torsion in more detail later. For now, we will follow the same idea as in \secref{Spin7inst} and consider the single-field ansatz~\eqref*{single_field_A_F}. This special case has already been studied in~\cite{Ivanova:2009yi} and can be solved explicitly. We find a point particle equation of motion of the form
\begin{equation}\eqlabel{YM_2nd_order_phi_eq}
 \ddot{\phi} = \sfrac12 (1+\alpha) \phi (\phi - 1) \left(\phi - \sfrac{(\kappa+2)\alpha - 1}{\alpha + 1}\right) \; ,
\end{equation}
after inserting the single-field ansatz~\eqref*{single_field_A_F} into~\eqref*{ddotX_eqn} and assuming that~\eqref*{YM_struct_const_formula} holds.

Different choices of the parameters $(\alpha, \kappa)$ correspond to different types of solutions of~\eqref*{YM_2nd_order_phi_eq}. This has been analyzed in detail in~\cite{Ivanova:2009yi}. For example, the choice $\alpha=0$ leads to
\begin{equation}\eqlabel{YM_2nd_order_phi4_kink_eq}
 \ddot{\phi} = \sfrac12 \phi (\phi^2 - 1) \; ,
\end{equation}
which can be integrated to $\dot{\phi} = \pm \sfr12 (1-\phi^2)$. The solutions, $\phi = \pm \tanh \sfrac{\tau-\tau_0}{2}$, are known as the $\phi^4$ kink $(+)$ and anti-kink $(-)$, respectively. Here, $\tau_0 \in \RR$ is an arbitrary integration constant fixing the position of the kink/anti-kink in the $\tau$ direction. On the other hand, for the choice $(\alpha, \kappa)=(3/5, 1)$ we find
\begin{equation}
 \ddot{\phi} = \sfrac45 \phi (\phi - 1) \left(\phi - \sfr12 \right) \; .
\end{equation}
This can be integrated to $\dot{\phi} = \pm \sfrac{1}{\sqrt{5}} \phi (\phi - 1)$. Hence, in this case we recover the tanh-kink-type instanton solutions presented in \secref{Spin7inst}.

\subsection{Explicit coset space constructions}\seclabel{G2structure_cosets}

To proceed further, we will now consider cylinders over explicit seven-dimensional coset spaces $G/H$ admitting a nearly parallel $G_2$-structure. Only a few such examples are known~\cite{Castellani:1983yg,Friedrich:1997,Acharya:1998db}. They comprise the Berger space $SO(5)/SO(3)_{\text{max}}$, the squashed 7-sphere $Sp(2)\times Sp(1)/Sp(1)^2$, and the Aloff-Wallach spaces $SU(3)/U(1)_{k,l}$, for a co-prime pair of integers $(k,l)$. We remark that $SO(5)/SO(3)_{\text{max}}$ is locally equivalent to $Sp(2)/Sp(1)$~\cite{Konishi:2001bd}.

\subsubsection{Cylinders over the Berger space \texorpdfstring{$SO(5)/SO(3)_{\text{max}}$}{}}\seclabel{SO5_SO3_max}

For the coset space\footnote{This is the unique maximal embedding of $SO(3)$ into $SO(5)$ taking the adjoint ${\bf 10}$ of $SO(5)$ to a ${\bf 3} \oplus {\bf 7}$ of $SO(3)$. The group $SO(5)$ has in fact two commuting $SO(3)$ subgroups. In \secref{SO5_SO3_AB} we will encounter another embedding of the two commuting $SO(3)$ subgroups into $SO(5)$.} $SO(5)/SO(3)_{\text{max}}$, we use the conventions of~\cite{Castellani:1983yg}, but with the generators and structure constants rescaled as $I_A \rightarrow \sfrac{1}{\sqrt{6}} I_A$, $f^C_{AB} \rightarrow \sfrac{1}{\sqrt{6}} f^C_{AB}$. This normalization is necessary in order for~\eqref*{g_CK} to hold. For reference, the rescaled structure constants used in this subsection can be found in full detail in \appref{struct_consts_SO5_SO3_max}.

We observe that the structure constants with all indices lowered, $f_{ABC} := \delta_{CD} f^D_{AB}$, are totally anti-symmetric, $f_{[ABC]}=f_{ABC}$, and that $f_{abc}$ defines a nearly parallel $G_2$-structure on $SO(5)/SO(3)_{\text{max}}$. We set $P_{abc} = \sqrt{30} f_{abc}$ to ensure the correct normalization such that \eqrangeref{PQrelns1}{PQrelns4} are satisfied. In the terminology of \secref{Spin7inst}, that means $\sigma=\sqrt{30}$. Moreover, we note that this coset space satisfies~\eqref*{YM_struct_const_formula} with $\alpha = 1/5$.

\paragraph{Instanton and Yang-Mills equations.}
In order to study the instanton and Yang-Mills equations, we first need to solve the $G$-invariance condition~\eqref*{Ginvcond_gaugefield}, where now $G=SO(5)$. We find $X_a = \phi(\tau) I_a$ and hence this case has already been covered by the general analyses in \secref{Spin7inst,YMsol}. In particular, the instanton equation yields, besides constant solutions, the standard tanh-kink-type solutions~\eqref*{kink_sol} with $(\alpha, \sigma) = (1/5, \sqrt{30})$.

Before moving on to the next coset space, we remark that it is not possible to consistently deform the $G_2$-structure on $SO(5)/SO(3)_{\text{max}}$ away from being nearly parallel. Indeed, after solving~\eqref{Ginvcondg} for this example, the only remaining admissible deformation parameter is an overall volume rescaling, which does not affect the nature of the $G_2$-structure.

\subsubsection{Cylinders over the squashed 7-sphere \texorpdfstring{$Sp(2)\times Sp(1)/Sp(1)^2$}{}}\seclabel{SquashedS7}

For the coset space $Sp(2)\times Sp(1)/Sp(1)^2$, we adopt the conventions of~\cite{Bais:1983wc}. We rescale the generators and structure constants according to
\begin{equation}\eqlabel{SquashedS7_rescalings}
  I_A \rightarrow c_A I_A \qquad \text{(no sum over $A$)} \; , \qquad
  f^A_{BC} \rightarrow \sfrac{c_B c_C}{c_A} f^A_{BC} \qquad \text{(no sum over $A, B, C$)} \; .
\end{equation}
In order for~\eqref*{g_CK} to hold, we need to choose $c_a = \pm\sfr32 \sqrt{\sfr35}$, $c_{\hat{i}} = \pm\sfrac{1}{\sqrt{3}}$, $c_{\tilde{i}} = \pm\sfrac{1}{\sqrt{5}}$. Here, we have subdivided the indices $i,j,\ldots$ according to $i=(\hat{i},\tilde{i})=(\{8,9,10\},\{11,12,13\})$, corresponding to the first and second summand in $\mathfrak{h}=\mathfrak{sp}(1)\oplus\mathfrak{sp}(1)$, respectively. The full set of rescaled structure constants is stated explicitly in \appref{struct_consts_SquashedS7}, for reference.

Also for this coset space, the structure constants with all indices lowered, $f_{ABC} := \delta_{CD} f^D_{AB}$, are totally anti-symmetric, $f_{[ABC]}=f_{ABC}$, and $f_{abc}$ defines a nearly parallel $G_2$-structure. We fix $P_{abc} = \sqrt{30} f_{abc}$ to ensure the correct normalization such that \eqrangeref{PQrelns1}{PQrelns4} are satisfied. In the terminology of \secref{Spin7inst}, that means $\sigma=\sqrt{30}$. The rescaled structure constants satisfy~\eqref*{YM_struct_const_formula} with $\alpha = 1/5$. 

The $G$-invariance condition~\eqref*{Ginvcond_gaugefield}, where now $G=Sp(2)\times Sp(1)$, is solved by
\begin{equation}\eqlabel{SquashedS7_X}
 X_{\hat{a}}^{\hat{b}} = \phi_1 (\tau) \delta_{\hat{a}}^{\hat{b}} \; , \qquad\text{and}\qquad
 X_{\tilde{a}}^{\tilde{b}} = \phi_2 (\tau) \delta_{\tilde{a}}^{\tilde{b}} \; ,
\end{equation}
where $\phi_1$, $\phi_2$ are two real-valued scalar fields and the seven-dimensional indices $a,b,\ldots$ are subdivided according to $a=(\hat{a},\tilde{a})=(\{1,2,3,4\},\{5,6,7\})$. The full expressions for the gauge connection $A$ and the corresponding curvature $F$ can consequently be expressed as
\begin{equation}\eqlabel{SquashedS7_A_F}
\begin{aligned}
  A &= e^i I_i + \phi_1 e^{\hat{a}} I_{\hat{a}} + \phi_2 e^{\tilde{a}} I_{\tilde{a}} \; , \\
  F &= \dot{\phi}_1 e^{0\hat{a}} I_{\hat{a}} + \dot{\phi}_2 e^{0\tilde{a}} I_{\tilde{a}} 
  - (\phi_1 - \phi_2) f^{\ah}_{i\bt} e^{i\bt} I_{\ah} + (\phi_1 - \phi_2) f^{\at}_{i\bh} e^{i\bh} I_{\at}
  \\ &+ \sfr12 (\phi_1^2 - 1) f^i_{\ah\bh} e^{\ah\bh} I_i + \sfr12 (\phi_2^2 - 1) f^i_{\at\bt} e^{\at\bt} I_i + (\phi_1 \phi_2 - 1) f^i_{\ah\bt} e^{\ah\bt} I_i
  \\ &+ \sfr12 \phi_1 (\phi_1 - 1) f^{\ch}_{\ah\bh} e^{\ah\bh} I_{\ch} + \sfr12 \phi_2 (\phi_2 - 1) f^{\ct}_{\at\bt} e^{\at\bt} I_{\ct}
  \\ &+ \sfr12 (\phi_1^2 - \phi_2) f^{\ct}_{\ah\bh} e^{\ah\bh} I_{\ct} + \sfr12 (\phi_2^2 - \phi_1) f^{\ch}_{\at\bt} e^{\at\bt} I_{\ch}
  \\ &+ \phi_1 (\phi_2 - 1) f^{\ch}_{\ah\bt} e^{\ah\bt} I_{\ch} + \phi_2 (\phi_1 - 1) f^{\ct}_{\ah\bt} e^{\ah\bt} I_{\ct} \; .
\end{aligned}
\end{equation}

\paragraph{Instanton equation.}
We now insert $f^A_{BC}$, $P_{abc}$ and $X_a^b$ into the instanton equation~\eqref*{insteqSpin7cosetX}. The quiver relations reduce to the quadratic constraint
\begin{equation}
 \phi_1^2 = \phi_2^2 \; .
\end{equation}
With $\phi_1 = \pm \phi_2 \equiv \pm\phi$, the first equation in~\eqref*{insteqSpin7cosetX} again yields the standard tanh-kink-type equation~\eqref*{kink_ODE} and solution~\eqref*{kink_sol} with $(\alpha, \sigma) = (1/5, \sqrt{30})$.

It is possible to deform the nearly parallel $G_2$-structure on $Sp(2)\times Sp(1)/Sp(1)^2$ in an $Sp(2)\times Sp(1)$-invariant fashion. Solving~\eqref{Ginvcondg}, we find the deformed quantities $\tilde{g}_{G/H}$ and $\tilde{P}$,
\begin{equation}
 \tilde{g}_{G/H} = R_1^2 \delta_{\hat{a}\hat{b}} e^{\hat{a}} \otimes e^{\hat{b}} + R_2^2 \delta_{\tilde{a}\tilde{b}} e^{\tilde{a}} \otimes e^{\tilde{b}} \; , \qquad\text{and}\qquad
 \tilde{P}_{\hat{a}\hat{b}\tilde{c}} = R_1^2 R_2 P_{\hat{a}\hat{b}\tilde{c}} \; , \quad
 \tilde{P}_{\tilde{a}\tilde{b}\tilde{c}} = R_2^3 P_{\tilde{a}\tilde{b}\tilde{c}} \; ,
\end{equation}
with two real deformation parameters $R_1$, $R_2$. The deformed $G_2$-structure satisfies~\eqrangeref*{PQrelns1}{PQrelns3}, after replacing $P\to\tilde{P}$, $Q\to\tilde{Q}$, $\delta_{ab} \to (\tilde{g}_{G/H})_{ab}$. The expressions for the exterior derivatives of the deformed $G_2$-structure forms are given by
\begin{equation}
 \dd\tilde{P} = \tilde{\tau}_0 \tilde{Q} + \tilde{\ast}_7 \tilde{\tau}_3 \; , \qquad
 \dd\tilde{Q} = 0 \; ,
\end{equation}
with $\tilde{Q} = \tilde{\ast}_7 \tilde{P}$ and
\begin{equation}
 \tilde{\tau}_0 = \frac{\sqrt{2}}{9} \frac{5 R_1^2 + R_2^2}{R_1^2 R_2} \; , \qquad
 \tilde{\tau}_3 = \frac{5\sqrt{2}}{9} \frac{R_2^2}{R_1^2} (R_1^2 - R_2^2) e^{567} \; .
\end{equation}
The symbol $\tilde{\ast}_7$ denotes the seven-dimensional Hodge star operator with respect to the deformed metric $\tilde{g}_{G/H}$. In the general classification of $G_2$-structures, this is called a cocalibrated (or semi-parallel) $G_2$-structure (see \secref{G2_structures}). For the special case $R_1 = R_2 \equiv R$, this reduces to the original nearly parallel $G_2$-structure,
\begin{equation}
 \tilde{\tau}_0 \big|_{R_1 = R_2 \equiv R} = \frac{2\sqrt{2}}{3 R} \; , \qquad
 \tilde{\tau}_3 \big|_{R_1 = R_2 \equiv R} = 0 \; ,
\end{equation}
up to an irrelevant volume rescaling. This bears a striking resemblance to the analogous case in one dimension lower, namely the deformed nearly K\"ahler structure on the six-dimensional coset space $Sp(2)/SU(2)\times U(1)$ considered, for example, in~\cite{Chatzistavrakidis:2009mh,Klaput:2011mz}.

In the deformed case, the quiver relations imply
\begin{equation}\eqlabel{SquashedS7_deformed_quivrel}
  R_1^2 \phi_1^2 - R_2^2 \phi_2^2 = R_1^2 - R_2^2 \; .
\end{equation}
On the other hand, the differential equation in~\eqref*{insteqSpin7cosetX} turns into
\begin{equation}\eqlabel{SquashedS7_deformed_ODEs}
\begin{aligned}
  \dot{\phi}_1 &= \sfrac{\sqrt{2}}{3} R_1^2 R_2 \phi_1 ( \phi_2 - 1 ) \; , \\
  \dot{\phi}_2 &= \sfrac{\sqrt{2}}{9} R_2 \left( 2 R_1^2 ( \phi_1^2 - \phi_2 ) + R_2^2 ( \phi_2^2 - \phi_2 ) \right) \; .
\end{aligned}
\end{equation}
By differentiating \eqref{SquashedS7_deformed_quivrel} and inserting \eqref{SquashedS7_deformed_ODEs}, we arrive at
\begin{equation}\eqlabel{SquashedS7_deformed_newrel}
  3 R_1^4 \phi _1^2 (\phi _2-1) - R_2^4 \phi _2^2 (\phi _2-1)  = 2 R_1^2 R_2^2 \phi _2 ( \phi _1^2 - \phi _2 ) \; .
\end{equation}
After inserting \eqref{SquashedS7_deformed_quivrel} into \eqref{SquashedS7_deformed_newrel}, we immediately learn that \emph{non-constant} solutions only exist for
\begin{equation}
  R_1 = R_2 \equiv R \; ,
\end{equation}
which corresponds to the original (that is, undeformed) nearly parallel $G_2$ case.

\paragraph{Yang-Mills equation.}
To compute the second-order equations of motion for $\phi_1$ and $\phi_2$, we start with~\eqref*{YMComp} and insert~\eqref*{SquashedS7_A_F}. We follow the philosophy of \secref{YMsol} and choose the same torsion~\eqref*{G_over_H_torsion_kappa}, affine spin connection~\eqref*{G_over_H_spin_connection} and 3-form $H$~\eqref*{YM_T_ansatz}. Restricting to the undeformed nearly parallel $G_2$ case, we then find two differential equations,
\begin{align}
  \ddot{\phi}_1 &= \sfr12 \phi_1^3 + \sfrac{1}{10} \phi_1 \phi_2^2 - \sfrac{\kappa+3}{10} \phi_1 \phi_2 + \sfrac{\kappa-3}{10} \phi_1 \; , \eqlabel{SquashedS7_YM_diffeq1} \\
  \ddot{\phi}_2 &= \sfrac{7}{15} \phi_2^3 + \sfrac{2}{15} \phi_1^2 \phi_2 - \sfrac{\kappa+3}{30} (2 \phi_1^2 +  \phi_2^2 ) + \sfrac{\kappa-3}{10} \phi_2 \; , \eqlabel{SquashedS7_YM_diffeq2}
\end{align}
and one algebraic equation,
\begin{equation}\eqlabel{SquashedS7_YM_algeq}
  (\phi_1^2 - \phi_2^2) (\kappa + 3 - 2\phi_2) = 0 \; .
\end{equation}
As a non-trivial consistency check, we remark that the same set of equations can also be obtained by inserting~\eqref*{SquashedS7_X} into eq.~(2.27) of ref.~\cite{Bauer:2010fia}.

\Eqref{SquashedS7_YM_algeq} has two branches of solutions, first $\phi_2 = (\kappa+3)/2$ and second $\phi_1 = \pm \phi_2 \equiv \pm\phi$. In the first case, \eqref{SquashedS7_YM_diffeq2} implies $\kappa\in\{-1, -3, -6 \}$ and \eqref{SquashedS7_YM_diffeq1} turns into,
\begin{equation}\eqlabel{rescaled_phi4_kink1}
  \ddot{\phi}_1 = \sfr12 \phi_1 \left( \phi_1^2 - c_\kappa \right) \; ,
\end{equation}
with $c_\kappa := (\kappa^2 + 2\kappa + 21)/20$. This can be integrated to $\dot{\phi}_1 = \pm \sfr12 \left( c_\kappa - \phi_1^2 \right)$, with solutions
\begin{equation}\eqlabel{rescaled_phi4_kink2}
  \phi_1 (\tau) = \pm \sqrt{c_\kappa} \tanh\left[ \sfrac{\sqrt{c_\kappa}}{2} (\tau - \tau_0) \right] \; .
\end{equation}
For the choice $\kappa=-1$, we have $c_{-1} = 1$ and we recover the $\phi^4$ kink/anti-kink solution of \secref{YMsol} (see~\eqref*{YM_2nd_order_phi4_kink_eq}).
For the other two choices, we have $c_{-3} = 6/5$, $c_{-6} = 9/4$ and the solutions are rescaled $\phi^4$ kinks/anti-kinks.

The second branch of solutions of \eqref{SquashedS7_YM_algeq} corresponds to $\phi_1 = \pm \phi_2 \equiv \pm\phi$. \Eqrangeref{SquashedS7_YM_diffeq1}{SquashedS7_YM_diffeq2} then become
\begin{equation}\eqlabel{SquashedS7_YM_1field}
  \ddot{\phi} = \sfrac{3}{5} \phi (\phi - 1) (\phi - \sfrac{\kappa - 3}{6}) \; .
\end{equation}
For the choice $\kappa=6$, we can integrate the equation to recover the standard tanh-kink-type instanton equation~\eqref*{kink_ODE} and solution~\eqref*{kink_sol} with $(\alpha, \sigma) = (1/5, \sqrt{30})$. Another case which can also be solved analytically is $\kappa=15$. \Eqref{SquashedS7_YM_1field} can then be integrated to $\dot{\phi} = \pm \sqrt{\sfrac{3}{10}} \phi (\phi - 2)$, which is solved by
\begin{equation}
  \phi (\tau) = 1 \mp \tanh \left[ \sqrt{\sfrac{3}{10}} (\tau - \tau_0) \right] \; .
\end{equation}

\subsubsection{Cylinders over Aloff-Wallach spaces \texorpdfstring{$SU(3)/U(1)_{k,l}$}{}}\seclabel{AW}

It is possible to define a $U(1)$ subgroup of $SU(3)$ given by matrices of the form
\begin{equation}\eqlabel{U_kl_generator}
 \begin{pmatrix}
\exp(\im(k+l)\vp)&0&0\\0&\exp(-\im k\vp)&0\\0&0&\exp(-\im l\vp)\end{pmatrix}\ ,
\end{equation}
where $0\le\vp\le 2\pi$. The integers $k$, $l$ parameterize the embedding of $U(1)$ into $SU(3)$ and as a mnemonic we write $U(1)_{k,l}$. For relatively prime integers $k$ and $l$, the coset spaces $SU(3)/U(1)_{k,l}$ are simply connected manifolds known as Aloff-Wallach spaces.

One can show that it is always possible to choose an orthonormal coframe $\{\et^a\}$ on $U\subset SU(3)/U(1)_{k,l}$ together with a connection one-form $\et^{8}$ defined as the dual of the generator $\It_8$ (a rescaled version of~\eqref*{U_kl_generator}) of the group $U(1)_{k,l}$, such that the coset space metric $g_{G/H} = \delta_{ab}\,\et^a \otimes \et^b$ is Einstein for a connection with a torsion 3-form $P$ given by
\begin{equation}\eqlabel{P_AW}
  P = e^{135} - e^{245} - e^{146} - e^{236} + e^{127} + e^{347} + e^{567} \; .
\end{equation}
Furthermore, the holonomy group of this connection is contained in $G_2$ and the 3-form $P$ defines a $G_2$-structure on $SU(3)/U(1)_{k,l}$. Here, we adopt the conventions\footnote{When comparing expressions in this subsection and in \appref{struct_consts_AW} with those in~\cite{Haupt:2011mg}, it should be kept in mind that we adjusted some notation in order to fit seamlessly into this paper. In particular, our $P$, $\{e^A\}$, $\{I_A\}$, $f^A_{BC}$ and $C^A_{BC}$ correspond to $\psi$, $\{\tilde{e}^A\}$, $\{\tilde{I}_A\}$, $\tilde{f}^A_{BC}$ and $\tilde{C}^A_{BC}$ in~\cite{Haupt:2011mg}, respectively.} of~\cite{Haupt:2011mg}. Note that in this subsection, the orientation is flipped compared to the rest of the paper. Hence, $\varepsilon^{(7)}_{1234567} = -1$ and $P\wedge Q = -7 \, \text{Vol}(G/H)$ (cf.~\eqref*{PQrelns4}), where $Q := \ast_7 P$. 

With the structure constants as defined in \appref{struct_consts_AW}, we find that $P_{abc} \not\propto f_{abc}$, even though the structure constants with all indices lowered, $f_{ABC} := (g_\mathfrak{g})_{CD} f^D_{AB}$, are totally anti-symmetric, $f_{[ABC]}=f_{ABC}$. We also note that generically~\eqref*{YM_struct_const_formula} does not hold in this case. The Killing-Cartan metric is given by
\begin{multline}\eqlabel{AW_gKC}
 (g_\mathfrak{g})_{AB} = - f^C_{AD} f^D_{BC} = 
 12 \zeta _1^2 (\delta_{A1} \delta_{B1} + \delta_{A2} \delta_{B2}) + 12 \zeta _2^2 (\delta_{A3} \delta_{B3} + \delta_{A4} \delta_{B4}) \\ + 12 \zeta _3^2 (\delta_{A5} \delta_{B5} + \delta_{A6} \delta_{B6}) + \sfrac{48 ( k^2-k l+l^2 )}{\Delta ^2 \mu ^2} \delta_{A7} \delta_{B7} \\ + \sfrac{16 ( k^2+k l+l^2 )}{\Delta ^2 \mu ^2} \delta_{A8} \delta_{B8} + \sfrac{16 \sqrt{3} ( l^2 - k^2 )}{\Delta ^2 \mu ^2} \delta_{7(A} \delta_{B)8} \; ,
\end{multline}
and it is used to raise, lower and contract Lie algebra indices. Here, $\Delta^2:=2(k^2+l^2)$ and $\zeta _1$, $\zeta _2$, $\zeta _3$, $\mu$ are four real parameters corresponding to the metric deformations allowed by~\eqref*{Ginvcondg}. They can be incorporated as rescalings of the generators $\{I_A\}$ and 1-forms $\{e^A\}$ as explained in~\cite{Haupt:2011mg}. (Note that for $k=\pm l$, there are four additional metric deformations, which will, however, not be considered in this paper.)

For the exterior derivatives of the structure forms, we find 
\begin{equation}\eqlabel{AW_dP}
  \dd P = \tau_0 \, Q + \ast_7 \tau_3 \; , \qquad\and\qquad \dd Q = 0 \; ,
\end{equation}
where 
\begin{equation}
\begin{aligned}
  \tau_0 &= - (\zeta _1^2 \zeta _2^2+\zeta _3^2 \zeta _2^2+\zeta _1^2 \zeta _3^2)/(\zeta _1 \zeta _2 \zeta _3) \; , \\
  \tau_3 &= [ ( \zeta _2 \zeta _3^{-1} + \zeta _2^{-1} \zeta _3 ) \zeta _1 - 3 \zeta _1^{-1} \zeta _2 \zeta _3 ]  e^{127} + 2 \mu \Delta^{-1} [\zeta _2^2 l - \zeta _3^2 (k+l) ]  e^{127} \\ &
    + [ ( \zeta _1 \zeta _3^{-1} + \zeta _1^{-1} \zeta _3 ) \zeta _2 - 3 \zeta _1 \zeta _2^{-1} \zeta _3 ] e^{347} + 2 \mu \Delta^{-1} [\zeta _1^2 k - \zeta _3^2 (k+l) ] e^{347} \\ &
    + [ ( \zeta _1 \zeta _2^{-1} + \zeta _1^{-1} \zeta _2 ) \zeta _3 - 3 \zeta _1 \zeta _2 \zeta _3^{-1} ] e^{567} + 2 \mu \Delta^{-1} [ \zeta _1^2 k + \zeta _2^2 l ] e^{567} \; .
\end{aligned}
\end{equation}
In general, $P$ thus defines a cocalibrated (or semi-parallel) $G_2$-structure on $SU(3)/U(1)_{k,l}$. However, for all $(k,l)$, it is possible to find an appropriate set of parameters $\s_1,\s_2,\s_3,\mu\in\RR$ such that $\tau_3 = 0$ (see~\cite{Cabrera:1996}). This justifies why we include $SU(3)/U(1)_{k,l}$ in the list of examples of coset spaces with nearly parallel $G_2$-structure. We emphasize however that most of the analysis carried out in this subsection not only applies to the special locus of parameter space where $P$ defines a nearly parallel $G_2$-structure, but also to the general case where $\tau_3 \neq 0$.

We will solve the $G$-invariance condition~\eqref*{Ginvcond_gaugefield}, where now $G=SU(3)$, in a slightly more general fashion as compared to the rest of this paper. Namely, we allow $X_a$ to also contain a piece valued in the Lie algebra $\mathfrak{h}=\mathfrak{u}(1)$. That is, $X_a = X_a^B I_B = X_a^b I_b + X_a^i I_i$. Usually, we set $X_a^i=0$, but in this subsection we consider $X_a^i \neq 0$, thereby ensuring full compatibility with~\cite{Haupt:2011mg}. We then find from the $G$-invariance condition~\eqref*{Ginvcond_gaugefield} eight $\tau$-dependent scalar degrees of freedom, arranged into three complex fields $\phi^\a$ $(\alpha=1,2,3)$ and two real fields $\chi^i$ $(i=7,8)$. The only non-vanishing components of $X_a^B$ are
\begin{equation}\eqlabel{AW_X}
\begin{aligned}
  &X_1^1 = X_2^2 = \Re(\phi^1) \; , \qquad &X_3^3 &= X_4^4 = \Re(\phi^2) \; , \qquad &X_5^5 &= X_6^6 = \Re(\phi^3) \; , \\
  &X_1^2 = -X_2^1 = \Im(\phi^1) \; , \qquad &X_3^4 &= -X_4^3 = \Im(\phi^2) \; , \qquad &X_5^6 &= -X_6^5 = -\Im(\phi^3) \; , \\
  &X_7^7 = \chi^7 \; , \;\; X_7^8 = \chi^8 \; , & &
\end{aligned}
\end{equation}
where $\Re(z):=(z+\bar{z})/2$ and $\Im(z):=(z-\bar{z})/(2i)$ for any $z\in\mathbb{C}$. We use the notation $\bar{(\cdot)}$ and $(\cdot) {}^\ast$ interchangeably to denote complex conjugation. Due to the complex nature of the fields  $\phi^\a$, it is beneficial to distinguish between upper and lower field indices in this subsection. This distinction is not necessary for the other coset spaces considered in this paper.

The structure of~\eqref*{AW_X} suggests to consider a complexified version of the Lie algebra $\mathfrak{g}=\mathfrak{su}(3)$, namely $\mathfrak{su}(3) \otimes\mathbb{C}$. As a basis for this Lie algebra, we take
\begin{equation}
\begin{aligned}
 &\J_1 :=\sfrac12(\It_{1}-\im\It_{2})\, ,\  & &\J_2 :=\sfrac12(\It_{3}-\im\It_{4})\, ,\ & &\J_3 :=\sfrac12(-\It_{5}-\im\It_{6})\, ,\\
 &\J_{\1} :=\sfrac12(\It_{1}+\im\It_{2})\, ,\ &  &\J_{\2} :=\sfrac12(\It_{3}+\im\It_{4})\, ,\ & &\J_{\3} :=\sfrac12(-\It_{5}+\im\It_{6})\, ,\\
 &\J_7:=-\im\It_{7}\, ,\ & &\J_8:=-\im\It_{8}\ .
\end{aligned}
\end{equation}
We denote the first set of three complex generators by $\{\J_\a\}_{\a=1,2,3}$, the second set of three complex generators by $\{\J_{\bar{\a}}\}_{\bar{\a}=\1,\2,\3}$, the third set of two generators by $\{\J_i\}_{i=7,8}$ and the whole set by $\{\J_A\}$. Note that, contrary to the rest of the paper, now the index $i$ runs not only over the $\mathfrak{h}$-directions (here, $i=8$), but also includes the value $i=7$. The complex generators satisfy commutation relations of the form $[\J_A, \J_B] = \Ct_{AB}^C \J_C$. The explicit expressions for the structure constants $\Ct^A_{BC}$ in the complex basis are listed in \appref{struct_consts_AW}, for completeness. Our real basis $\{I_A\}$ is chosen such that $\J_{\ab} = - \J_\a^\dagger$ and $\J_i^\dagger = \J_i^\top = \J_i^\ast = \J_i$.

The dynamical matrix $X_a (\tau)$ can be expanded in the complex basis as
\begin{equation}
 Y_1 :=\sfrac{1}{2} (X_1-\im \, X_2),  \quad 
 Y_2 :=\sfrac{1}{2} (X_3-\im \, X_4),\quad
 Y_3 :=\sfrac{1}{2} (-X_5-\im \, X_6), \quad 
 Y_{\ab} = - Y_\a^\dagger, \and X_7 \; .
\end{equation}
After solving the $G$-invariance condition~\eqref*{Ginvcond_gaugefield}, we find
\begin{equation}
 Y_{\a}=\phi^\a\J_{\a}\ ,\quad  Y_{\ab}=\bar{\phi}^{\ab}\J_{\ab}\ , \quad X_7=i\chi^7\J_7 + i\chi^8\J_8\ , \qquad\text{(no sum over $\a$, $\ab$)}\; ,
\end{equation}
which is the complex version of~\eqref*{AW_X}. Here, we introduced the notation $\bar{\phi}^{\ab} := (\phi^\a)^\ast$ to denote the complex conjugated field.

\paragraph{Instanton equation.}
This case has already been analyzed in~\cite{Haupt:2011mg}. For completeness, we briefly review the key results relevant to the present paper. From the instanton equation~\eqref*{insteqSpin7cosetX}, one obtains a set of coupled first-order ordinary differential equations,
\begin{equation}\eqlabel{AW_inst_eq_scalar_fields}
\begin{aligned}
\dot\phi^1&=(2\Ct^{\1}_{2\3}+\Ct^{\1}_{7\1}-\chi^7\Ct^{\1}_{7\1}-\chi^8\Ct^{\1}_{8\1})\phi^1-2
\Ct^{\1}_{2\3}\bar\phi^{\2}\phi^{3}\ ,\\
\dot\phi^2&=(2\Ct^{\2}_{\31}+\Ct^{\2}_{7\2}-\chi^7\Ct^{\2}_{7\2}-\chi^8\Ct^{\2}_{8\2})\phi^2-2
\Ct^{\2}_{\31}\bar\phi^{\1}\phi^{3}\ ,\\
\dot\phi^3&=(2\Ct^{\3}_{\1\2}-\Ct^{\3}_{7\3}+\chi^7\Ct^{\3}_{7\3}+\chi^8\Ct^{\3}_{8\3})\phi^3-2
\Ct^{\3}_{\1\2}\phi^{1}\phi^{2}\ ,\\
\dot \chi^7&=2\Ct^7\chi^7 -2\Ct^7_{1\1}|\phi^1|^2-2\Ct^7_{2\2}|\phi^2|^2
+2\Ct^7_{3\3}|\phi^3|^2\ ,\\
\dot \chi^8&=2\Ct^8+2\Ct^7\chi^8-2\Ct^8_{1\1}|\phi^1|^2-2\Ct^8_{2\2}|\phi^2|^2
+2\Ct^8_{3\3}|\phi^3|^2\ ,
\end{aligned}
\end{equation}
where $\Ct^i:=\Ct^i_{1\1}+\Ct^i_{2\2}-\Ct^i_{3\3}$.

The set of equations~\eqref*{AW_inst_eq_scalar_fields} can be succinctly expressed by means of a superpotential $W$, schematically
\begin{equation}\eqlabel{insteqs_spot}
	\dot{\phi}^\a = K^{\a\bb} \frac{\partial W}{\partial \bar{\phi}^{\bb}} \; , \qquad\qquad 
	\dot{\chi}^i = -2 K^{ij} \frac{\partial W}{\partial \chi^j} \; ,
\end{equation}
where $K^{AB}$ is the inverse of $K_{AB} := - \tr(J_A J_B)$, which is the Killing-Cartan metric in the complex basis. The superpotential $W$ is a cubic function of the scalar fields $\{\phi^{\a}, \chi^i\}$,
\begin{multline}\eqlabel{AW_spot}
 W = \zeta _1^2 ( 2 \Ct_{2\3}^\1 + \Ct_{7\1}^\1 ) |\phi^1|^2 + \zeta _2^2 ( 2 \Ct_{\3 1}^\2 + \Ct_{7\2}^\2 ) |\phi^2|^2 + \zeta _3^2 ( 2 \Ct_{\1\2}^\3 - \Ct_{7\3}^\3 ) |\phi^3|^2 \\ 
   - 2 \zeta _1 \zeta _2 \zeta _3 (\phi^1 \phi^2 \bar{\phi}^{\3}+\bar{\phi}^{\1} \bar{\phi}^{\2} \phi^3) - ( \zeta _1^2 \Ct_{i\1}^\1 |\phi^1|^2 + \zeta _2^2 \Ct_{i\2}^\2 |\phi^2|^2 - \zeta _3^2 \Ct_{i\3}^\3 |\phi^3|^2 ) \chi^i \\
   - \Ct^8 K_{8i} \chi^i - \sfrac{1}{2} \Ct^7 K_{ij} \chi^i \chi^j \; .
\end{multline}

\Eqref{AW_inst_eq_scalar_fields} is a complicated system of coupled, non-linear first-order ordinary differential equations and finding the general solution is a considerable task. Nonetheless, a particularly simple yet important special solution is readily found:
\begin{equation}\eqlabel{AW_chi8only_inst}
\begin{aligned}
 &\phi^1 = \phi^2 = \phi^3 = \chi^7 = 0 \; ,\\
 &\chi^8 (\tau) = \begin{cases}
                   -\frac{\Ct^8}{\Ct^7} + c_1\,e^{2 \Ct^7 \tau} & \text{if $\Ct^7 \neq 0$} \; ,\\
                   2 \Ct^8\,\tau + c_2 & \text{if $\Ct^7 = 0$} \; .
                  \end{cases}
\end{aligned}
\end{equation}
Here, $c_1,c_2\in\mathbb{R}$ are constants of integration. For $\Ct^7 \neq 0$, $c_1=0$, this solution is stationary and corresponds to the abelian (rescaled, if $\Ct^8\neq 0$) canonical connection on a line bundle over $SU(3)/U(1)_{k,l}$. The same is also true for $\Ct^7 = \Ct^8 = 0$ with the rescaled canonical connection corresponding to the case $c_2\neq 0$. This is arguably the simplest example of a $G_2$ instanton on $SU(3)/U(1)_{k,l}$.

Another special case, which allows a more thorough analysis, is the choice $k=l=1$. In addition, we fix $\s_1^2 = \s_2^2 = 2 \s_3^2 = 2\a^2$, $\mu=1/\a^2$, where $\a\in\RR$ is a residual free parameter.\footnote{The free parameter $\alpha$ introduced here should not be confused with the constant $\alpha$ used in~\eqref*{YM_struct_const_formula}.} The instanton equation then becomes
\begin{equation}\eqlabel{3.20}
	2\a^2\dot\phi^1 = \frac{\partial W}{\partial \bar\phi^\1} \; , \;\;\;
	2\a^2\dot\phi^2 = \frac{\partial W}{\partial \bar\phi^\2} \; , \;\;\;
	\a^2\dot\phi^3 = \frac{\partial W}{\partial \bar\phi^\3} \; , \;\;\;
	\a^4\dot\chi^7 = \frac{\partial W}{\partial \chi^7} \; , \;\;\;
	\a^4\dot\chi^8 = \frac{\partial W}{\partial \chi^8} \; ,
\end{equation}
with superpotential
\begin{multline}\eqlabel{spot11}
	\a^{-3} W = 2(2 - \a) ( |\phi^1|^2 + |\phi^2|^2 ) + 2(2 + \a) |\phi^3|^2 - 4 ( \phi^1 \phi^2 \bar{\phi}^{\3} + \bar{\phi}^{\1} \bar{\phi}^{\2} \phi^3 ) \\ + 2 \a ( |\phi^1|^2 + |\phi^2|^2 - |\phi^3|^2 )\chi^7 - 2 \sqrt{3} \a ( |\phi^1|^2 - |\phi^2|^2 )\chi^8 - \a ((\chi^7)^2 + (\chi^8)^2 ) .
\end{multline}
When searching for the critical points of $W$, one may use its symmetries to argue that it suffices to consider the case $\phi^\a \in \RR$, $\phi^1 \geq 0$ and $\phi^2 \geq 0$, without loss of generality. 

As an example, we consider the case $\a=\pm 2$. In this case, $SU(3)/U(1)_{1,1}$ admits a cocalibrated (or semi-parallel) $G_2$-structure. In other words, $P$ satisfies~\eqref*{AW_dP} with $\tau_0 \neq 0$ and $\tau_3 \neq 0$. The two signs correspond to the choice of orientation. For $\a = +2$, the critical points of $W$ are listed in the following table:
\begin{center}
\begin{tabular}[h]{c|c|c|c|c|c|c}
$\phi^1$& $\phi^2$	&	$\phi^3$	&	$\chi^7$	&	$\chi^8$&	Eigenvalues of Hessian & $W\vphantom{\frac{\frac{1}{2}}{2}}$ \\\hline
$0$	& $0$		&	$0$		&	$0$		&	$0$	&	$(+,-,-,0,0)$ & $0$ \\
$1$	& $1$		&	$\pm 1$		&	$1$		&	$0$ 	&	$(+,+,-,-,-)$ & $16$ \\
\end{tabular}
\end{center}
On the other hand, for $\a = -2$, the table of critical points of $W$ is given by:
\begin{center}
\begin{tabular}[h]{c|c|c|c|c|c|c}
$\phi^1$	& $\phi^2$	& $\phi^3$	& $\chi^7$	& $\chi^8$ 	& Eigenvalues of Hessian & $W\vphantom{\frac{\frac{1}{2}}{2}}$ \\\hline
$0$	& $0$	&	$0$	&	$0$	&	$0$	&	$(-,-,-,-,0)$ & $0$ \\
$1/\sqrt{2}$ & $0$ &	$0$	&	$1/2$	&	$-\sqrt{3}/2$ &	$(+,-,-,-,-)$ & $-16$ \\
$0$ & $1/\sqrt{2}$ &	$0$	&	$1/2$	&	$\sqrt{3}/2$ &	$(+,-,-,-,-)$ & $-16$ \\
$c_+$	& $c_-$ & 	$\pm 2/3$&	$1/6$	&	$-\sqrt{35}/6$ & $(+,+,-,-,-)$ & $-496/27$ \\
$(9 c_+)^{-1}$	& $c_+$ & 	$\pm 2/3$&	$1/6$	&	$\sqrt{35}/6$ & $(+,+,-,-,-)$ & $-496/27$ \\
$2\sqrt{2}/3$ & $2\sqrt{2}/3$ & $\pm 2/3$ & $4/3$ &	$0$	&	$(+,+,-,-,-)$ & $-1280/27$ \\
$1$	& $1$	&	$\pm 1$	&	$1$	&	$0$	&	$(+,+,+,-,-)$ & $-48$ \\
\end{tabular}
\end{center}
where $c_\pm = \sqrt{11\pm\sqrt{105}}/6$. There is a kink solution for $\a = -2$ flowing from $W = -16$ at $\tau=-\infty$ to $W = 0$ as $\tau\rightarrow\infty$. The numerical solution for this case is shown in \figref{flow_eqs_minussqrt2_ex1}.
\begin{figure}[t]
	\centering
	\includegraphics[width=0.75\textwidth,trim=20 35 20 50,clip]{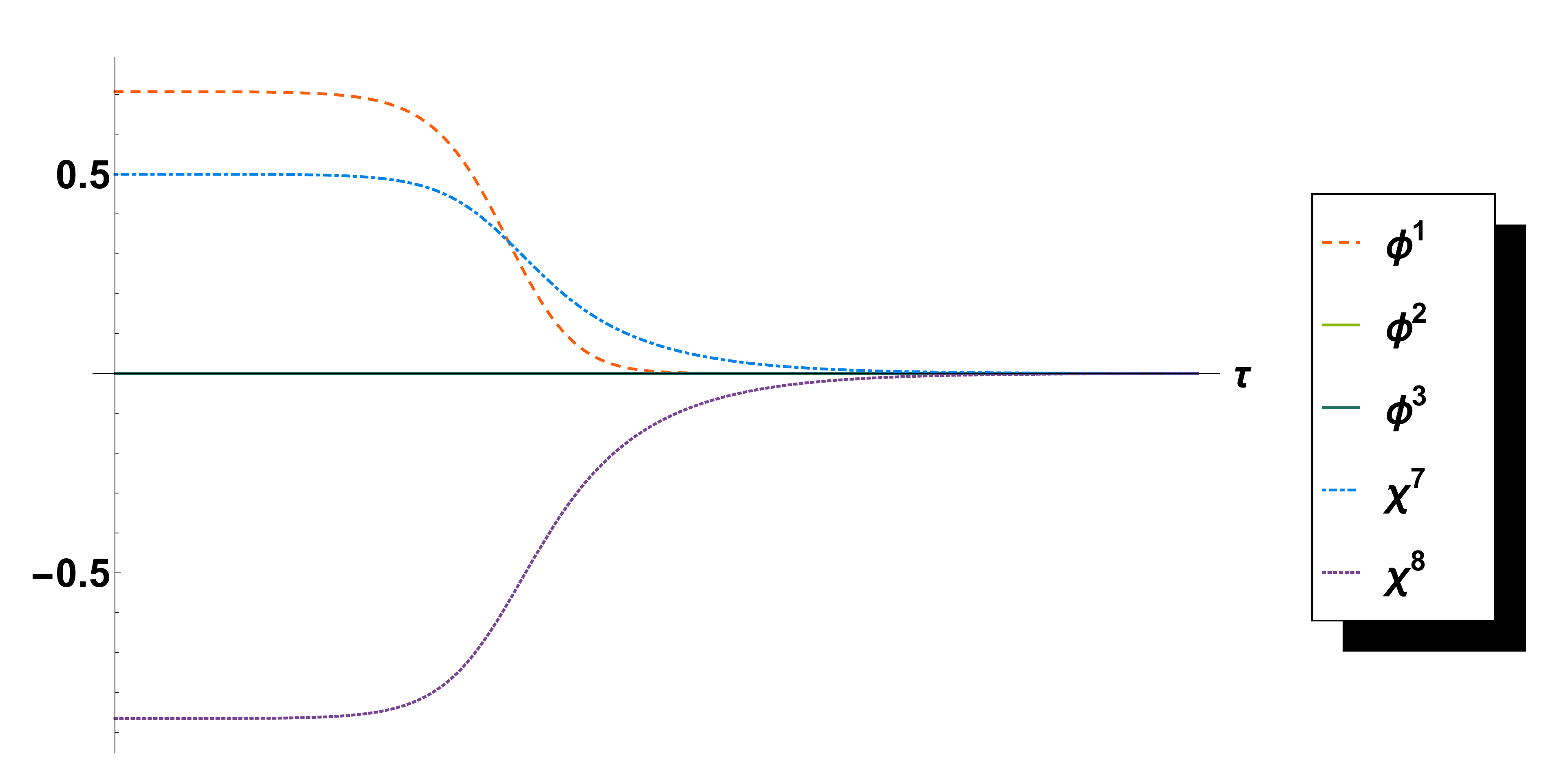}
	\caption{Numerical kink solution on $\RR \times SU(3)/U(1)_{1,1}$ for $\a = -2$ flowing from $W = -16$ at $\tau=-\infty$ to $W = 0$ as $\tau\rightarrow\infty$. $\phi^2$ and $\phi^3$ are zero everywhere and thus their plot coincides with the $\tau$-axis. (Plot adapted from~\cite{Haupt:2011mg}.)}
	\figlabel{flow_eqs_minussqrt2_ex1}
\end{figure}
The shape of these curves resembles that of a hyperbolic tangent type kink. The maximal deviation of a fitted hyperbolic tangent from the numerical solution is of the order of $2 \%$.

\paragraph{Yang-Mills equation.}
The second-order equations of motion for $\{\phi^{\a}, \chi^i\}$ can be worked out from~\eqref*{YMComp}, taking over the ans\"atze~\eqrangeref*{G_over_H_torsion_kappa}{G_over_H_spin_connection} for $T$ and $\omega$ from \secref{YMsol}. For $H$ we choose
\begin{equation}\eqlabel{AW_H_ansatz}
 H = - \lambda \, \ast\dd\ast Q_\M = - \lambda \, \ast_7 \dd P \; ,
\end{equation}
with $Q_\M = \Psi = P \wedge\dd\tau -Q$ and $\lambda\in\RR$ being some new free parameter. This ansatz for $H$ differs from~\eqref*{YM_T_ansatz} and is a consequence of $P_{abc}\not\propto f_{abc}$. This is a precursor of the situation encountered in \secref{G_over_H_SU3}, where the appropriate choice of $H$ will be discussed and motivated further.

We now insert~\eqrangeref*{G_over_H_torsion_kappa}{G_over_H_spin_connection},~\eqref*{AW_H_ansatz} together with~\eqref*{Aansatz},~\eqref*{Fansatz} and~\eqref*{AW_X} into~\eqref*{YMComp}. First, we obtain the Gauss-law constraint~\eqref*{gauss_law}, which now becomes,
\begin{align}
 & \zeta _1^2 k ( \dot{\phi}^1 \bar{\phi}^\1 - \phi^1 \dot{\bar{\phi}}^\1 ) + \zeta _2^2 l ( \dot{\phi}^2 \bar{\phi}^\2 - \phi^2 \dot{\bar{\phi}}^\2 ) + \zeta _3^2 (k+l) ( \dot{\phi}^3 \bar{\phi}^\3 - \phi_3 \dot{\bar{\phi}}^\3 ) = 0 \; , \\
 & \zeta _1^2 l ( \dot{\phi}^1 \bar{\phi}^\1 - \phi^1 \dot{\bar{\phi}}^\1 ) - \zeta _2^2 k ( \dot{\phi}^2 \bar{\phi}^\2 - \phi^2 \dot{\bar{\phi}}^\2 ) + \zeta _3^2 (l-k) ( \dot{\phi}^3 \bar{\phi}^\3 - \phi^3 \dot{\bar{\phi}}^\3 ) = 0 \; .
\end{align}
Taking suitable linear combinations yields
\begin{equation}\eqlabel{gauss_law_AW}
 ( \dot{\phi}^3 \bar{\phi}^\3 - \phi^3 \dot{\bar{\phi}}^\3 ) = - \sfrac{\zeta _1^2}{\zeta _3^2} ( \dot{\phi}^1 \bar{\phi}^\1 - \phi^1 \dot{\bar{\phi}}^\1 ) = - \sfrac{\zeta _2^2}{\zeta _3^2} ( \dot{\phi}^2 \bar{\phi}^\2 - \phi^2 \dot{\bar{\phi}}^\2 ) \; .
\end{equation}

In addition, we obtain three complex and two real second-order equations expressible as a gradient system of the form,
\begin{equation}\eqlabel{AW_eom}
 \ddot{\phi}^\a = K^{\a\bb} \frac{\partial V}{\partial \bar{\phi}^{\bb}} \; , \qquad\qquad
 \ddot{\chi}^i = - 2 K^{ij} \frac{\partial V}{\partial \chi^j} \; ,
\end{equation}
for some potential $V$ defined below. These are the Euler-Lagrange equations derived from the action~\eqref*{YMaction}, with the replacement $Q_\M \to \lambda Q_\M$. Indeed, the action $S$ in this case turns into 
\begin{equation}
 S = -24 \, \mathrm{Vol}(G/H) \int\limits_{-\infty}^{\infty} \mathcal{E} \,\dd\tau \; .
\end{equation}
The energy density $\mathcal{E} = T+V$ comprises the kinetic energy $T= K_{\a\bb} \dot{\phi}^\a \dot{\bar{\phi}}^{\bb} - \sfrac{1}{4} K_{ij} \dot{\chi}^i \dot{\chi}^j$ and the potential $V$ given by
\begin{align}
 V &= V_0 + V_1 + V_2 + V_3 + V_4 \; , \nonumber \\
 V_0 &= -\left[ (\Ct^8_{1\1})^2 + (\Ct^8_{2\2})^2 + (\Ct^8_{3\3})^2 + 2\lambda( \Ct^8_{1\1} \Ct^8_{2\2} - \Ct^8_{1\1} \Ct^8_{3\3} - \Ct^8_{2\2} \Ct^8_{3\3} ) \right] K_{88} \; , \nonumber \\
 V_1 &= -2 \left[ \Ct^7_{1\1} \Ct^8_{1\1} + \Ct^7_{2\2} \Ct^8_{2\2} + \Ct^7_{3\3} \Ct^8_{3\3} + 2 \lambda ( \Ct^{(7}_{1\1} \Ct^{8)}_{2\2} - \Ct^{(7}_{1\1} \Ct^{8)}_{3\3} - \Ct^{(7}_{2\2} \Ct^{8)}_{3\3} ) \right] K_{8i} \chi^i \; , \nonumber \\
 V_2 &=+ \left[ 2 \zeta _2^2 \zeta _3^2 (1+\lambda) + 4 \zeta _1 \zeta _2 \zeta _3 \lambda  \Ct^\1_{7\1} + \zeta _1^2 \left((\Ct^\1_{7\1})^2 - 2 \Ct^\1_{8\1} (\Ct^8_{1\1} + \lambda \Ct^8_{2\2} - \lambda \Ct^8_{3\3} )\right) \right] |\phi^1|^2 \nonumber \\ & \quad
 + \left[ 2 \zeta _1^2 \zeta _3^2 (1+\lambda) + 4 \zeta _1 \zeta _2 \zeta _3 \lambda  \Ct^\2_{7\2} + \zeta _2^2 \left((\Ct^\2_{7\2})^2 - 2 \Ct^\2_{8\2} (\lambda \Ct^8_{1\1} + \Ct^8_{2\2} - \lambda \Ct^8_{3\3} )\right) \right] |\phi^2|^2 \nonumber \\ & \quad
 + \left[ 2 \zeta _1^2 \zeta _2^2 (1+\lambda) - 4 \zeta _1 \zeta _2 \zeta _3 \lambda \Ct^\3_{7\3} + \zeta _3^2 \left( (\Ct^\3_{7\3})^2 + 2 \Ct^\3_{8\3} (\lambda \Ct^8_{1\1} + \lambda \Ct^8_{2\2} - \Ct^8_{3\3} ) \right) \right] |\phi^3|^2 \nonumber \\ & \quad
 -\left[ (\Ct^7_{1\1})^2 + (\Ct^7_{2\2})^2 + (\Ct^7_{3\3})^2 + 2 \lambda ( \Ct^7_{1\1} \Ct^7_{2\2} - \Ct^7_{1\1} \Ct^7_{3\3} - \Ct^7_{2\2} \Ct^7_{3\3} ) \right] K_{ij} \chi^i \chi^j \; , \nonumber \\
 V_3 &= -2 (1+\lambda) ( \zeta _1^2 \zeta _2^2 + \zeta _1^2 \zeta _3^2 + \zeta _2^2 \zeta _3^2 ) ( \bar{\phi }^{\bar{1}} \bar{\phi }^{\bar{2}} \phi^3 + \phi^1 \phi^2 \bar{\phi }^{\bar{3}} ) \nonumber \\ & \quad
 + \left[ 2 \zeta_1^2 \Ct^\1_{i\1} \chi^i (\Ct^1_{71} - \Ct^7_{1\1} - \lambda \Ct^7_{2\2} + \lambda \Ct^7_{3\3} + 2 \lambda \Ct^1_{3\2}) |\phi^1|^2 \right. \nonumber \\ & \quad\;\;\left.
        + 2 \zeta_2^2 \Ct^\2_{i\2} \chi^i (\Ct^2_{72} - \lambda \Ct^7_{1\1} - \Ct^7_{2\2} + \lambda \Ct^7_{3\3} - 2 \lambda \Ct^2_{3\1}) |\phi^2|^2 \right. \nonumber \\ & \quad\;\;\left.
        + 2 \zeta_3^2 \Ct^\3_{i\3} \chi^i (\Ct^3_{73} + \lambda \Ct^7_{1\1} + \lambda \Ct^7_{2\2} - \Ct^7_{3\3} + 2 \lambda \Ct^3_{12}) |\phi^3|^2 \right] \; , \nonumber \\ 
 V_4 &= 2 \zeta _1^4 |\phi^1|^4 + 2 \zeta _2^4 |\phi^2|^4 + 2 \zeta _3^4 |\phi^3|^4 + 2 \zeta _1^2 \zeta _2^2 |\phi^1|^2 |\phi^2|^2 + 2 \zeta _1^2 \zeta _3^2 |\phi^1|^2 |\phi^3|^2 + 2 \zeta _2^2 \zeta _3^2 |\phi^2|^2 |\phi^3|^2 \nonumber \\ & \quad +\left[ \zeta _1^2 (\Ct^\1_{7\1} \chi^7 + \Ct^\1_{8\1} \chi^8 )^2 |\phi^1|^2 + \zeta _2^2 ( \Ct^\2_{7\2} \chi^7 + \Ct^\2_{8\2} \chi^8 )^2 |\phi^2|^2 + \zeta _3^2 ( \Ct^\3_{7\3} \chi^7 + \Ct^\3_{8\3} \chi^8 )^2 |\phi^3|^2 \right] \; .
\end{align}
A full analysis of this large system of equations, for generic $k$, $l$, $\zeta _1$, $\zeta _2$, $\zeta _3$, $\mu$ and $\lambda$, is beyond the scope of the present paper. However, we shall now discuss some special cases, where simplifications occur and thus analytical methods apply.

First, we study the case $\lambda = 1$. For this value of $\lambda$, it is possible to write $V$ as the square of the superpotential $W$ defined in~\eqref*{AW_spot},
\begin{equation}\eqlabel{pot_spot}
 V = K^{\a\bb} W_{\a} W_{\bb} - K^{ij} W_i W_j \; ,
\end{equation}
where $W_\a = \partial W / \partial \phi^\a$, $W_{\bb} = \partial W / \partial \bar{\phi}^{\bb}$ and $W_i = \partial W / \partial \chi^i$. The gradient of $V$ thus becomes
\begin{equation}\eqlabel{pot_spot_grad}
\begin{aligned}
  V_\a &= K^{\b\bar\gamma} (W_{\a\b} W_{\bar\gamma} + W_{\a\bar\gamma} W_{\b}) - 2 K^{ij} W_{\a i} W_j \; , \\
  V_i &= K^{\a\bb} (W_{i\a} W_{\bb} + W_{i\bb} W_{\a}) - 2 K^{jk} W_{ij} W_k \; .
\end{aligned}
\end{equation}
From this and~\eqref*{pot_spot} we learn that critical points of the superpotential are both zeros and critical points of the potential. We are interested in solutions with finite total energy, 
\begin{equation}
 E = \int\limits_{-\infty}^\infty {\cal E}\, \dd\tau = \int\limits_{-\infty}^\infty (T+V)\, \dd\tau < \infty \; . 
\end{equation}
Such solutions necessarily interpolate between zero-potential critical points. For $\lambda = 1$, they are critical points of the superpotential, and the second-order equations of motion~\eqref*{AW_eom} can be integrated to the first-order instanton equations~\eqref*{insteqs_spot}. Hence, we have reduced this case to the construction of instanton solutions discussed in~\cite{Haupt:2011mg} and reviewed above.\footnote{This result fits well into the general analysis recently carried out in~\cite{Chen:2015bek}, where it was carefully examined under which circumstances the first-order system $\dot{\phi} = \nabla W$ is equivalent to the second-order system $\ddot{\phi} = \nabla V$ with $V=||\nabla W||^2 / 2$.}

Second, we consider the single-field reduction $\phi^1 = \phi^2 = \phi^3 = \chi^7 = 0$. The only remaining degree of freedom is $\chi^8$ and the Gauss-law constraint~\eqref*{gauss_law_AW} is trivially satisfied. The equations of motion~\eqref*{AW_eom} collapse to a single linear equation,
\begin{equation}\eqlabel{AW_chi8only}
 \ddot{\chi}^8 = A \chi^8 + B \; ,
\end{equation}
where $A = 4 [ (\Ct^7_{1\1})^2 + (\Ct^7_{2\2})^2 + (\Ct^7_{3\3})^2 + 2 \lambda ( \Ct^7_{1\1} \Ct^7_{2\2} - \Ct^7_{1\1} \Ct^7_{3\3} - \Ct^7_{3\3} \Ct^7_{2\2} ) ]$ and $B = 4 [ \Ct^7_{1\1} \Ct^8_{1\1} + \Ct^7_{2\2} \Ct^8_{2\2} + \Ct^7_{3\3} \Ct^8_{3\3} + 2 \lambda (\Ct^{(7}_{1\1} \Ct^{8)}_{2\2} - \Ct^{(7}_{1\1} \Ct^{8)}_{3\3} - \Ct^{(7}_{2\2} \Ct^{8)}_{3\3} ) ]$. The solution reads
\begin{equation}
 \chi^8 (\tau) = \begin{cases}
                   -\frac{B}{A} + c_1\,e^{\sqrt{A}\,\tau} + c_2\,e^{-\sqrt{A}\,\tau}  & \text{if $A \neq 0$} \; ,\\
                   \frac{B}{2}\,\tau^2 + c_1\,\tau + c_2 & \text{if $A = 0$} \; .
                  \end{cases}
\end{equation}
Here, $c_1,c_2\in\RR$ are constants of integration. For $\lambda=1$, the coefficients $A$ and $B$ factorize, $A = 4 (C^7)^2$, $B = 4 C^7 C^8$, and~\eqref*{AW_chi8only} can be integrated to
\begin{equation}
 \dot{\chi}^8 = \pm 2 (C^7 \chi^8 + C^8) \; .
\end{equation}
This corresponds to the instanton case, whose solution was given in~\eqref*{AW_chi8only_inst}. We remark that the only finite-energy solution is $\chi^8 = \text{const.}$, which corresponds to the abelian (rescaled, if $\Ct^8\neq 0$) canonical connection on a line bundle over $SU(3)/U(1)_{k,l}$. This requires either $C^7 = C^8 = 0$ or $C^7 \neq 0$, $c_1=0$.

Our last example is the choice $k=l=1$, together with $\s_1^2 = \s_2^2 = 2 \s_3^2 = 2\a^2$, $\mu=1/\a^2$, $\a = -2$, $\lambda = 3/2$, $\phi^2 = \phi^3 = \chi^8 = 0$, $\chi^7 = 11/4$ and $\phi^1$ real. This solves the Gauss-law constraint~\eqref*{gauss_law_AW}, and the equations of motion~\eqref*{AW_eom} reduce to
\begin{equation}
 \ddot{\phi}^1 = A\, \phi^1 \left( (\phi^1)^2 - B \right) \; ,
\end{equation}
where $A = 32$ and $B = 25/32$. This equation is a rescaled version of~\eqref*{rescaled_phi4_kink1}. It can be integrated to $\dot{\phi}^1 = \pm \sqrt{\sfrac{A}{2}} \left( B - (\phi^1)^2 \right)$, with solutions
\begin{equation}\eqlabel{AW_rescaled_phi4_kink}
  \phi^1 (\tau) = \pm \sqrt{B} \tanh\left[ \sqrt{\sfrac{AB}{2}} (\tau - \tau_0) \right] \; ,
\end{equation}
describing rescaled $\phi^4$ kinks ($+$) and anti-kinks ($-$), respectively.

\section[Seven-dimensional coset spaces with \texorpdfstring{$SU(3)$}{SU(3)}-structure]{Seven-dimensional coset spaces\\ with \texorpdfstring{$SU(3)$}{SU(3)}-structure}\seclabel{G_over_H_SU3}

Another class of coset spaces well-suited for the methods developed in this paper are seven-dimensional coset spaces admitting an $SU(3)$-structure. This scenario is the subject of the present section.

\subsection{\texorpdfstring{$SU(3)$}{SU(3)}-structure in seven dimensions}\seclabel{SU3_struct}

$SU(3)$-structure is often studied on \emph{six}-dimensional manifolds (see for example~\cite{Grana:2005jc} for a review). Important examples are Calabi-Yau three-folds, which have vanishing intrinsic torsion and thus $SU(3)$ holonomy.

In this section, we consider compact \emph{seven}-dimensional coset spaces $G/H$ with $SU(3)$-structure. Following the useful references~\cite{DallAgata:2003ir,Behrndt:2005im,Micu:2006ey}, we will review here some key properties of this slightly unusual set-up, in order to equip us with the necessary knowledge to study the instanton equation and the Yang-Mills equation with torsion on (cylinders over) these spaces. 

Under the branching $SO(7) \to SU(4) \to SU(3)$, the space of seven-dimensional 1-, 2- and 3-forms, $\Lambda^1$, $\Lambda^2$, $\Lambda^3$,  decomposes according to:
\begin{center}
\begin{tabular}{c|c|l}
   & $SO(7)$ & $SU(3)$ \\\hline
  $\Lambda^1$ & ${\bf 7}$ & ${\bf 1}\oplus{\bf 3}\oplus\bar{{\bf 3}}$ \\
  $\Lambda^2$ & ${\bf 21}$ & ${\bf 1}\oplus2({\bf 3}\oplus\bar{{\bf 3}})\oplus{\bf 8}$ \\
  $\Lambda^3$ & ${\bf 35}$ & $3({\bf 1})\oplus2({\bf 3}\oplus\bar{{\bf 3}})\oplus{\bf 6}\oplus\bar{{\bf 6}}\oplus{\bf 8}$
\end{tabular}
\end{center}
An $SU(3)$-structure in seven dimensions is thus characterized by an $SU(3)$-invariant real 1-form $V$, an $SU(3)$-invariant real 2-form $J$ and an $SU(3)$-invariant complex 3-form $\Omega=\Omega^+ + i\Omega^-$. The three singlets in the decomposition of $\Lambda^3$ are the real ($\Omega^+$) and imaginary ($\Omega^-$) parts of $\Omega$, as well as $V\wedge J$, which is not independent.

The appearance of $V$ is a crucial difference compared to six-dimensional $SU(3)$-structures. The invariant 1-form (or rather, its dual vector field) does not define a Killing direction, unless all torsion classes vanish, in which case the manifold has $SU(3)$ holonomy and is hence a direct product of a Calabi-Yau three-fold with a circle. The 1-form $V$ does however provide a foliation of the seven-dimensional manifold by a six-dimensional base-manifold, denoted $X_6$.

The forms $(V,J,\Omega)$ satisfy seven-dimensional $SU(3)$-structure relations given by
\begin{equation}\eqlabel{SU3_structure_relations}
\begin{aligned}
  & J\wedge J\wedge J = \frac{3i}{4} \Omega\wedge \bar\Omega \; ,\\
  & \Omega\wedge J = V\lrcorner J = V \lrcorner \Omega = 0 \; .
\end{aligned}
\end{equation}
Here, we introduced the symbol $\lrcorner$ to denote the interior product of differential forms on $G/H$. It is defined as $\omega\lrcorner\eta := \ast_7 (\ast_7 \omega\wedge\eta)$. It is possible to show that a number of additional useful relations hold, namely
\begin{eqnarray}
  V \lrcorner V &=& 1 \nonumber \; , \\
  J^a {}_{c} J^c {}_{b}&=&-\delta^a_b+V^a V_b \nonumber \; ,\\
  J_a {}^{d} \Omega_{\pm dbc}&=&\mp\Omega_{\mp abc} \; , \eqlabel{SU3_structure_relations_extra} \\
  \ast_7 \Omega^{\pm} &=& \pm \Omega^{\mp} \wedge V  \; , \nonumber \\
  \ast_7 \left( J\wedge V \right) &=& \sfr12 J \wedge J \; . \nonumber
\end{eqnarray}

The intrinsic torsion $T^0_{ab} {}^c$ decomposes as follows under $SU(3)$,
\begin{equation}
\begin{aligned}
 T^0_{ab} {}^c \in \Lambda^1 \otimes \mathfrak{su}(3)^\perp &= ({\bf 1}\oplus{\bf 3}\oplus\bar{{\bf 3}}) \otimes ({\bf 1}\oplus 2({\bf 3}\oplus\bar{{\bf 3}})) \\
 & = 5 ({\bf 1}) \oplus 5 ({\bf 3}\oplus\bar{{\bf 3}}) \oplus 2 ({\bf 6}\oplus\bar{{\bf 6}}) \oplus 4 ({\bf 8}) \; .
\end{aligned}
\end{equation}
From this we can read off the torsion classes. The five singlets are conveniently arranged into one real and two complex 0-forms, denoted $R$ (real) and ${\cal W}_1$, $E$ (both complex), respectively. The five 1-forms are denoted $V_{1,2}$, $W_0$ and ${\cal W}_{4,5}$. The four 2-forms are arranged into two real ones, denoted $T_{1,2}$, and a complex one, denoted ${\cal W}_2$. Finally, there are two 3-forms, denoted ${\cal W}_3$ and $S$. 

When arranged in such a way, the exterior derivatives of the forms $(V,J,\Omega)$ can be succinctly expressed in terms of these torsion classes~\cite{DallAgata:2003ir,Behrndt:2005im,Micu:2006ey},
\begin{eqnarray}
  \dd V &=& R \, J + \bar{V}_1 \lrcorner \; \Omega + V_1 \lrcorner\;\bar{\Omega} + T_1  + V \wedge W_0 \, , \eqlabel{SU3_dV}\\
  \dd J &=& \sfrac{3 i}{4}\left(\bar {\cal W}_1\, \Omega -  {\cal W}_1\, \bar{\Omega}\right) + {\cal W}_3 + J\wedge {\cal W}_4 \nonumber \\
           &+& V \wedge \left[\sfrac13 (E + \bar{E})  J + \bar{V}_2 \lrcorner\; \Omega + V_2 \lrcorner\; \bar{\Omega} + T_2\right] \, , \eqlabel{SU3_dJ}\\
  \dd \Omega &=& {\cal W}_1  J\wedge J +  J\wedge {\cal W}_2 + \Omega\wedge {\cal W}_5 + V \wedge \left(E\, \Omega - 4\, J \wedge V_2 + S\right) \; . \eqlabel{SU3_dOmega}
\end{eqnarray}
The numerical coefficients in these expressions are fixed by demanding compatibility with the seven-dimensional $SU(3)$-structure relations~\eqref*{SU3_structure_relations}. We remark that the torsion classes have been arranged such that the subset $\{ {\cal W}_1, \ldots, {\cal W}_5 \}$ precisely equals the set of torsion classes of a six-dimensional $SU(3)$-structure (see, for example,~\cite{Grana:2005jc}). This can be understood as the $SU(3)$-structure that fixes the geometry of the six-dimensional base-manifold $X_6$ corresponding to the foliation defined by $V$.

To make contact with the previous section, we note that $SU(3)\subset G_2$ and thus an $SU(3)$-structure automatically implies the existence of a $G_2$-structure on the seven-dimensional manifold. Actually, an $SU(3)$-structure on a seven-dimensional manifold can be used to define \emph{two independent} $G_2$-structures via
\begin{equation}\eqlabel{G2_SU3_struct_forms_rel}
  P^\pm = \pm \Omega^- - J \wedge V \; .
\end{equation}
The intersection of the two $G_2$-structures $P^\pm$ is precisely the $SU(3)$-structure. Henceforth, we will mostly be working with the $G_2$-forms $P^\pm$, instead of the $SU(3)$-forms $(V,J,\Omega)$. It should be kept in mind though that the two formulations are equivalent.

\subsection{Explicit coset space constructions}\seclabel{SU3structure_cosets}

To proceed further, we will now consider cylinders over explicit seven-dimensional coset spaces $G/H$ that admit an $SU(3)$-structure. Specifically, we shall examine four cases, namely cylinders over $SO(5)/SO(3)_{A+B}$, over $N^{pqr} = (SU(3)\times U(1)) / (U(1) \times U(1))$, over $M^{pqr} = (SU(3)\times SU(2)\times U(1)) / (SU(2)\times U(1) \times U(1))$ and over $Q^{pqr} = (SU(2)\times SU(2)\times SU(2)) / (U(1) \times U(1))$, where $(p,q,r)$ is a triple of mutually co-prime integers.

\subsubsection{Cylinders over \texorpdfstring{$SO(5)/SO(3)_{A+B}$}{SO5/SO3 subscript (A+B)}}\seclabel{SO5_SO3_AB}

The coset space $SO(5)/SO(3)$ has already been studied in \secref{SO5_SO3_max}. However, as mentioned there, the group $SO(5)$ has two commuting $SO(3)$ subgroups, denoted $SO(3)_A$ and $SO(3)_B$. These subgroups may be embedded in various ways into $SO(5)$ and when quotiented out, this leads to different coset spaces. In this subsection we consider the case where the group $H$ appearing in the quotient $G/H$ is a linear combination of the two commuting $SO(3)$'s.

As a starting point, we use the conventions of~\cite{Micu:2006ey} in order to define the coset space $SO(5)/SO(3)_{A+B}$. Although the structure constants are not written down explicitly, they can be extracted straightforwardly from eq. (D.12) in~\cite{Micu:2006ey}. We work with rescaled generators and structure constants, $I_A \rightarrow \sfrac{1}{\sqrt{6}} I_A$, $f^C_{AB} \rightarrow \sfrac{1}{\sqrt{6}} f^C_{AB}$, to ensure the correct normalization of~\eqref*{g_CK}. Our rescaled structure constants are listed in \appref{struct_consts_SO5_SO3_AB} for completeness.

The structure constants with all indices lowered, $f_{ABC} := \delta_{CD} f^D_{AB}$, are totally anti-symmetric, $f_{[ABC]}=f_{ABC}$, as before. However, we will now encounter a crucial difference in comparison to the coset spaces studied in \secref{SO5_SO3_max, SquashedS7}, namely that $P_{abc} \propto f_{abc}$ does not lead to a well-defined $G_2$-structure on $SO(5)/SO(3)_{A+B}$. This assertion can be verified \emph{inter alia} by computing the so-called \emph{associated metric} (see for example~\cite{Bryant:1987,Hitchin:2000,Grigorian:2015,Reidegeld:2010}),
\begin{equation}\eqlabel{g_P}
  (g_P)_{ab} = B_{ab} \det(B)^{-1/9} \; , \qquad\quad\text{with}\quad B_{ab} = - \sfrac{1}{144} P_{a c_1 c_2} P_{b c_3 c_4} P_{c_5 c_6 c_7} \varepsilon^{(7)} {}^{c_1 \ldots c_7} \; .
\end{equation}
For $P_{abc} \propto f_{abc}$ this metric is singular, that is $\det g_P = 0$.

To resolve this problem, recall that there is an $SU(3)$-structure on this coset space. Hence, we may instead use~\eqref*{G2_SU3_struct_forms_rel} to define a $G_2$-structure. On $SO(5)/SO(3)_{A+B}$ and with the correct normalization such that \eqrangeref{PQrelns1}{PQrelns4} are satisfied, we then find
\begin{equation}\eqlabel{SO5_SO3_P}
\begin{aligned}
  P^\pm &= \pm \sfrac{1}{\sqrt{2}} \Omega^- - \sqrt{6} f \; , \qquad\quad &\text{with} \\
  \Omega^- &:= e^{123} - e^{127} + e^{136} - e^{167} - e^{235} + e^{257} - e^{356} + e^{567} \; , \qquad\quad &\text{and} \\
  f &:= \sfrac{1}{3!} f_{abc} e^{abc} = - \sfrac{1}{\sqrt{6}} (e^{145} + e^{246} + e^{347}) \; . &
\end{aligned}
\end{equation}
The corresponding 4-form $Q^\pm := \ast_7 P^\pm$ is explicitly given by
\begin{multline}\eqlabel{SO5_SO3_Q}
  Q^\pm = \pm \sfrac{1}{\sqrt{2}} \left( e^{1234} - e^{1247} + e^{1346} - e^{1467} - e^{2345} + e^{2457} - e^{3456} + e^{4567} \right) \\ + \left( e^{1256} + e^{1357} + e^{2367} \right) \; .
\end{multline}
Note that it is closed, $\dd Q^\pm = 0$. The exterior derivative of $P^\pm$ can be written as,
\begin{equation}\eqlabel{SO5_SO3_dP}
  \dd P^\pm = \tau_0 Q^\pm + \ast_7 \tau_3^\pm \; ,
\end{equation}
with $\tau_0 = - \sqrt{2/3}$ and $\tau_3^\pm = \pm (-\sqrt{3}/6) \Omega^-$. This shows that the $SU(3)$-structure induces two cocalibrated (or semi-parallel) $G_2$-structures on $SO(5)/SO(3)_{A+B}$.

The induced Spin$(7)$-structure on the cylinder is fixed by $\Psi^\pm = P^\pm \wedge\dd\tau - Q^\pm$, according to~\eqref*{PsiPQ}. What kind of Spin$(7)$-structure does this define? We find $\Theta \propto \dd\tau \neq 0$, $\dd\Psi^\pm \neq 0$ and $\dd\Psi^\pm + \sfr17 \Theta\wedge\Psi^\pm \neq 0$. Therefore, this is a general Spin$(7)$-structure where both torsion classes $W_8$ and $W_{48}$ are turned on.

The $G$-invariance condition~\eqref*{Ginvcond_gaugefield}, where now $G=SO(5)$, is solved by
\begin{equation}\eqlabel{SO5_SO3_AB_X}
\begin{aligned}
  &X_1^1 = X_2^2 = X_3^3 = \phi_1 (\tau) \; , \qquad X_4^4 = \phi_2 (\tau) \; , \qquad &X_5^5 = X_6^6 = X_7^7 = \phi_3 (\tau) \; , \\
  &X_1^5 = X_2^6 = X_3^7 = \phi_4 (\tau) \; , \qquad &X_5^1 = X_6^2 = X_7^3 = \phi_5 (\tau) \; .
\end{aligned}
\end{equation}
Hence, we need to deal with \emph{a priori} five real $\tau$-dependent scalar degrees of freedom. The full expressions for the gauge connection $A$ and the corresponding curvature $F$ can be obtained straightforwardly from~\eqref*{Aansatz},~\eqref*{Fansatz} and are rather lengthy. We thus omit them here.

\paragraph{Instanton equation.}
We have now collected all necessary information to compute the system of equations for $\{\phi_1, \ldots,\phi_5\}$ following from the instanton equation~\eqref*{insteqSpin7cosetX}. It turns out that the form of the resulting equations is the same for both choices, $P^+$ or $P^-$. The quiver relations yield
\begin{align}
  \phi _1^2+\phi _4^2 = \phi _3^2+2 \phi _3 \phi _4+2 \phi _1 \phi _5+\phi _5^2 \; , \eqlabel{SO5_SO3_AB_quivrel1} \\
  \phi _1^2+\phi _4^2 = \phi _3^2-2 \phi _3 \phi _4-2 \phi _1 \phi _5+\phi _5^2 \; , \eqlabel{SO5_SO3_AB_quivrel2}
\end{align}
which is solved by $\left(\phi _3,\phi _5\right) = \left(\pm \phi_1, \mp \phi_4\right)$. This effectively reduces the number of degrees of freedom to three, namely $\{\phi_1, \phi_2 ,\phi_4\}$. 

The first-order ordinary differential equations obtained from~\eqref*{insteqSpin7cosetX} are given by
\begin{equation}\eqlabel{SO5_SO3_AB_DE1}
\begin{aligned}
  \dot{\phi _1} &= \sfrac{1}{\sqrt{6}} \phi _1 \left(1 \mp \phi _2\right) \; , \\
  \dot{\phi _4} &= \sfrac{1}{\sqrt{6}} \phi _4 \left(1 \mp \phi _2\right) \; , \\
  \dot{\phi _2} &= \sqrt{\sfrac{3}{2}} \left( \phi _2 \mp \phi _1^2 \mp \phi _4^2 \right) \; ,
\end{aligned}
\end{equation}
where the reduction of the number of degrees of freedom, $\left(\phi _3,\phi _5\right) = \left(\pm \phi_1, \mp \phi_4\right)$, has already been taken into account. Albeit its apparent simplicity,~\eqref*{SO5_SO3_AB_DE1} is still a coupled non-linear system of first-order ordinary differential equations and hence finding the general solution is out of reach with current methods. One may however readily list some special solutions. Besides the trivial case $\phi_1=\phi_2=\phi_4=0$, there are static solutions such as $\phi_1=\phi_2=\pm 1$, $\phi_4 = 0$ and $\phi_1=\phi_4=\pm 1/\sqrt{2}$, $\phi_2 = \pm 1$. A simple non-static solution is $\phi_1=\phi_4=0$, $\phi_2 (\tau) = c \exp(\sqrt{3/2}\, \tau)$ with a constant of integration $c\in\RR$.

\paragraph{Yang-Mills equation.}
We will now turn to the second-order equations of motion for $\{\phi_1, \ldots,\phi_5\}$. Our starting point is the Yang-Mills equation in components~\eqref*{YMComp}. First, we need to choose ans\"atze for the torsion $T$, affine spin connection $\omega$ and 3-form $H$. For $T$ and $\omega$ we take over the expressions from \secref{YMsol}, that is~\eqref*{cylinder_torsion_spin_conn} and
\begin{equation}\eqlabel{SU3_T_ansatz}
 T^a_{bc} = \kappa f^a_{bc} \; , \qquad\qquad \omega^a {}_b = f^a_{ib} e^i + \sfr12 (\kappa+1) f^a_{cb} e^c \; ,
\end{equation}
depending on the real parameter $\kappa\in\RR$.

The ansatz for $H$ is more delicate, however. In most of \secref{G_over_H_G2}, we took $H_{bc}^a = - T_{bc}^a$. Since $H_{bc}^a \propto T_{bc}^a \propto f_{bc}^a \propto P^a {}_{bc}$, this choice ensured that we could recover the instanton solutions, for which it is required that $H \propto P$ (see \fref{H_propto_P}), as special cases of the second-order equations of motion. Here, $P_{abc}\not\propto f_{abc}$ and hence this logic breaks down. Nevertheless, we aim at keeping the property that the instanton solutions can be recovered as special cases. We are thus led to setting
\begin{equation}\eqlabel{SU3_H_ansatz1}
 H = - \ast\dd\ast Q_{\M,\lambda} \; .
\end{equation}
with ${\cal M}=Z(G/H)$ (in this subsection, $G/H = SO(5)/SO(3)_{A+B}$) and $Q_{\M,\lambda} := \lambda Q_\M$, where $\lambda\in\RR$ is some new free parameter. The instanton case is then contained as a special case, namely $\lambda = 1$. After inserting $Q_\M = \Psi^\pm = P^\pm \wedge\dd\tau - Q^\pm$ into~\eqref*{SU3_H_ansatz1} and using~\eqrangeref{SO5_SO3_P}{SO5_SO3_dP}, we obtain\footnote{An alternative way to compute $H$ from~\eqref*{SU3_H_ansatz1} is to use $\ast\,\dd\ast Q_{\M,\lambda} = (1/3!) \nabla^A (Q_{\M,\lambda})_{ABCD} e^{BCD}$ and~\eqref*{SU3_T_ansatz}. This yields $H^\pm = - \lambda \sfrac{\kappa+1}{4} f_{ab}^e (Q^\pm) {}^a {}_{cde} e^{bcd}$, which shows that our choice of $H$ is very similar to those considered in related works (for example, see~\cite{Tormahlen:2014zia} eq.~(3.8)). As a side result, we learn that $\kappa=1$ in order to achieve compatibility with~\eqref*{SU3_H_ansatz2}.}
\begin{equation}\eqlabel{SU3_H_ansatz2}
 H^\pm = - \lambda \ast_7 \dd P^\pm = \lambda \left( \pm \sfrac{\sqrt{3}}{2} \Omega^- - 2 f \right) .
\end{equation}
Note that, although allowed by~\eqref*{SU3_H_ansatz1}, $H^\pm$ has no components with legs in the $\tau$-direction. This is a consequence of $\dd Q^\pm = 0$. 

We now insert the above ans\"atze~\eqref*{SU3_T_ansatz},~\eqref*{SU3_H_ansatz2} together with~\eqref*{Aansatz},~\eqref*{Fansatz} and~\eqref*{SO5_SO3_AB_X} into~\eqref*{YMComp}. First, we obtain \eqrangeref{SO5_SO3_AB_quivrel1}{SO5_SO3_AB_quivrel2} again, provided $\lambda \neq 0$ and irrespective of the choice $P^+$ or $P^-$. Hence, $\left(\phi _3,\phi _5\right) = \left(\pm \phi_1, \mp \phi_4\right)$, as in the case of the instanton equation. Second, we obtain a system of differential equations. With the reduction of the number of degrees of freedom taken into account, the system of differential equations becomes 
\begin{align}
 \ddot{\phi}_1 &= \sfrac{1}{2} \phi_1^3 + \sfrac{1}{6} \phi_1 ( \phi_2^2  + 3 \phi_4^2 ) \mp \sfrac{2\lambda + 3}{6} \phi_1 \phi_2 + \sfrac{2 \lambda - 1}{6} \phi_1 \; , \eqlabel{SO5_SO3_AB_2ndordereqs1} \\
 \ddot{\phi}_4 &= \sfrac{1}{2} \phi_4^3 + \sfrac{1}{6} \phi_4 ( \phi_2^2  + 3 \phi_1^2 ) \mp \sfrac{2\lambda + 3}{6} \phi_4 \phi_2 + \sfrac{2 \lambda - 1}{6} \phi_4 \; , \eqlabel{SO5_SO3_AB_2ndordereqs2} \\
 \ddot{\phi}_2 &= \phi_2 (\phi_1^2 + \phi_4^2) \mp \sfrac{2 \lambda + 3}{2} (\phi_1^2 + \phi_4^2) + \sfrac{2 \lambda + 1}{2} \phi_2 \; , \eqlabel{SO5_SO3_AB_2ndordereqs3} \\
 0 &= \phi_1 \dot{\phi_4} - \dot{\phi_1} \phi_4 \; . \eqlabel{SO5_SO3_AB_2ndordereqs4}
\end{align}
The sign ambiguity in \eqrangeref{SO5_SO3_AB_2ndordereqs1}{SO5_SO3_AB_2ndordereqs3} is not related to the choice $P^+$ or $P^-$, but rather a consequence of the identification $\left(\phi _3,\phi _5\right) = \left(\pm \phi_1, \mp \phi_4\right)$. Note that one may switch between the two sign choices by sending $\phi_2 \to -\phi_2$. Since this does not lead to qualitatively new solutions, we will henceforth ignore the sign choice and restrict to the upper sign, without loss of generality.

\Eqref{SO5_SO3_AB_2ndordereqs4} is a consequence of the Gauss-law constraint~\eqref*{gauss_law}. It is solved by $\phi_4 = c\, \phi_1$ for some $c\in\RR$ (or $\phi_1=0$, which is however equivalent to the case $c=0$, due to the symmetry of \eqrangeref{SO5_SO3_AB_2ndordereqs1}{SO5_SO3_AB_2ndordereqs4} under interchanging $\phi_1 \leftrightarrow \phi_4$). The remaining \eqrangeref{SO5_SO3_AB_2ndordereqs1}{SO5_SO3_AB_2ndordereqs3} form an \emph{a priori} over-determined system of non-linear second-order ordinary differential equations. We refrain from solving the most general case here. Instead, we shall now examine the two obvious special choices $c=0$ and $c=1$.

Upon setting $c=0$, that is $\phi_4=0$, we are left with the following reduced system of differential equations in the two remaining fields $\{\phi_1, \phi_2\}$,
\begin{align}
 \ddot{\phi_1} &= \sfrac{1}{2} \phi_1^3 + \sfrac{1}{6} \phi_1 \phi_2^2 - \sfrac{2\lambda + 3}{6} \phi_1 \phi_2 + \sfrac{2 \lambda - 1}{6} \phi_1 \; , \eqlabel{SO5_SO3_AB_2ndordereqs_reduced1} \\
 \ddot{\phi_2} &= \phi_2 \phi_1^2 - \sfrac{2 \lambda + 3}{2} \phi_1^2 + \sfrac{2 \lambda + 1}{2} \phi_2 \; . \eqlabel{SO5_SO3_AB_2ndordereqs_reduced2}
\end{align}
At this point, it is useful to note that \eqrangeref{SO5_SO3_AB_2ndordereqs_reduced1}{SO5_SO3_AB_2ndordereqs_reduced2} can also be obtained from the action~\eqref*{YMaction} by directly inserting the ansatz and computing the Euler-Lagrange equations. Indeed, the action~\eqref*{YMaction} can then be written as
\begin{equation}\eqlabel{SO5_SO3_AB_action}
 S = - 12\, \mathrm{Vol}(G/H) \, E = - 12\, \mathrm{Vol}(G/H) \, \int\limits_{-\infty}^\infty {\cal E}\, \dd\tau \; ,
\end{equation}
with energy density\footnote{The appearance of the energy density rather than the Lagrangian in~\eqref*{SO5_SO3_AB_action} is due to the fact that our cylinder metric~\eqref*{metric} has Euclidean signature.} ${\cal E}$ given by
\label{SO5_SO3_AB_E_start}
\begin{equation}\eqlabel{SO5_SO3_AB_E}
 {\cal E} = T + V = \sfrac{1}{2} \dot{\phi}_1^2 + \sfrac{1}{12} \dot{\phi}_2^2 + \sfrac{1}{8} \phi_1^4 + \sfrac{2\lambda - 1}{12} \phi_1^2 + \sfrac{1}{12} \phi_1^2 \phi_2^2 - \sfrac{2\lambda+3}{12} \phi_1^2 \phi_2 + \sfrac{2 \lambda + 1}{24} \phi_2^2 + \sfrac{1 - \lambda}{12} \; ,
\end{equation}
plus a total derivative, which is omitted. Here, we used the fact that $\tr(I_A I_B) = f^C_{AD} f^D_{BC} = -\delta_{AB}$, as explained below~\eqref*{g_CK}, to resolve the trace appearing in the action~\eqref*{YMaction}. We also replaced $Q_\M$ by $Q_{\M,\lambda}$ in the action to ensure compatibility with~\eqref*{SU3_H_ansatz1}.

\Eqrangeref{SO5_SO3_AB_2ndordereqs_reduced1}{SO5_SO3_AB_2ndordereqs_reduced2} can be re-written as a gradient system of the form,
\begin{equation}
 \ddot{\phi}_1 = \partial_1 V \; , \qquad\qquad \ddot{\phi}_2 = 6\, \partial_2 V \; ,
\end{equation}
where $\partial_\alpha := \partial / \partial\phi_\alpha$, $\alpha=1,2$, and the potential $V$ is determined in~\eqref*{SO5_SO3_AB_E}. In order to analyze this gradient system further, we study the critical points of $V$, that is field values for which $\partial_\alpha V = 0$ holds. We find seven distinct critical points and list them, together with some important properties, in \tabref{SO5_SO3_AB_critpts}.
\begin{table}[t]\centering
\begin{tabular}{c|c||c|c|c}
  \triplerowheight $(\phi_1, \phi_2)$ & $V |_{\text{crit. pt.}}$ & $\lambda_0$ & $(\phi_1, \phi_2)|_{\lambda_0}$ & $\text{Eigenvalues of Hessian}|_{\lambda_0}$ \\ \hline\hline
  \triplerowheight $(0,0)$ & $(1-\lambda)/12$ & $1$ & $(0,0)$ & $(+,+)$ \\ \hline
  \triplerowheight $(\pm 1,1)$ & $0$ & $\RR$ & $(\pm 1,1)$ & \begin{tabular}{@{}l@{\hskip 1ex}l@{}} \vspace{-2pt}$(+,+)$ & if $c_4^- <\lambda < c_4^+$, \\ \vspace{-2pt}$(+,0)$ & if $\lambda = c_4^\pm$, \\ $(+,-)$ & otherwise \end{tabular} \\ \hline
  \triplerowheight $(\pm c_1^+, c_2^-)$ & $c_3^-$ & \begin{tabular}{c} \vspace{-2pt}$c_4^+$ \\ $-3/2$ \end{tabular} & \begin{tabular}{c} \vspace{-2pt} $(\pm 1, 1)$ \\ $(\pm 1, -1)$ \end{tabular} & \begin{tabular}{c} $(+,0)$ \\ $(+,-)$ \end{tabular} \\ \hline
  \triplerowheight $(\pm c_1^-, c_2^+)$ & $c_3^+$ & $c_4^-$ & $(\pm 1, 1)$ & $(+,0)$ 
\end{tabular}
\caption{Critical points of the potential $V$ and some of their properties for $G/H = SO(5)/SO(3)_{A+B}$ with $\phi_4=0$. In the first two columns we list the values of the critical points and of the potential at the corresponding critical point, respectively. In the right part of the table (columns 3-5), we demand that $V |_{\text{crit. pt.}}\stackrel{!}{=}0$ and solve for $\lambda$. The resulting $\lambda_0$ is shown in column three. Column four contains the values of the critical points with $\lambda=\lambda_0$ inserted. In column five we summarize the signs of the eigenvalues of the Hessian matrix, again with $\lambda=\lambda_0$ inserted. This allows us to draw further conclusions regarding the different types of critical points. Throughout the table, we set $c_1^\pm := \sqrt{(1+2 \lambda ) (-11+2 \lambda \pm \sqrt{73+4 \lambda  (13+\lambda )})/ 24}$, $c_2^\pm := (7+6 \lambda \pm \sqrt{73+4 \lambda  (13+\lambda )})/4$, $c_3^\pm := (827 + 1696 \lambda + 1464 \lambda^2 + 448 \lambda^3 - 16 \lambda ^4 \pm \sqrt{73+52 \lambda +4 \lambda ^2} (73+198 \lambda +108 \lambda ^2+8 \lambda ^3))/2304$ and $c_4^\pm := (1\pm\sqrt{33})/4$.}\tablabel{SO5_SO3_AB_critpts}
\end{table}

Below we are interested in solutions with finite total energy, 
\begin{equation}\eqlabel{energy}
 E = \int\limits_{-\infty}^\infty {\cal E}\, \dd\tau = \int\limits_{-\infty}^\infty (T+V)\, \dd\tau < \infty \; . 
\end{equation}
Such solutions necessarily interpolate between zero-potential critical points. As an example, consider the case $\phi_2 = 1$ and $\lambda=-1/2$. Then, \eqref*{SO5_SO3_AB_2ndordereqs_reduced2} becomes trivial, whereas \eqref*{SO5_SO3_AB_2ndordereqs_reduced1} reduces to \eqref*{YM_2nd_order_phi4_kink_eq}, implying that $\phi_1$ is a $\phi^4$ kink or anti-kink. The solution in this case is a closed-form expression and interpolates between $(\phi_1, \phi_2) = (-1,1)$ and $(1,1)$ as $\tau\to\pm\infty$. These two points in field space are indeed zero-potential critical points, and correspond moreover to local minima of the potential, as can be read off from \tabref{SO5_SO3_AB_critpts}. The total energy in this case attains the value
\begin{equation}\eqlabel{total_energy}
 E = \int\limits_{-\infty}^\infty (\sfrac{1}{2} \dot{\phi}_1^2 + \sfrac{1}{8} (\phi_1^2 - 1)^2)\, \dd\tau = \sfrac{2}{3} \; ,
\end{equation}
as was shown for example in~\cite{Ivanova:2009yi,Manton:2004tk}. A plot of this solution in field space can be found in \figref{SO5_SO3_AB_plot}.
\begin{figure}[t]
	\centering
	\includegraphics[width=0.5\textwidth]{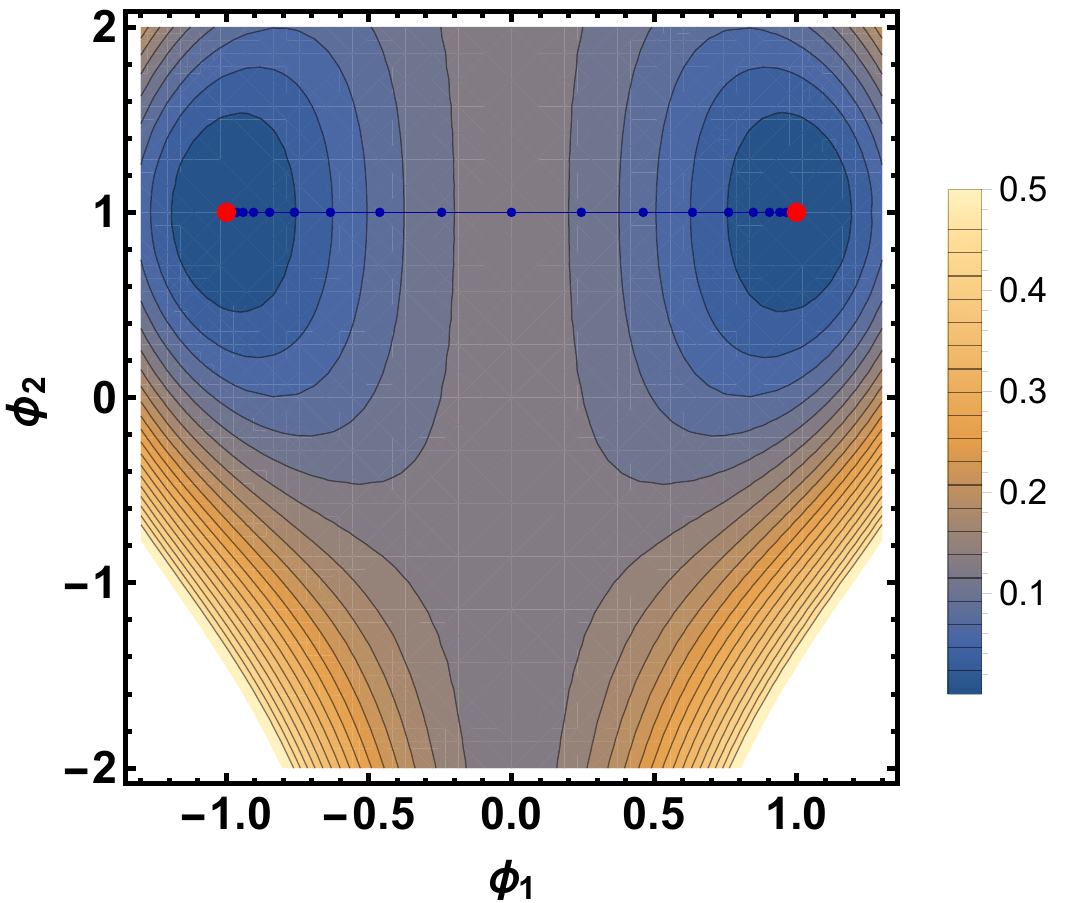}
	\caption{A plot of the potential $V$ for $G/H = SO(5)/SO(3)_{A+B}$ with $\phi_4=0$ and $\lambda=-1/2$. The two red points at $(\phi_1, \phi_2) = (\pm 1,1)$ correspond to the two local minima. The blue line represents the analytical $\phi^4$ kink/anti-kink solution interpolating between the two minima.}
	\figlabel{SO5_SO3_AB_plot}
\end{figure}

Solutions connecting other pairs of zero-potential critical points generally have both $\phi_1$ and $\phi_2$ non-constant, and thus lie beyond the scope of analytical methods. We will not present the respective numerical solutions here, albeit they may be constructed straightforwardly (see~\cite{Bauer:2010fia} for analogous constructions in one dimension lower).
\label{SO5_SO3_AB_E_end}
Another option is to further reduce the number of fields by setting $\phi_2 = \pm \sqrt{3} \phi_1$ and $\lambda = -1$. The single remaining field $\phi_1$ is governed by the following equation of motion,
\begin{equation}\eqlabel{SO5_SO3_AB_specialsol1}
 \ddot{\phi}_1 = \phi_1^3 \mp \sfrac{1}{2 \sqrt{3}} \phi_1^2 - \sfrac{1}{2} \phi_1 \; .
\end{equation}
The corresponding potential $V$ derived from~\eqref*{SO5_SO3_AB_E} \emph{cannot} be written as a square $V(\phi_1) = f(\phi_1)^2$. Hence, \eqref*{SO5_SO3_AB_specialsol1} \emph{cannot} be integrated straightforwardly to a first-order equation $\dot{\phi}_1 = 2 f(\phi_1)$. The potential $V$ has three critical points $0$, $\mp 1/\sqrt{3}$ and $\pm \sqrt{3}/2$, with values $V|_{\phi_1=0} = 1/6$, $V|_{\phi_1=\mp 1/\sqrt{3}} = 1/9$ and $V|_{\phi_1=\pm \sqrt{3}/2} = 1/384$, respectively. Numerical methods yield solutions which display approximate oscillatory behavior, albeit with unbounded amplitude and thus diverging total energy.

Finally, we shall examine the case $c=1$, that is $\phi_4=\phi_1$. The action in this case becomes
\begin{equation}
 S = - 24\, \mathrm{Vol}(G/H) \, \int\limits_{-\infty}^\infty {\cal E}\, \dd\tau \; ,
\end{equation}
with energy density given by
\begin{equation}
 {\cal E} = T+V = \sfrac{1}{2} \dot{\phi}_1^2 + \sfrac{1}{24} \dot{\phi}_2^2 + \sfrac{1}{4} \phi_1^4 + \sfrac{2\lambda - 1}{12} \phi_1^2 + \sfrac{1}{12} \phi_1^2 \phi_2^2 - \sfrac{2\lambda + 3}{12} \phi_1^2 \phi_2 + \sfrac{2\lambda + 1}{48} \phi_2^2 + \sfrac{1-\lambda}{24}\; .
\end{equation}
Up to a field rescaling of the form $\phi_1 \to \phi_1 / \sqrt{2}$, this expression is identical to~\eqref*{SO5_SO3_AB_E}. Hence, the case $c=1$ reduces to the case $c=0$, which has been discussed above.

\subsubsection{Cylinders over \texorpdfstring{$N^{pqr} = (SU(3)\times U(1)) / (U(1) \times U(1))$}{Npqr}}\seclabel{Npqr}

We now turn to studying the coset space $N^{pqr} = (SU(3)\times U(1)) / (U(1) \times U(1))$ and cylinders thereover. For the description of $N^{pqr}$, we will closely follow~\cite{Castellani:1983tc}, and begin with the observation that the group $G=SU(3)\times U(1)$ has a subgroup $H=U(1) \times U(1)$ which can be embedded into $G$ in various ways. The different embeddings are parameterized by three integers $p$, $q$ and $r$. Without loss of generality, they can be taken to be relatively prime.

To proceed further, we adopt the conventions of~\cite{Castellani:1983tc}. We now encounter an important difference compared to the coset spaces studied above. Previously, we used, where possible, an orthonormal basis of $\mathfrak{g}$ (orthonormal in the sense of~\eqref*{g_CK}), which is then neatly compatible with the choice of metric~\eqref*{metric}. In the present case, the Lie algebra $\mathfrak{g}=\mathfrak{su}(3)\oplus \mathfrak{u}(1) \simeq \mathfrak{u}(3)$ is not semisimple and hence, there is no orthonormal basis.

However, there is a basis in which at least the \emph{coset space components} of the Killing-Cartan metric are equal to the Kronecker delta, $(g_\mathfrak{g})_{ab} = \delta_{ab}$. This can be achieved by taking the conventions of~\cite{Castellani:1983tc} with the following rescalings of the generators $I_A$ and structure constants $f^A_{BC}$,
\begin{equation}\eqlabel{Npqr_rescalings}
 I_A \rightarrow \begin{cases} \sfrac{\zeta}{\sqrt{3}\, \eta} I_A & \text{if $A=7$} \\ \sfrac{1}{\sqrt{3}} I_A & \text{if $A\neq 7$} \end{cases} \; , \qquad\qquad
 f^A_{BC} \rightarrow \begin{cases} \sfrac{\eta}{\sqrt{3}\, \zeta} f^A_{BC} & \text{if $A=7$} \\ \sfrac{\zeta}{\sqrt{3}\, \eta} f^A_{BC} & \text{if $B=7$ or $C=7$} \\ \sfrac{1}{\sqrt{3}} f^A_{BC} & \text{otherwise} \end{cases} \; ,
\end{equation}
where $\zeta := \sqrt{3 p^2+q^2+2 r^2}$ and $\eta := \sqrt{3 p^2+q^2}$. We will henceforth work in this basis. The full set of structure constants is listed in \appref{struct_consts_Npqr}, in order for this paper to be self-contained. We emphasize that the cylinder metric is still taken to be~\eqref*{metric}. However, it is necessary to use the full Killing-Cartan metric, 
\begin{equation}\eqlabel{Npqr_gKC}
 (g_\mathfrak{g})_{AB} = - f^C_{AD} f^D_{BC} = 
 \delta_{A=a,B=b} + \sfrac{2 r^2}{\zeta^2} \delta_{A8}\delta_{B8} + \delta_{A9}\delta_{B9} + \sfrac{\sqrt{2} r}{\zeta} (\delta_{A8}\delta_{B9} + \delta_{A9}\delta_{B8}) \; ,
\end{equation}
in order to raise, lower and contract Lie algebra indices. We will highlight places where this is relevant in due course. For example, the structure constants with all indices lowered, $f_{ABC} := (g_\mathfrak{g})_{CD} f^D_{AB}$, are totally anti-symmetric, $f_{[ABC]}=f_{ABC}$.

As in the case $G/H=SO(5)/SO(3)_{A+B}$ studied in the previous subsection, the choice $P_{abc} \propto f_{abc}$ does not lead to a well-defined $G_2$-structure on $N^{pqr}$. Instead, one may proceed by constructing a suitable 3-form $P$, subject to the $G$-invariance condition~\eqref*{GinvcondP} and~\eqref*{PQrelns1}. As solutions, we find\footnote{In principle, there are six additional solutions related to~\eqref*{Npqr_P} by sign flips. However, we demand also that the associated metric $g_P$, as defined in~\eqref*{g_P}, be positive definite. This singles out the two choices $P^+$ and $P^-$ given in~\eqref*{Npqr_P}.}
\begin{equation}\eqlabel{Npqr_P}
  P^\pm = e^{127} \mp e^{136} \pm e^{145} \mp e^{235} \mp e^{246} - e^{347} + e^{567} \; ,
\end{equation}
with corresponding 4-form $Q^\pm := \ast_7 P^\pm$ given by
\begin{equation}\eqlabel{Npqr_Q}
  Q^\pm = e^{1234} - e^{1256} \mp e^{1357} \mp e^{1467} \pm e^{2367} \mp e^{2457} + e^{3456} \; .
\end{equation}
Note that it is closed, $\dd Q^\pm = 0$, whereas the exterior derivative of $P^\pm$ can be expressed as,
\begin{equation}\eqlabel{Npqr_dP}
  \dd P^\pm = \tau_0^\pm Q^\pm + \ast_7 \tau_3^\pm \; ,
\end{equation}
with $\tau_0^\pm = - (\pm 1) \sqrt{3}/2$ and $( 2 \sqrt{3} \zeta^2 )\,\tau_3^\pm = - ( 2 q \eta \pm \zeta^2 ) e^{127} - ( 3 p \eta+q \eta \mp \zeta^2 ) e^{347} - ( 3 p \eta-q \eta \pm \zeta^2 ) e^{567}$. Hence, the $SU(3)$-structure defines two cocalibrated (or semi-parallel) $G_2$-structures on $N^{pqr}$.

The induced Spin$(7)$-structure on the cylinder is fixed by $\Psi^\pm = P^\pm \wedge\dd\tau - Q^\pm$, according to~\eqref*{PsiPQ}. Using that $\tau_0^\pm \neq 0$, $\tau_3^\pm \neq 0$, $\dd\Psi^\pm \neq 0$ and the results of \secref{Spin7_structures}, we immediately conclude that this is a general Spin$(7)$-structure where both torsion classes $W_8$ and $W_{48}$ are turned on.

The $G$-invariance condition~\eqref*{Ginvcond_gaugefield}, where now $G=SU(3)\times U(1)$, is solved by
\begin{equation}\eqlabel{Npqr_X}
\begin{aligned}
  &X_1^1 = X_2^2 = \phi_1 (\tau) \; , &\; &X_3^3 = X_4^4 = \phi_2 (\tau) \; , &\; &X_5^5 = X_6^6 = \phi_3 (\tau) \; , &\; &X_7^7 = \phi_4 (\tau) \; , \\
  &X_1^2 = -X_2^1 = \phi_5 (\tau) \; , &\; &X_3^4 = -X_4^3 = \phi_6 (\tau) \; , &\; &X_5^6 = -X_6^5 = \phi_7 (\tau) \; . &\; &
\end{aligned}
\end{equation}
Hence, the dynamical degrees of freedom are in this case \emph{a priori} seven real $\tau$-dependent scalar fields. By means of~\eqref*{Aansatz},~\eqref*{Fansatz}, it is straightforward to compute the explicit expressions for the gauge connection $A$ and the corresponding curvature $F$.

\paragraph{Instanton equation.}
In order to compute the system of equations for $\{\phi_1, \ldots,\phi_7\}$ imposed by the instanton equation, we insert $f^A_{BC}$, $P^\pm$ and $X_a^b$ into~\eqref*{insteqSpin7cosetX}. Independent of the choice $P^+$ or $P^-$, the quiver relations yield
\begin{align}
  q r ( 2 \phi _1^2-\phi _2^2-\phi _3^2+2 \phi _5^2-\phi _6^2-\phi _7^2 ) + 3 p r ( -\phi _2^2+\phi _3^2-\phi _6^2+\phi _7^2 ) &= 0 \; , \\
  - p ( 2 \phi _1^2-\phi _2^2-\phi _3^2+2 \phi _5^2-\phi _6^2-\phi _7^2 ) + q ( -\phi _2^2+\phi _3^2-\phi _6^2+\phi _7^2 ) &= 0 \; .
\end{align}
This pair of equations has two branches of solutions, $r=0$ and $r\neq 0$. For $r=0$, it becomes $( 2 \phi _1^2-\phi _2^2-\phi _3^2+2 \phi _5^2-\phi _6^2-\phi _7^2 ) = \sfrac{q}{p} ( -\phi _2^2+\phi _3^2-\phi _6^2+\phi _7^2 )$, thereby removing one degree of freedom. For $r\neq 0$, we have two conditions, namely $p ( 2 \phi _1^2-\phi _2^2-\phi _3^2+2 \phi _5^2-\phi _6^2-\phi _7^2 ) = 0$ and $q ( -\phi _2^2+\phi _3^2-\phi _6^2+\phi _7^2 ) = 0$. This removes two degrees of freedom if $pq\neq 0$, and one, otherwise. (Note that $p=q=0$ is excluded since the triple $(p,q,r)$ is pairwise coprime, by assumption.)

The first-order ordinary differential equations derived from~\eqref*{insteqSpin7cosetX} are given by
\begin{align}
 \dot{\phi _1} &= \sfrac{\pm\eta - q}{\sqrt{3} \eta} \phi_1 + \sfrac{q}{\sqrt{3} \eta} \phi_1 \phi_4 \mp \sfrac{1}{\sqrt{3}} (\phi_2 \phi_3 + \phi_6 \phi_7) \; , \\
 \dot{\phi _5} &= \sfrac{\pm\eta - q}{\sqrt{3} \eta} \phi_5 + \sfrac{q}{\sqrt{3} \eta} \phi_4 \phi_5 \pm \sfrac{1}{\sqrt{3}} (\phi_2 \phi_7 - \phi_3 \phi_6) \; , \\
 \dot{\phi _2} &= \sfrac{3 p + q \pm 2 \eta}{2 \sqrt{3} \eta} \phi_2 - \sfrac{3 p + q}{2 \sqrt{3} \eta} \phi_2 \phi_4 \pm \sfrac{1}{\sqrt{3}} (\phi_5 \phi_7 - \phi_1 \phi_3) \; , \\
 \dot{\phi _3} &= \sfrac{-3 p+ q \pm 2 \eta}{2 \sqrt{3} \eta} \phi_3 + \sfrac{3 p - q}{2 \sqrt{3} \eta} \phi_3 \phi_4 \mp \sfrac{1}{\sqrt{3}} (\phi_1 \phi_2 + \phi_5 \phi_6) \; , \\
 \dot{\phi _6} &= \sfrac{3 p + q \pm 2 \eta}{2 \sqrt{3} \eta} \phi_6 - \sfrac{3 p + q}{2 \sqrt{3} \eta} \phi_4 \phi_6 \mp \sfrac{1}{\sqrt{3}} (\phi_3 \phi_5 + \phi_1 \phi_7) \; , \\
 \dot{\phi _7} &= \sfrac{-3 p+ q \pm 2 \eta}{2 \sqrt{3} \eta} \phi_7 + \sfrac{3 p - q}{2 \sqrt{3} \eta} \phi_4 \phi_7 \pm \sfrac{1}{\sqrt{3}} (\phi_2 \phi_5 - \phi_1 \phi_6) \; , \\
 \dot{\phi _4} &= \sfrac{3 p \eta}{2 \sqrt{3} \zeta^2} \left( - \phi_2^2 + \phi_3^2 - \phi_6^2 + \phi_7^2 \right) +\sfrac{q \eta}{2 \sqrt{3} \zeta^2} \left(2 \phi _1^2-\phi _2^2-\phi _3^2+2 \phi _5^2-\phi _6^2-\phi _7^2\right) \; . \eqlabel{Npqr_insteq_phi4}
\end{align}
The sign ambiguity is due to the choice $P^+$ or $P^-$. 

Finding the general solution of this system of equations and constraints is a formidable task, even for fixed values of $(p,q,r)$. One may however study special sub-sectors of field space, where simplifications occur. For example, let us consider the case $\phi \equiv \phi_1 = \phi_2 = \phi_3$, $\phi_4 = 1$, $\phi_5 = \phi_6 = \phi_7 = 0$ and $(p,q,r)$ arbitrary. This solves the quiver relations identically. The differential equations collapse to the standard tanh-kink-type instanton~\eqref*{kink_sol} for the single remaining degree of freedom $\phi$ with $\alpha\sigma = \mp 2/\sqrt{3}$.

\paragraph{Yang-Mills equation.}
Our next goal is to compute the second-order equations of motion for $\{\phi_1, \ldots,\phi_7\}$. We will use the same ans\"atze for the torsion $T$, affine spin connection $\omega$ and 3-form $H$ as in the case $G/H=SO(5)/SO(3)_{A+B}$ presented in \secref{SO5_SO3_AB}. In particular, from~\eqref*{SU3_H_ansatz1} we find the explicit expression
\begin{equation}
  H^\pm = \lambda \left( \pm \sfrac{\sqrt{3}}{2} P^\pm - \tau_3^\pm \right) \; .
\end{equation}
We have now collected all necessary ingredients to compute the equations of motion following from~\eqref*{YMComp}. We first obtain two algebraic constraints,
\begin{multline}\eqlabel{Npqr_EOM_algconstr1}
    4 p \left\{ q \left(2 \zeta ^2 + (1 - \lambda) \eta^2 \right) \mp 2\eta \zeta ^2 \lambda \right\} (\phi_1^2 + \phi_5^2) \\
  + (p-q) \left\{\left(2 \zeta ^2+\eta ^2\right) (3 p+q)-\eta  \lambda  \left(\eta  (3 p+q) \mp 4 \zeta ^2\right)\right\} (\phi_2^2 + \phi_6^2) \\
  - (p+q) \left\{\left(2 \zeta ^2+\eta ^2\right) (3 p-q)-\eta  \lambda  \left(\eta  (3 p-q) \pm 4 \zeta ^2\right)\right\} (\phi_3^2 + \phi_7^2) \\
  - 2 \zeta^2 \left\{ 4 p q (\phi_1^2 + \phi_5^2) + (p-q) (3 p+q) (\phi_2^2 + \phi_6^2) - (p+q) (3 p-q) (\phi_3^2 + \phi_7^2) \right\} \phi_4 = 0 \; ,
\end{multline}
\vspace*{-1.6\baselineskip}
\begin{multline}\eqlabel{Npqr_EOM_algconstr2}
  6 \eta^4  (1 - \lambda) r + 4 q r \left\{\eta  \lambda  \left(\eta  q \pm 2 \zeta ^2\right)-q \left(2 \zeta ^2+\eta ^2\right)\right\} (\phi_1^2 + \phi_5^2) \\ 
  +r (3 p+q) \left\{\eta  \lambda  \left(\eta  (3 p+q) \mp 4 \zeta ^2\right) - \left(2 \zeta ^2+\eta ^2\right) (3 p+q)\right\} (\phi_2^2 + \phi_6^2) \\ 
  +r (3 p-q) \left\{\eta  \lambda  \left(\eta  (3 p-q) \pm 4 \zeta ^2\right) - \left(2 \zeta ^2+\eta ^2\right) (3 p-q)\right\} (\phi_3^2 + \phi_7^2) \\ 
  +2\zeta^2 r \left\{ 4 q^2 (\phi _1^2 + \phi_5^2) + (3 p+q)^2 (\phi_2^2 + \phi_6^2) + (3p-q)^2 (\phi_3^2 + \phi_7^2) \right\} \phi_4 = 0 \; ,
\end{multline}
as well as two first-order constraints,
\begin{align}
            2 p (\dot{\phi _1} \phi _5 - \phi _1 \dot{\phi _5}) + (  p - q) (\dot{\phi _2} \phi _6 - \phi _2 \dot{\phi _6}) - (  p + q) (\dot{\phi _3} \phi _7 - \phi _3 \dot{\phi _7}) = 0 \; , \eqlabel{Npqr_EOM_diffconstr1} \\
  r \left\{ 2 q (\dot{\phi _1} \phi _5 - \phi _1 \dot{\phi _5}) + (3 p + q) (\dot{\phi _2} \phi _6 - \phi _2 \dot{\phi _6}) + (3 p - q) (\dot{\phi _3} \phi _7 - \phi _3 \dot{\phi _7}) \right\} = 0 \; . \eqlabel{Npqr_EOM_diffconstr2}
\end{align}
The first-order equations~\eqrangeref*{Npqr_EOM_diffconstr1}{Npqr_EOM_diffconstr2} originate from the Gauss-law constraint~\eqref*{gauss_law}. Note that~\eqref*{Npqr_EOM_algconstr2,Npqr_EOM_diffconstr2} become trivial for $r=0$. 

In addition, we obtain seven second-order equations expressible as a gradient system of the form,
\begin{equation}\eqlabel{Npqr_gradsys}
  \ddot{\phi}_\alpha = \eta_{\alpha\beta} \frac{\partial V}{\partial \phi_\beta} \; , \qquad\qquad\alpha,\beta=1,\ldots,7 \; ,
\end{equation}
where $\eta_{\alpha\beta} := \diag(1,1,1,2,1,1,1)_{\alpha\beta}$ and the potential $V$ is defined below. This gradient system corresponds to the Euler-Lagrange equations obtained from the action~\eqref*{YMaction} with the reduction to scalar fields,~\eqref*{Npqr_X}, inserted. Indeed, the action $S$ in this case turns into
\begin{equation}
  S = - 4\, \mathrm{Vol}(N^{pqr}) \, \int\limits_{-\infty}^\infty {\cal E}\, \dd\tau \; ,
\end{equation}
where we used $\tr(I_A I_B) = -(g_\mathfrak{g})_{AB}$ and the replacement $Q_\M \to Q_{\M,\lambda}$. The energy density ${\cal E} = T + V$ comprises the standard kinetic term $T = \sfrac{1}{2} \eta^{\alpha\beta} \dot{\phi}_\alpha \dot{\phi}_\beta$, where $\eta^{\alpha\beta}$ is the matrix inverse of $\eta_{\alpha\beta}$, and the potential $V$, given by
\begin{multline}\eqlabel{Npqr_V}
  V = \sfrac{1}{12} ( \phi_1^4 + \phi_2^4 + \phi_3^4 + \phi_5^4 + \phi_6^4 + \phi_7^4 )
  + \sfrac{\lambda + 1}{2} ( \phi _2 \phi _5 \phi _7 - \phi _3 \phi _5 \phi _6 - \phi _1 \phi _2 \phi _3 - \phi _1 \phi _6 \phi _7 ) \\
  + \sfrac{1}{12} ( \phi_6^2 \phi_7^2 + \phi_5^2 \phi_6^2 + \phi_5^2 \phi_7^2 + \phi_1^2 \phi_6^2 + \phi_1^2 \phi_7^2 + \phi_1^2 \phi_2^2 + \phi_1^2 \phi_3^2 + \phi_2^2 \phi_3^2 + \phi_2^2 \phi_5^2 + \phi_2^2 \phi_7^2 + \phi_3^2 \phi_5^2 \\
  + \phi_3^2 \phi_6^2 + 2 \phi_1^2 \phi_5^2 + 2 \phi_2^2 \phi_6^2 + 2 \phi_3^2 \phi_7^2 ) 
  + \sfrac{q^2 }{6 \eta ^2} ( \phi _1^2 + \phi _5^2 ) \phi_4^2 + \sfrac{(3 p+q)^2}{24 \eta ^2} ( \phi _2^2 + \phi _6^2 ) \phi_4^2 + \sfrac{(3 p-q)^2}{24 \eta ^2} ( \phi _3^2 + \phi _7^2 ) \phi_4^2 \\
  + \sfrac{C_1^\pm}{12 \zeta ^4 \eta ^2} (\phi_1^2 + \phi_5^2) + \sfrac{C_2^\pm}{24 \zeta ^2 \eta ^2} ( \phi_2^2 + \phi_6^2 ) + \sfrac{C_3^\pm}{24 \zeta ^2 \eta ^2} ( \phi_3^2 + \phi_7^2 ) + \sfrac{C_4^\pm}{6 \zeta ^2 \eta ^2} ( \phi_1^2 + \phi_5^2 ) \phi_4 \\
  + \sfrac{C_5^\pm}{24 \zeta ^2 \eta ^2} ( \phi_2^2 + \phi_6^2 ) \phi_4 + \sfrac{C_6^\pm}{24 \zeta ^2 \eta ^2} ( \phi_3^2 + \phi_7^2 ) \phi_4 
  + \sfrac{\eta ^4 (1 - \lambda)}{8 \zeta ^4} \phi_4^2 + \sfrac{r^2 \eta^2 (1-\lambda)}{2 \zeta ^4} \phi_4 + \sfrac{(1-\lambda) ( \zeta ^4+4 r^4 )}{8 \zeta ^4} \; .
\end{multline}
Here, we introduced the following coefficients,
\begin{align}
 C_1^\pm &:= \zeta ^4 \left(\eta ^2 (3 \lambda +1)-6 p^2\right)+2 q^2 \left(\zeta ^4-\zeta ^2 \eta ^2 \lambda -4 r^4-2 \eta ^2 r^2\right) \mp 4 \zeta ^4 \eta  \lambda  q \; , \nonumber \\ 
 C_2^\pm &:= 2 \eta ^2 \left(2 \zeta ^2+3 p q-q^2-3 r^2\right)+\eta  \lambda  \left(\eta  \left(3 \zeta ^2-6 p q+2 q^2+6 r^2\right) \right. \nonumber \\ & \left. \pm 4 \zeta ^2 (3 p+q)\right)+2 \zeta ^2 q (3 p-q) \; , \nonumber \\
 C_3^\pm &:= 2 \eta ^2 \left(2 \zeta ^2-3 p q-q^2-3 r^2\right)+\eta  \lambda  \left(\eta  \left(3 \zeta ^2+6 p q+2 q^2+6 r^2\right) \right. \nonumber \\ & \left. \mp 4 \zeta ^2 (3 p-q)\right)-2 \zeta ^2 q (3 p+q) \; , \\
 C_4^\pm &:= q \left(\eta  \lambda  \left(\eta  q \pm 2 \zeta ^2\right)-q \left(2 \zeta ^2+\eta ^2\right)\right) \; , \nonumber \\
 C_5^\pm &:= (3 p+q) \left(\eta  \lambda  \left(\eta  (3 p+q) \mp 4 \zeta^2\right)-(3 p+q) \left(3 \eta ^2+4 r^2\right)\right) \; , \nonumber \\
 C_6^\pm &:= (3 p-q) \left(\eta  \lambda  \left(\eta  (3 p-q) \pm 4 \zeta^2\right)-(3 p-q) \left(3 \eta ^2+4 r^2\right)\right)\; . \nonumber
\end{align}
Note that these coefficients depend on $(p,q,r)$, $\lambda$ and the choice of $P^+$ or $P^-$. We also remark that in order to show the equivalence of the second-order equation for $\phi_4$ derived from~\eqref*{Npqr_gradsys} and the respective equation obtained from~\eqref*{YMComp}, one needs to make us of the algebraic constraints~\eqrangeref*{Npqr_EOM_algconstr1}{Npqr_EOM_algconstr2}.

Instead of further considering this system of equations in full generality, we shall now briefly discuss a special case. Namely, let us consider the ansatz $\phi_1 = \phi_2 = \phi_3$, $\phi_5 = \phi_6 = \phi_7 = 0$ and $\lambda=1$. The system of equations then enforces $\phi_1 = 0$ or $\phi_4 = 1$. In the former case, we find $\phi_4 (\tau) = a\, \tau + b$ and all other $\phi$'s vanishing. This solution is unphysical, since the total energy~\eqref*{energy} is divergent. In the latter case, $\phi_4 = 1$ and we recover the tanh-kink-type instanton solution for $\phi_1$ that has already been found above (see below~\eqref*{Npqr_insteq_phi4}).

\subsubsection{Cylinders over \texorpdfstring{$M^{pqr} = (SU(3)\times SU(2)\times U(1)) / (SU(2)\times U(1) \times U(1))$}{Mpqr}}\seclabel{Mpqr}

This case can be studied in close analogy to $N^{pqr}$, which was presented in the previous subsection. For the sake of brevity, we thus condense the discussion as much as possible and only highlight relevant differences. Of course, the main difference is the group structure, which, for our purpose, manifests itself in a different set of structure constants.

The group $G=SU(3)\times SU(2)\times U(1)$ admits various embeddings of the subgroup $H=SU(2)\times U(1) \times U(1)$. The different embeddings are parameterized by three coprime integers $p$, $q$ and $r$. The Lie algebra $\mathfrak{g}=\mathfrak{su}(3)\oplus \mathfrak{su}(2)\oplus \mathfrak{u}(1)$ is not semisimple and hence, does not have an orthonormal basis. Instead, we use a basis in which at least the \emph{coset space components} of the Killing-Cartan metric are equal to the Kronecker delta, $(g_\mathfrak{g})_{ab} = \delta_{ab}$. This can be achieved by taking the conventions of~\cite{Karthauser:2006wb} with the following rescalings of the generators $I_A$ and structure constants $f^A_{BC}$,
\begin{equation}\eqlabel{Mpqr_rescalings}
  I_A \rightarrow c_A I_A \qquad \text{(no sum over $A$)} \; , \qquad
  f^A_{BC} \rightarrow \sfrac{c_B c_C}{c_A} f^A_{BC} \qquad \text{(no sum over $A, B, C$)} \; ,
\end{equation}
where
\begin{equation}
 c_A = \begin{cases} 1/\sqrt{3} & \text{if $A=1,2,3,4$} \\ 1/\sqrt{2} & \text{if $A=5,6$} \\ \zeta / \rho & \text{if $A=7$} \\ 1 & \text{otherwise} \end{cases} \; .
\end{equation}
Here, we set $\zeta := \sqrt{3 p^2+q^2+2 r^2}$, $\eta := \sqrt{3 p^2+q^2}$ and $\rho := \sqrt{9 p^2+2 q^2}$. The precise expressions for the structure constants in the new basis can be found in \appref{struct_consts_Mpqr}. The Killing-Cartan metric, $(g_\mathfrak{g})_{AB} = - f^C_{AD} f^D_{BC}$, in this basis is given by,
\begin{multline}\eqlabel{Mpqr_gKC}
 (g_\mathfrak{g})_{AB} = \delta_{A=a,B=b} + 3 (\delta_{A8}\delta_{B8} + \delta_{A9}\delta_{B9} + \delta_{A,10}\delta_{B,10}) 
 + \sfrac{2 \rho ^2 r^2}{\zeta ^2 \eta ^2} \delta_{A,11}\delta_{B,11} + \sfrac{3 ( 2 p^2 + q^2 )}{\eta ^2} \delta_{A,12}\delta_{B,12} \\
 + \sfrac{2\sqrt{2} \rho r}{\zeta  \eta } (\delta_{7(A}\delta_{B),11}) - \sfrac{2\sqrt{3} p q}{\rho \eta} (\delta_{7(A}\delta_{B),12}) - \sfrac{2\sqrt{6} p q r}{\zeta  \eta ^2} (\delta_{11,(A}\delta_{B),12}) \; .
\end{multline}
It is used to raise, lower and contract Lie algebra indices. The structure constants with all indices lowered, $f_{ABC} := (g_\mathfrak{g})_{CD} f^D_{AB}$, are totally anti-symmetric, $f_{[ABC]}=f_{ABC}$.

Next, we need to construct a 3-form $P$ defining the $G_2$-structure on $M^{pqr}$. As before, the 3-form $P$ is, up to discrete sign choices, fully determined by~\eqref*{GinvcondP},~\eqref*{PQrelns1} and the requirement that the associated metric $g_P$ (see~\eqref*{g_P}) be positive definite. At this point, we encounter a crucial novelty when compared to $N^{pqr}$. Namely, a 3-form $P$ on $M^{pqr}$ that solves the aforementioned requirements only exists for $p=\pm q$, $r=0$ and without loss of generality, we may assume $p=q=1$. This observation is a manifestation of the well-known fact that only $M^{110}$ admits an $SU(3)\times SU(2)\times U(1)$-invariant $G_2$-structure~\cite{Castellani:1983mf,Reidegeld:2010}. 

For the rest of this subsection, we thus restrict to $M^{110}$. There are two independent $G_2$-structures on $M^{110}$ defined by
\begin{equation}\eqlabel{M110_P}
  P^\pm = - e^{127} \mp e^{135} \pm e^{146} \pm e^{236} \pm e^{245} - e^{347} - e^{567} \; .
\end{equation}
Comparing with~\eqref*{G2_SU3_struct_forms_rel} and using the fourth relation in~\eqref*{SU3_structure_relations_extra}, we immediately obtain the underlying $SU(3)$-structure, namely
\begin{equation}\eqlabel{M110_SU3_struct}
  V=e^7 \; , \quad J = e^{12} + e^{34} + e^{56} \; , \quad \Omega = (e^{136}+e^{145}+e^{235}-e^{246}) + i (-e^{135}+e^{146}+e^{236}+e^{245}) \; .
\end{equation}
A straightforward computation shows that $(V,J,\Omega)$ indeed satisfy the $SU(3)$-structure relations~\eqrangeref*{SU3_structure_relations}{SU3_structure_relations_extra}. The 4-form $Q^\pm := \ast_7 P^\pm$ is given by
\begin{equation}\eqlabel{M110_Q}
  Q^\pm = \mp\Omega^+ \wedge V - \sfrac{1}{2} J\wedge J 
        = - e^{1234} - e^{1256} \mp e^{1367} \mp e^{1457} \mp e^{2357} \pm e^{2467} - e^{3456} \; .
\end{equation}
It is closed, $\dd Q^\pm = 0$, whereas the exterior derivative of $P^\pm$ can be expressed as,
\begin{equation}\eqlabel{M110_dP}
  \dd P^\pm = \tau_0 Q^\pm + \ast_7 \tau_3 \; ,
\end{equation}
with $\tau_0 = - 4/\sqrt{11}$ and $\tau_3 = - \sfrac{5}{4 \sqrt{11}} J\wedge V$. Hence, the $SU(3)$-structure~\eqref*{M110_SU3_struct} defines two cocalibrated (or semi-parallel) $G_2$-structures on $M^{110}$.

The induced Spin$(7)$-structure on the cylinder is fixed by $\Psi^\pm = P^\pm \wedge\dd\tau - Q^\pm$, according to~\eqref*{PsiPQ}. Using that $\tau_0 \neq 0$, $\tau_3 \neq 0$, $\dd\Psi^\pm \neq 0$ and the results of \secref{Spin7_structures}, we immediately conclude that this is a general Spin$(7)$-structure where both torsion classes $W_8$ and $W_{48}$ are turned on.

The $G$-invariance condition~\eqref*{Ginvcond_gaugefield}, where now $G=SU(3)\times SU(2)\times U(1)$, is solved by
\begin{equation}\eqlabel{Mpqr_X}
\begin{aligned}
  &X_1^1 = X_2^2 = X_3^3 = X_4^4 = \phi_1 (\tau) \; , &\; &X_5^5 = X_6^6 = \phi_2 (\tau) \; , &\; &X_7^7 = \phi_3 (\tau) \; , \\
  &X_1^2 = -X_2^1 = X_3^4 = -X_4^3 = \phi_4 (\tau) \; , &\; &X_5^6 = -X_6^5 = \phi_5 (\tau) \; . &\; &
\end{aligned}
\end{equation}
Hence, the dynamical degrees of freedom are in this case \emph{a priori} five real $\tau$-dependent scalar fields.

\paragraph{Instanton equation.}
We are interested in computing the system of equations for $\{\phi_1, \ldots,\phi_5\}$ imposed by the instanton equation on $Z(M^{110})$. To achieve this, we insert $f^A_{BC}$, $P^\pm$ and $X_a^b$ into~\eqref*{insteqSpin7cosetX}. We find a coupled system of first-order ordinary differential equations,
\begin{equation}
\begin{aligned}
 \dot{\phi}_1 &= \sfrac{3}{2 \sqrt{11}} \phi_1 ( 1 - \phi_3 ) \; , \qquad & \dot{\phi}_4 &= \sfrac{3}{2 \sqrt{11}} \phi_4 ( 1 - \phi_3 ) \; , \\
 \dot{\phi}_2 &= \sfrac{1}{  \sqrt{11}} \phi_2 ( 1 - \phi_3 ) \; , \qquad & \dot{\phi}_5 &= \sfrac{1}{  \sqrt{11}} \phi_5 ( 1 - \phi_3 ) \; , \\ 
 \dot{\phi}_3 &= -\sfrac{\sqrt{11}}{8} \rlap{$( 2 \phi _1^2+\phi _2^2+2 \phi _4^2+\phi _5^2-3 \phi _3 ) \; , $} & &
\end{aligned}
\end{equation}
supplemented by an algebraic constraint,
\begin{equation}
  2 \phi_1^2 + 2 \phi_4^2 + 1 = 3 \phi_2^2 + 3 \phi_5^2 \; ,
\end{equation}
which follows from the quiver relation. Note that the entire system of equations is independent of the choice $P^+$ or $P^-$.

Before moving on to the second-order Yang-Mills equations, we mention a special solution which is analytical. This is the static solution, $\phi_1=\phi_4=\pm 1/\sqrt{2}$, $\phi_2=\phi_5=\pm 1/\sqrt{2}$ and $\phi_3=1$.

\paragraph{Yang-Mills equation.}
In this part, we shall compute the equations of motion for $\{\phi_1, \ldots,\phi_5\}$, using the same ans\"atze for the torsion $T$, affine spin connection $\omega$ and 3-form $H$ as before. In particular, from~\eqref*{SU3_H_ansatz1} we find the explicit expression
\begin{equation}
 H^\pm = - \lambda \ast_7 \dd P^\pm = \lambda \left( \pm \sfrac{4}{\sqrt{11}} \Omega^- - \sfrac{\sqrt{11}}{4} J\wedge V \right) \; .
\end{equation}
We have now collected all necessary ingredients to derive the equations of motion on $Z(M^{110})$ from~\eqref*{YMComp}. We obtain one algebraic constraint,
\begin{multline}\eqlabel{M110_constr}
  2 (22 \lambda + 35) (\phi _1^2 + \phi _4^2) - 3 (22 \lambda + 27) (\phi _2^2 + \phi _5^2) \\ - 48 \phi _3 \left( \phi _1^2 + \phi _4^2 - \phi _2^2 - \phi _5^2\right) + 11 (2 \lambda + 1) = 0 \; ,
\end{multline}
as well as two first-order constraints,
\begin{equation}
 \phi_1 \dot{\phi}_4 = \dot{\phi}_1 \phi_4 \; , \qquad\qquad 
 \phi_2 \dot{\phi}_5 = \dot{\phi}_2 \phi_5 \; .
\end{equation}
The latter are a consequence of the Gauss-law constraint~\eqref*{gauss_law} and are solved by\footnote{Note that without loss of generality, we may assume that $\phi_1 \neq 0$, $\phi_2 \neq 0$. Indeed, since the equations of motion are symmetric under interchanging $\phi_1 \leftrightarrow \phi_4$, $\phi_2 \leftrightarrow \phi_5$, the cases where $\phi_1 = 0$ or $\phi_2 = 0$ correspond to $c_1 = 0$ or $c_2 = 0$, respectively.} $\phi_4 = c_1 \phi_1$, $\phi_5 = c_2 \phi_2$ for some $c_1, c_2\in\RR$. \Eqref{YMComp} also yields a set of second-order differential equations, which can equivalently be obtained as the Euler-Lagrange equations of the action~\eqref*{YMaction} with the reduction to scalar fields,~\eqref*{Mpqr_X}, inserted. After using that $\tr(I_A I_B) = -(g_\mathfrak{g})_{AB}$ and replacing $Q_\M$ by $Q_{\M,\lambda}$, the action becomes
\begin{equation}\eqlabel{M110_action}
  S = - 16\, \mathrm{Vol}(M^{110}) \, \int\limits_{-\infty}^\infty {\cal E}\, \dd\tau \; .
\end{equation}
Here, the energy density ${\cal E} = T + V$ comprises the standard kinetic term 
\begin{equation}
  T = \sfrac{1}{2} \dot{\phi}_1^2 + \sfrac{1}{2} \dot{\phi}_2^2 + \sfrac{1}{2} \dot{\phi}_3^2 \; ,
\end{equation}
and the potential $V$, given by
\begin{multline}\eqlabel{M110_V}
  V = \sfrac{1}{4} \phi_1^4 + \sfrac{1}{2} \phi_2^4 + \sfrac{66 \lambda - 19}{352} \phi_1^2 + \sfrac{22 \lambda - 25}{176} \phi_2^2 + \sfrac{33(2 \lambda +1)}{128} \phi_3^2 + \sfrac{1}{11} ( 9 \phi_1^2 + 4 \phi_2^2 ) \phi_3^2 \\ 
  - \sfrac{1}{88 \sqrt{2}} \left\{ 3 (22 \lambda +35) \phi _1^2 + 2 (22 \lambda +27) \phi _2^2 \right\} \phi _3 - \sfrac{2 \lambda +1}{128 \sqrt{2}} \phi _3 + \sfrac{65-62 \lambda }{1024} \; .
\end{multline}
In order to bring the expressions for $T$ and $V$ in the above form, we performed a field redefinition, $\phi_1 \to \phi_1 \sqrt{(1+c_1^2)/2}$, $\phi_2 \to \phi_2 \sqrt{(1+c_2^2)}/2$, $\phi_3 \to \phi_3/(2\sqrt{2})$. Note that the equivalence of the second-order equation for $\phi_3$ derived from~\eqref*{M110_action} and the respective equation obtained from~\eqref*{YMComp} holds up to a term proportional to the algebraic constraint, which reads as follows,
\begin{equation}
 4 (22 \lambda + 35) \phi _1^2 - 12 (22 \lambda + 27) \phi _2^2 - 192 \sqrt{2} ( \phi_1^2 - 2 \phi_2^2 ) \phi _3 + 11 (2 \lambda + 1) = 0 \; ,
\end{equation}
after the field redefinition. We also remark that all equations starting from~\eqref*{M110_constr} and below are independent of the choice $P^+$ or $P^-$.

We shall now consider an interesting scenario, where the constraint can be easily satisfied, namely $\lambda = -1/2$. The constraint then implies either $\phi _1^2 = 2 \phi _2^2$ or $\phi_3 = 1/(2 \sqrt{2})$. The first case, $\phi _1^2 = 2 \phi _2^2$, can be quickly dealt with, for it leads to $\phi_1 = \phi_2 = 0$ and $\phi_3 = a\tau + b$ with integration constants $a,b\in\RR$. This solution is unphysical since the value of the total energy is divergent, $E = \int_{-\infty}^\infty {\cal E}\, \dd\tau = (1/32) \int_{-\infty}^\infty (16 a^2 + 3)\, \dd\tau \to \infty$.

The second case, namely $\phi_3 = 1/(2 \sqrt{2})$, turns out to be more fruitful however. The energy density then reduces to
\begin{equation}\eqlabel{M110_E}
  {\cal E} = T + V = \sfrac{1}{2} \dot{\phi}_1^2 + \sfrac{1}{2} \dot{\phi}_2^2 + \sfrac{1}{4} \phi_1^4 + \sfrac{1}{2} \phi_2^4 - \sfrac{1}{4} \phi_1^2 - \sfrac{1}{4} \phi_2^2 + \sfrac{3}{32} \; ,
\end{equation}
and the equations of motion become
\begin{equation}\eqlabel{M110_EOM}
\begin{aligned}
  \ddot{\phi}_1 &= \sfrac{1}{2} \phi_1 ( 2 \phi_1^2 - 1 ) \; , \\
  \ddot{\phi}_2 &= \sfrac{1}{2} \phi_2 ( 4 \phi_2^2 - 1 ) \; .
\end{aligned}
\end{equation}
The critical points of the potential $V$ in the field space $(\phi_1,\phi_2)$ are listed in \tabref{M110_critpts}. In total, there are nine distinct critical points, subdivided into a local maximum at $(0,0)$, four saddle points $(0,\pm 1/2)$, $(\pm 1/\sqrt{2},0)$, and four local minima at $(\pm 1/\sqrt{2}, \pm 1/2)$.
\begin{table}[t]\centering
\begin{tabular}{c|c|c}
  $(\phi_1, \phi_2)$ & $V |_{\text{crit. pt.}}$ & Eigenvalues of Hessian \\ \hline
  $(0,0)$ & $3/32$ & $(-,-)$ \\ 
  $(0,\pm 1/2)$ & $1/16$ & $(+,-)$ \\ 
  $(\pm 1/\sqrt{2},0)$ & $1/32$ & $(+,-)$ \\ 
  $(\pm 1/\sqrt{2}, \pm 1/2)$ & $0$ & $(+,+)$ \\ 
\end{tabular}
\caption{Critical points of the potential $V$ and some of their properties on $Z(M^{110})$ with $\lambda=-1/2$ and $\phi_3 = 1/(2 \sqrt{2})$. In the first two columns we list the values of the critical points and of the potential at the corresponding critical point, respectively. In the third column we summarize the signs of the eigenvalues of the Hessian matrix at the respective critical point. In the last row, the signs are mutually independent in the entries of $(\phi_1, \phi_2)$.}
\tablabel{M110_critpts}
\end{table}

There are different types of non-constant solutions, depending on whether one allows both fields to have non-trivial $\tau$-dependence or not. In the former case, the equations of motions~\eqref*{M110_EOM} can be integrated to
\begin{equation}
\begin{aligned}
  \dot{\phi}_1 &= \pm\sfrac{1}{2\sqrt{2}} (1 - 2 \phi_1^2) \; , \\
  \dot{\phi}_2 &= \pm\sfrac{1}{4} (1 - 4 \phi_2^2) \; .
\end{aligned}
\end{equation}
The solution is a system of two uncoupled rescaled $\phi^4$ kinks/anti-kinks,
\begin{equation}\eqlabel{M111_specialsol1}
\begin{aligned}
  \phi_1 (\tau) &= \pm\sfrac{1}{\sqrt{2}} \tanh\left[ \pm \sfrac{\tau-\tau_{0,1}}{2} \right] \; , \\
  \phi_2 (\tau) &= \pm\sfrac{1}{2} \tanh\left[ \pm \sfrac{\tau-\tau_{0,2}}{2} \right] \; ,
\end{aligned}
\end{equation}
where $\tau_{0,1}, \tau_{0,2} \in \RR$ are integration constants determining the positions of the (anti-)kinks in the $\tau$ direction. The four signs in~\eqref*{M111_specialsol1} are mutually independent. The different types of non-trivial solutions are summarized in \tabref{M110_energies}.
\begin{table}[t]\centering
\begin{tabular}{c|c}
  $(\phi_1, \phi_2)$ & $E$ \\ \hline
  $(\tanh, \tanh)$ & $1/2$ \\ 
  $(\pm 1/\sqrt{2}, \tanh)$ & $1/6$ \\ 
  $(\tanh, \pm 1/2)$ & $1/3$ \\ 
  $(0, \tanh)$ & $\infty$ \\ 
  $(\tanh, 0)$ & $\infty$ \\ 
\end{tabular}
\caption{Different types of non-trivial analytical solutions on $Z(M^{110})$ with $\lambda=-1/2$ and $\phi_3 = 1/(2 \sqrt{2})$. Here, ``$\tanh$'' refers to a rescaled $\phi^4$ kink/anti-kink solution of the form~\eqref*{M111_specialsol1}. The second column holds the value of the total energy $E$ of the respective solution. The last two rows correspond to unphysical solutions with divergent total energy.}
\tablabel{M110_energies}
\end{table}
There, we also list each solution's total energy
\begin{equation}
  E = \int\limits_{-\infty}^\infty {\cal E}\, \dd\tau \; ,
\end{equation}
with ${\cal E}$ given by~\eqref*{M110_E}. The last two solutions in \tabref{M110_energies} each interpolate between two saddle points. They have divergent total energy and are thus unphysical. On the other hand, the first three rows do correspond to non-trivial, analytical and physically well-behaved solutions. We end this subsection by showing in \figref{M110_plot} contour plots of the different types of non-trivial solutions.
\begin{figure}[t]
\centering
\includegraphics[width=\textwidth]{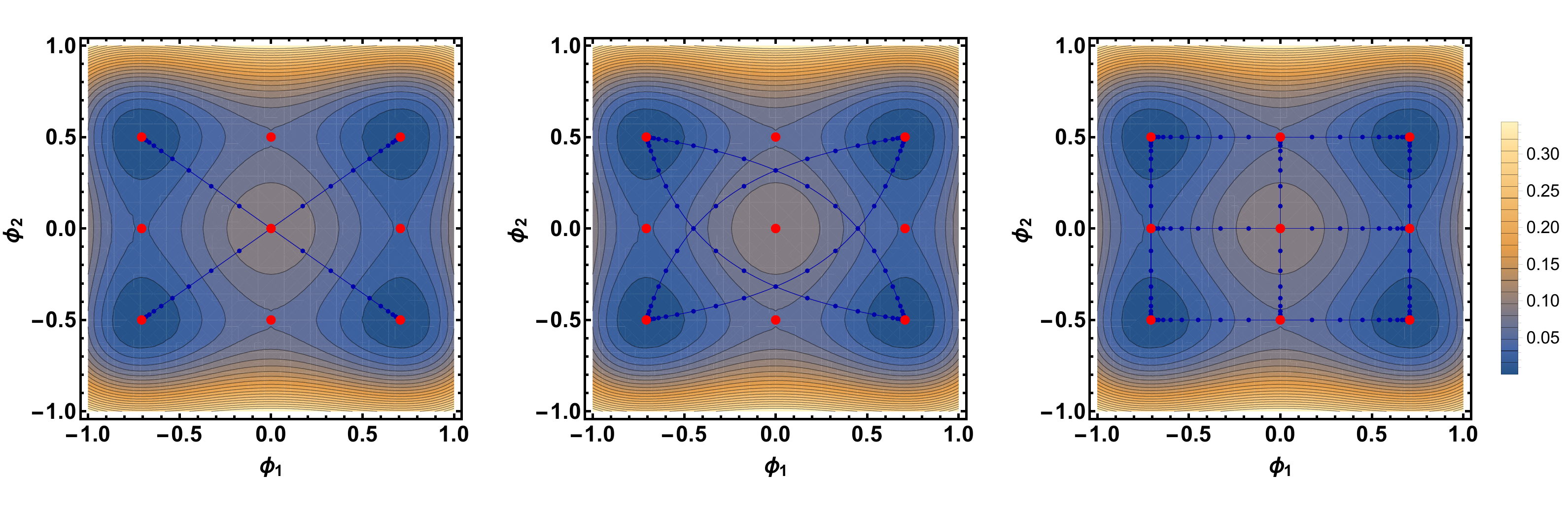}
\caption{Contour plots of the potential $V$ and different types of analytical solutions on $Z(M^{110})$ with $\lambda=-1/2$ and $\phi_3 = 1/(2 \sqrt{2})$. Marked in red are the critical points of the potential as listed in \tabref{M110_critpts}. The blue lines in the left and middle plot represent the $(\tanh,\tanh)$-solution~\eqref*{M111_specialsol1} with $\tau_{0,1} = \tau_{0,2}$ and $\tau_{0,1} \neq \tau_{0,2}$, respectively. The blue lines in the right plot correspond to solutions where one of the $\phi$'s is kept constant. The two blue lines in the right plot that pass through the origin are unphysical for they have divergent total energy (see \tabref{M110_energies}).}
\figlabel{M110_plot}
\end{figure}

\subsubsection{Cylinders over \texorpdfstring{$Q^{pqr} = (SU(2)\times SU(2)\times SU(2)) / (U(1) \times U(1))$}{Qpqr}}\seclabel{Qpqr}

The last seven-dimensional coset space we consider in this paper is $Q^{pqr} = (SU(2)\times SU(2)\times SU(2)) / (U(1) \times U(1))$. The treatment is similar to $N^{pqr}$ and $M^{pqr}$, which is why we refer to \secref{Npqr,Mpqr} for the technical details. Before discussing the instanton and Yang-Mills equations on $Z(Q^{pqr})$, we list the coset space's defining data, which is necessary to carry out our computations. 

The group $G=SU(2)\times SU(2)\times SU(2)$ admits various embeddings of the subgroup $H=U(1) \times U(1)$. The different embeddings are parameterized by three coprime integers $p$, $q$ and $r$. In order to define the coset space, we start with the generators $I_A$ and structure constants $f^A_{BC}$ from~\cite{Karthauser:2006wb}, and perform a rescaling, $I_A \rightarrow \sfrac{1}{\sqrt{2}} I_A$, $f^C_{AB} \rightarrow \sfrac{1}{\sqrt{2}} f^C_{AB}$, so as to ensure the correct normalization of~\eqref*{g_CK}. Our rescaled structure constants are listed in \appref{struct_consts_Qpqr} for completeness. 

The structure constants with all indices lowered, $f_{ABC} := \delta_{CD} f^D_{AB}$, are totally anti-symmetric, $f_{[ABC]}=f_{ABC}$. However, taking $P_{abc} \propto f_{abc}$ does not yield a well-defined $G_2$-structure on $Q^{pqr}$ (for example, the associated metric $g_P$ is singular for this choice). Instead, we construct the 3-form $P$ subject to~\eqref*{GinvcondP},~\eqref*{PQrelns1} and the requirement that $g_P$ be positive definite. We find that such a $P$ only exists if $p=\pm q$, $q=\pm r$ and without loss of generality, we may restrict to the case $p=q=r=1$. This observation is a manifestation of the well-known fact that only $Q^{111}$ admits an $(SU(2))^3$-invariant $G_2$-structure~\cite{D'Auria:1983vy,Reidegeld:2010}. 

For the rest of this subsection, we thus restrict to $Q^{111}$. There are two independent $G_2$-structures on $Q^{111}$ defined by
\begin{equation}\eqlabel{Q111_P}
  P^\pm = - e^{127} \pm e^{136} \pm e^{145} \pm e^{235} \mp e^{246} - e^{347} - e^{567} \; .
\end{equation}
Comparing with~\eqref*{G2_SU3_struct_forms_rel} and using the fourth relation in~\eqref*{SU3_structure_relations_extra}, we immediately obtain the underlying $SU(3)$-structure, namely
\begin{equation}\eqlabel{Q111_SU3_struct}
  V=e^7 \; , \quad J = e^{12} + e^{34} + e^{56} \; , \quad \Omega = (e^{135}-e^{146}-e^{236}-e^{245}) + i (e^{136}+e^{145}+e^{235}-e^{246}) \; .
\end{equation}
A straightforward computation shows that $(V,J,\Omega)$ indeed satisfy the $SU(3)$-structure relations~\eqrangeref*{SU3_structure_relations}{SU3_structure_relations_extra}. The 4-form $Q^\pm := \ast_7 P^\pm$ is given by
\begin{equation}\eqlabel{Q111_Q}
  Q^\pm = \mp\Omega^+ \wedge V - \sfrac{1}{2} J\wedge J 
        = - e^{1234} - e^{1256} \mp e^{1357} \pm e^{1467} \pm e^{2367} \pm e^{2457} - e^{3456} \; .
\end{equation}
It is closed, $\dd Q^\pm = 0$, whereas the exterior derivative of $P^\pm$ can be expressed as,
\begin{equation}\eqlabel{Q111_dP}
  \dd P^\pm = \tau_0 Q^\pm + \ast_7 \tau_3 \; ,
\end{equation}
with $\tau_0 = - \sqrt{3/2}$ and $\tau_3 = - \sfrac{1}{\sqrt{6}} J\wedge V$. Hence, the $SU(3)$-structure~\eqref*{Q111_SU3_struct} defines two cocalibrated (or semi-parallel) $G_2$-structures on $Q^{111}$.

The induced Spin$(7)$-structure on the cylinder is fixed by $\Psi^\pm = P^\pm \wedge\dd\tau - Q^\pm$, according to~\eqref*{PsiPQ}. Using that $\tau_0 \neq 0$, $\tau_3 \neq 0$, $\dd\Psi^\pm \neq 0$ and the results of \secref{Spin7_structures}, we immediately conclude that this is a general Spin$(7)$-structure where both torsion classes $W_8$ and $W_{48}$ are turned on.

The $G$-invariance condition~\eqref*{Ginvcond_gaugefield}, where now $G=(SU(2))^3$, is solved by
\begin{equation}\eqlabel{Qpqr_X}
\begin{aligned}
  &X_1^1 = X_2^2 = \phi_1 (\tau) \; , &\; &X_3^3 = X_4^4 = \phi_2 (\tau) \; , &\; &X_5^5 = X_6^6 = \phi_3 (\tau) \; , &\; &X_7^7 = \phi_4 (\tau) \; , \\
  &X_1^2 = -X_2^1 = \phi_5 (\tau) \; , &\; &X_3^4 = -X_4^3 = \phi_6 (\tau) \; , &\; &X_5^6 = -X_6^5 = \phi_7 (\tau) \; . &\; &
\end{aligned}
\end{equation}
Hence, the dynamical degrees of freedom are in this case \emph{a priori} seven real $\tau$-dependent scalar fields.

\paragraph{Instanton equation.}
We are interested in computing the system of equations for $\{\phi_1, \ldots,\phi_7\}$ imposed by the instanton equation on $Z(Q^{111})$. To achieve this, we insert $f^A_{BC}$, $P^\pm$ and $X_a^b$ into~\eqref*{insteqSpin7cosetX}. We find a coupled system of first-order ordinary differential equations,
\begin{equation}
\begin{aligned}
  \dot{\phi}_\alpha &= \sfrac{1}{\sqrt{6}} \phi_\alpha ( 1 - \phi_4 ) \; , \qquad\qquad\qquad \alpha \in \{ 1,2,3,5,6,7 \} \; , \\
  \dot{\phi}_4 &= -\sfrac{1}{\sqrt{6}} (\phi _1^2+\phi _2^2+\phi _3^2+\phi _5^2+\phi _6^2+\phi _7^2-3 \phi _4) \; ,
\end{aligned}
\end{equation}
supplemented by two algebraic constraints,
\begin{equation}
\begin{aligned}
 \phi _1^2 - \phi _2^2 + \phi _5^2 - \phi _6^2 &= 0 \; , \\
 \phi _1^2 + \phi _2^2 + \phi _5^2 + \phi _6^2 &= 2 (\phi _3^2+\phi _7^2) \; .
\end{aligned}
\end{equation}
which follow from the quiver relation. Note that the entire system of equations is independent of the choice $P^+$ or $P^-$.

Before studying the second-order Yang-Mills equations, we list some special solutions. First, there is the trivial case $\phi_1=\ldots=\phi_7=0$. There are also static but non-zero solutions, such as $\phi_4 = 1$, $\phi_\alpha = \pm 1/\sqrt{2}$, $\forall \alpha\in \{ 1,2,3,5,6,7 \}$. Lastly, there are simple non-static, albeit unphysical (diverging) solutions, for example $\phi_4 = c \exp((3\tau)/\sqrt{6})$, with a constant of integration $c\in\RR$, and $\phi_\alpha = 0$, $\forall \alpha\in \{ 1,2,3,5,6,7 \}$.

\paragraph{Yang-Mills equation.}
Our next goal is to compute the second-order equations of motion for $\{\phi_1, \ldots,\phi_7\}$, using the same ans\"atze for the torsion $T$, affine spin connection $\omega$ and 3-form $H$ as before. In particular, from~\eqref*{SU3_H_ansatz1} we find the explicit expression
\begin{equation}
 H^\pm = - \lambda \ast_7 \dd P^\pm = \lambda \left( \pm \sqrt{\sfrac{3}{2}} \Omega^- - \sqrt{\sfrac{2}{3}} J\wedge V \right) \; .
\end{equation}
With all the necessary ingredients at hand, we are in a position to compute the equations of motion on $Z(Q^{111})$ following from~\eqref*{YMComp}. It turns out that they are independent of the choice $P^+$ or $P^-$. We obtain two algebraic constraints,
\begin{equation}\eqlabel{Q111_constr}
\begin{aligned}
 (\phi _1^2-\phi _2^2+\phi _5^2-\phi _6^2 ) (2 \lambda -2 \phi _4+3) &= 0 \; , \\
 (\phi _1^2+\phi _2^2-2 \phi _3^2+\phi _5^2+\phi _6^2-2 \phi _7^2 ) (2 \lambda -2 \phi _4+3) &= 0 \; ,
\end{aligned}
\end{equation}
as well as three first-order constraints (originating from the Gauss-law constraint~\eqref*{gauss_law}),
\begin{equation}
\begin{aligned}
 -\dot{\phi}_5 \phi _1+\dot{\phi}_6 \phi _2+\dot{\phi}_1 \phi _5-\dot{\phi}_2 \phi _6 &= 0 \; , \\
 -\dot{\phi}_5 \phi _1-\dot{\phi}_6 \phi _2+2 \dot{\phi}_7 \phi _3+\dot{\phi}_1 \phi _5+\dot{\phi}_2 \phi _6-2 \dot{\phi}_3 \phi _7 &= 0 \; , \\
 -\dot{\phi}_5 \phi _1-\dot{\phi}_6 \phi _2-\dot{\phi}_7 \phi _3+\dot{\phi}_1 \phi _5+\dot{\phi}_2 \phi _6+\dot{\phi}_3 \phi _7 &= 0 \; .
\end{aligned}
\end{equation}
After taking suitable linear combinations, the latter become
\begin{equation}
 \phi_1 \dot{\phi}_5 = \dot{\phi}_1 \phi_5 \; , \qquad\quad
 \phi_2 \dot{\phi}_6 = \dot{\phi}_2 \phi_6 \; , \qquad\quad
 \phi_3 \dot{\phi}_7 = \dot{\phi}_3 \phi_7 \; ,
\end{equation}
and thus, $\phi_5 = c_1 \phi_1$, $\phi_6 = c_2 \phi_2$, $\phi_7 = c_3 \phi_3$ for some $c_1, c_2, c_3 \in \RR$ (without loss of generality, we may assume that $\phi_1 \neq 0$, $\phi_2 \neq 0$, $\phi_3 \neq 0$, since the equations of motion are symmetric under interchanging $\phi_1 \leftrightarrow \phi_5$, $\phi_2 \leftrightarrow \phi_6$, $\phi_3 \leftrightarrow \phi_7$). \Eqref{YMComp} also yields a set of second-order equations, which can equivalently be obtained as the Euler-Lagrange equations of the action~\eqref*{YMaction} with the reduction to scalar fields,~\eqref*{Qpqr_X}, inserted. After using that $\tr(I_A I_B) = -\delta_{AB}$ and replacing $Q_\M$ by $Q_{\M,\lambda}$, the action becomes
\begin{equation}\eqlabel{Q111_action}
  S = - 4\, \mathrm{Vol}(Q^{111}) \, \int\limits_{-\infty}^\infty {\cal E}\, \dd\tau \; .
\end{equation}
Here, the energy density ${\cal E} = T + V$ comprises the standard kinetic term 
\begin{equation}
  T = \sfrac{1}{2} \dot{\phi}_1^2 + \sfrac{1}{2} \dot{\phi}_2^2 + \sfrac{1}{2} \dot{\phi}_3^2 + \sfrac{1}{2} \dot{\phi}_4^2 \; ,
\end{equation}
and the potential $V$, given by
\begin{multline}\eqlabel{Q111_V}
  V = \sfrac{1}{8} (\phi _1^4 + \phi _2^4 + \phi _3^4) + \sfrac{1}{6} ( \phi _1^2 + \phi _2^2 + \phi _3^2 ) \phi _4^2 - \sfrac{2 \lambda +3}{6 \sqrt{2}} ( \phi _1^2 + \phi _2^2 + \phi _3^2 ) \phi _4 \\ 
    + \sfrac{2 \lambda -1}{12} (\phi _1^2 + \phi _2^2 + \phi _3^2) + \sfrac{2 \lambda +1}{4} \phi _4^2 + \sfrac{1-\lambda }{4} \; .
\end{multline}
In order to bring the expressions for $T$ and $V$ in the above form, we performed a field redefinition, $\phi_1 \to \phi_1 \sqrt{(1+c_1^2)}$, $\phi_2 \to \phi_2 \sqrt{(1+c_2^2)}$, $\phi_3 \to \phi_3 \sqrt{(1+c_3^2)}$, $\phi_4 \to \phi_4/\sqrt{2}$. These expressions are still supplemented by the two algebraic constraints~\eqref*{Q111_constr}, which are given by
\begin{equation}
\begin{aligned}
  (\phi _1^2-\phi _2^2) (2 \lambda +3 -2 \sqrt{2} \phi _4) &= 0 \; , \\
  (\phi _1^2+\phi _2^2-2 \phi _3^2) (2 \lambda +3 -2 \sqrt{2} \phi _4) &= 0 \; ,
\end{aligned}
\end{equation}
after the field redefinition. These constraints imply either $\phi_1^2 = \phi_2^2 = \phi_3^2$ or $\phi_4 = (2 \lambda +3)/(2 \sqrt{2})$.

In the first case, that is $\phi_1^2 = \phi_2^2 = \phi_3^2$, the energy density reduces to
\begin{equation}\eqlabel{Q111_E1}
  {\cal E} = \sfrac{3}{2} \dot{\phi}_1^2 + \sfrac{1}{2} \dot{\phi}_4^2 + \sfrac{3}{8} \phi _1^4 + \sfrac{1}{2} \phi _1^2 \phi _4^2 - \sfrac{2 \lambda +3}{2 \sqrt{2}} \phi _1^2 \phi _4 + \sfrac{2 \lambda -1}{4} \phi _1^2 + \sfrac{2 \lambda +1}{4} \phi _4^2 + \sfrac{1-\lambda }{4} \; .
\end{equation}
This case has already been analyzed in \secref{SO5_SO3_AB}. Indeed, after identifying $\phi_4 = \phi_2 / \sqrt{2}$, the energy density~\eqref*{Q111_E1} is equal to three times the energy density~\eqref*{SO5_SO3_AB_E} and the solutions are thus the same as those discussed in detail on pages~\pageref{SO5_SO3_AB_E_start}-\pageref{SO5_SO3_AB_E_end}.

\begin{table}[t]\centering
\begin{tabular}{c|c||c|c|c}
  \doublerowheight $\boldsymbol{\mathbf{\phi}}^\top$ & $V |_{\text{crit. pt.}}$ & $\lambda_0$ & $\boldsymbol{\mathbf{\phi}}^\top|_{\lambda_0}$ & $\text{Eigenvalues of Hessian}|_{\lambda_0}$ \\ \hline\hline
  \doublerowheight $(0,0,0)_1$ & $c_2$ & $\approx-2.79$ & $(0,0,0)$ & $(-,-,-)$ \\ \hline
  \doublerowheight $(0,0,c_1^\pm)_6$ & $c_3$ & \begin{tabular}{c} \vspace{-2pt}$\approx-2.30$ \\ $\approx 19.0$ \end{tabular} & \begin{tabular}{c} \vspace{-2pt}$\approx(0,0,1.44)$ \\ $\approx(0,0,11.3)$ \end{tabular} & \begin{tabular}{c} \vspace{-2pt}$(+,-,-)$ \\ $(+,-,-)$ \end{tabular} \\ \hline
  \doublerowheight $(0,c_1^\pm,c_1^\pm)_{12}$ & $c_4$ & \begin{tabular}{c} \vspace{-2pt}$\approx-1.88$ \\ $\approx 9.68$ \end{tabular} & \begin{tabular}{c} \vspace{-2pt} $\approx(0,1.28,1.28)$ \\ $\approx(0,5.96,5.96)$ \end{tabular} & \begin{tabular}{c} $(+,+,-)$ \\ $(+,+,-)$ \end{tabular} \\ \hline
  \doublerowheight $(c_1^\pm,c_1^\pm,c_1^\pm)_8$ & $c_5$ & \begin{tabular}{c} \vspace{-2pt}$-1/2$ \\ $c_6^\pm$ \end{tabular} & \begin{tabular}{c} \vspace{-2pt}$(\pm 1, \pm 1, \pm 1)$ \\ $(\pm c_7^\pm, \pm c_7^\pm, \pm c_7^\pm)$ \end{tabular} & \begin{tabular}{c} \vspace{-2pt}$(+,+,+)$ \\ $(+,+,+)$ \end{tabular}
\end{tabular}
\caption{Critical points of $V$ and some of their properties on $Z(Q^{111})$ with $\phi_4 = (2 \lambda +3)/(2 \sqrt{2})$. Column one contains the positions of the critical points. The subscript denotes the multiplicities of each point viewed as equivalence classes under $S_3 \times (Z_2)^3$-transformations.\protect\footnotemark~In total, there are 27 distinct critical points. In the second column we list the value of the potential at the corresponding critical point. In the right part of the table (columns 3-5), we demand that $V |_{\text{crit. pt.}}\stackrel{!}{=}0$ and solve for $\lambda$. The resulting $\lambda_0$ is shown in column three. For the sake of brevity, we refrain from showing lengthy analytical expressions for $\lambda_0$ in some unimportant cases and merely state their approximate numerical value indicated by $\approx$. The exact values are known and can be obtained straightforwardly as real roots of cubic/quartic polynomials, if necessary. In column five we summarize the signs of the eigenvalues of the Hessian matrix, again with $\lambda=\lambda_0$ inserted. Throughout the table, we set $c_1^\pm := \pm \sqrt{4 \lambda  (\lambda +1)+13}/(2 \sqrt{3})$, $c_2 := \left(8 \lambda ^3+28 \lambda ^2+22 \lambda +17\right)/32$, $c_3 := (-16 \lambda ^4+256 \lambda ^3+888 \lambda ^2+688 \lambda +443)/1152$, $c_4 := (-16 \lambda ^4+112 \lambda ^3+384 \lambda ^2+292 \lambda +137)/576$, $c_5 := (-16 \lambda ^4+64 \lambda ^3+216 \lambda ^2+160 \lambda +35)/384$, $c_6^\pm := (5 \pm 2 \sqrt{15})/2$ and $c_7^\pm := \sqrt{9 \pm 2 \sqrt{15}}$.}
\tablabel{Q111_critpts}
\end{table}
\footnotetext{For example, the second entry describes besides $(0,0,c_1^+)$ also $(0,c_1^+,0)$, $(c_1^+,0,0)$, $(0,0,c_1^-)$, $(0,c_1^-,0)$, $(c_1^-,0,0)$ and thus comes with multiplicity six. Note also that the sign of $c_1^+$ or $c_1^-$ in each entry of the vector $\boldsymbol{\mathbf{\phi}}^\top$ can be chosen independently of the other entries.}
The second case is $\phi_4 = (2 \lambda +3)/(2 \sqrt{2})$, implying
\begin{equation}\eqlabel{Q111_O3_model}
  {\cal E} = \sum_{\alpha=1}^3 \left\{ \sfrac{1}{2} \dot{\phi}_\alpha^2 + \sfrac{1}{8} \phi_\alpha^4 - \sfrac{4 \lambda  (\lambda +1)+13}{48} \phi_\alpha^2 + \sfrac{1}{96} (8 \lambda ^3+28 \lambda ^2+22 \lambda +17) \right\} \; .
\end{equation}
This is a three-component $\phi^4$ theory invariant under $S_3 \times (Z_2)^3$. The different types of critical points (modulo $S_3 \times (Z_2)^3$-transformations) together with some of their properties are listed in \tabref{Q111_critpts}, where we introduced the 3-vector $\boldsymbol{\mathbf{\phi}} = (\phi_1, \phi_2, \phi_3)^\top$. In total, we find 27 distinct critical points, which fall into four classes. 

As an example, we consider $\lambda = -1/2$ and construct the solutions which connect the critical points $(\pm 1, \pm 1, \pm 1)$ (upper part of row four in \tabref{Q111_critpts}). The energy density~\eqref*{Q111_O3_model} then becomes
\begin{equation}
  {\cal E} = \sum_{\alpha=1}^3 \left\{ \sfrac{1}{2} \dot{\phi}_\alpha^2 + \sfrac{1}{8} (\phi_\alpha^2 - 1)^2 \right\} \; .
\end{equation}
The model describes a 3-vector of independent $\phi^4$ kinks-/anti-kinks,
\begin{equation}
  \boldsymbol{\mathbf{\phi}} = (\pm \tanh\sfrac{\tau-\tau_{0,1}}{2}, \pm \tanh\sfrac{\tau-\tau_{0,2}}{2}, \pm \tanh\sfrac{\tau-\tau_{0,3}}{2})^\top \; .
\end{equation}
The positions ($\tau_{0,1}, \tau_{0,2}, \tau_{0,3} \in \RR$) and nature (kink (+) or anti-kink (-)) can be chosen independently in each entry of the 3-vector. This solution is physically allowed, since the total energy of the configuration, given by three times that of a single $\phi^4$ (anti-)kink (cf.~\eqref*{total_energy}),
\begin{equation}
   E = \int\limits_{-\infty}^\infty {\cal E}\, \dd\tau = 2 \; ,
\end{equation}
is finite.

Another example, which is rather similar to the previous one, is $\lambda = c_6^\pm$ (lower part of row four in \tabref{Q111_critpts}). The energy density turns into,
\begin{equation}
  {\cal E} = \sum_{\alpha=1}^3 \left\{ \sfrac{1}{2} \dot{\phi}_\alpha^2 + \sfrac{1}{8} \left(\phi_\alpha^2 - (c_7^\pm)^2 \right)^2 \right\} \; .
\end{equation}
This model describes a 3-vector of independent \emph{rescaled} $\phi^4$ kinks-/anti-kinks (cf.~\eqrangeref*{rescaled_phi4_kink1}{rescaled_phi4_kink2}),
\begin{equation}
  \boldsymbol{\mathbf{\phi}} = c_7^\pm \begin{pmatrix} \pm \tanh\left[ \sfrac{c_7^\pm}{2} (\tau - \tau_{0,1}) \right] \\ \pm \tanh\left[ \sfrac{c_7^\pm}{2} (\tau - \tau_{0,2}) \right] \\ \pm \tanh\left[ \sfrac{c_7^\pm}{2} (\tau - \tau_{0,3}) \right] \end{pmatrix} \; .
\end{equation}
Note that the choice of overall sign in each entry of $\boldsymbol{\mathbf{\phi}}$ is independent of the other entries and of the choice $c_7^+$ or $c_7^-$. The solution interpolates between $(\pm c_7^\pm, \pm c_7^\pm, \pm c_7^\pm)$ as $\tau\to\pm\infty$. The total energy of the configuration attains the value
\begin{equation}
   E = \int\limits_{-\infty}^\infty {\cal E}\, \dd\tau = 2 (c_7^\pm)^3 \; .
\end{equation}
Since $E<\infty$, the solution is also physically allowed.

\section{Conclusions and outlook}\seclabel{conclusions}

In this paper, we studied the instanton equation,~\eqref*{insteq}, and the Yang-Mills equation with torsion,~\eqref*{YM}, on cylinders $Z(G/H)=\RR\times G/H$ over seven-dimensional coset spaces $G/H$ with $G_2$-structure. The $G_2$-structure on $G/H$ lifts naturally to a Spin$(7)$-structure on $Z(G/H)$.

In order to be able to construct instanton and Yang-Mills solutions, we chose an ansatz for the gauge connection $A$, which can be written as
\begin{equation}
 A = e^i I_i + e^a X_a^b (\tau) I_b
\end{equation}
after gauge fixing. The dynamical degrees of freedom are carried by the $\tau$-dependent $7\times 7$-matrix $X_a^b (\tau)$, where $\tau$ is the cylinder coordinate. The matrix $X_a^b (\tau)$ is constrained by the requirement that $A$ be $G$-invariant, which translates into the condition $X_a^b f^c_{ib} = f^b_{ia} X_b^c$.

The simplest solution of this condition is given by a single-field ansatz $X_a^b (\tau) = \phi(\tau) \delta_a^b$. This common solution exists on any $G/H$ and has already been found in~\cite{Ivanova:2009yi}. 
Under certain additional assumptions, the instanton equation reduces to 
\begin{equation}\eqlabel{summary_insteq}
 \dot\phi = 2 \phi (\phi-1) \; ,
\end{equation}
modulo a trivial overall rescaling. \Eqref{summary_insteq} has two static solutions, $\phi = 0$, $1$, and the well-known interpolating kink solution $\phi(\tau) = \sfrac12 \left( 1 - \tanh\left[\tau - \tau_0 \right] \right)$, where $\tau_0 \in \RR$ is an arbitrary integration constant fixing the position of the instanton in the $\tau$ direction. 

The Yang-Mills equation with torsion reduces to 
\begin{equation}\eqlabel{summary_eom}
 \ddot{\phi} = \sfrac12 (1+\alpha) \phi (\phi - 1) \left(\phi - \sfrac{(\kappa+2)\alpha - 1}{\alpha + 1}\right) \; ,
\end{equation}
where $\alpha$ and $\kappa$ are two real parameters. The choice $\alpha=0$ leads to
\begin{equation}
 \ddot{\phi} = \sfrac12 \phi (\phi^2 - 1) \; ,
\end{equation}
which can be integrated to $\dot{\phi} = \pm \sfr12 (1-\phi^2)$. The non-trivial solution $\phi = \varpm \tanh \sfrac{\tau-\tau_0}{2}$ is a $\phi^4$ (anti-)kink with $\tau_0 \in \RR$ being an arbitrary integration constant fixing its position in the $\tau$ direction. For the choice $(\alpha, \kappa)=(3/5, 1)$, we find
\begin{equation}
 \ddot{\phi} = \sfrac45 \phi (\phi - 1) \left(\phi - \sfr12 \right) \; .
\end{equation}
This can be integrated to $\dot{\phi} = \pm \sfrac{1}{\sqrt{5}} \phi (\phi - 1)$, which is, up to rescaling, equivalent to the tanh-kink-type instanton~\eqref*{summary_insteq}.

In the general case, there are multiple $\tau$-dependent scalar fields given by the matrix components $X_a^b (\tau)$ and restricted by $X_a^b f^c_{ib} = f^b_{ia} X_b^c$. This restriction involves the structure constants $f^A_{BC}$ of $G$ and hence, the solutions of this condition will depend on the choice of coset space $G/H$. In particular, even the number of independent components of $X_a^b$, or in other words, the number of scalar degrees of freedom, is different for different coset spaces. This forbids a unified treatment. Instead, we had to consider the multi-field set-up on a case-by-case basis.

In total, we considered seven explicit coset space constructions that fall into two families, namely those with nearly parallel $G_2$-structure and those with $SU(3)$-structure. The first family comprises the Berger space $SO(5)/SO(3)_{\text{max}}$, the squashed 7-sphere $Sp(2)\times Sp(1)/Sp(1)^2$, and the Aloff-Wallach spaces $SU(3)/U(1)_{k,l}$, for a co-prime pair of integers $(k,l)$. All three seven-dimensional coset spaces admit a nearly parallel $G_2$-structure and an induced Spin$(7)$-structure on the cylinder thereover.

For $G/H$ equal to $SO(5)/SO(3)_{\text{max}}$, the $G$-invariance condition forces $X_a^b (\tau) = \phi(\tau) \delta_a^b$. Thus, this case reduces to the single-field set-up already discussed above. In the second example of this group, namely $Sp(2)\times Sp(1)/Sp(1)^2$, we a priori have two real scalar fields $\phi_1 (\tau)$ and $\phi_2 (\tau)$, after solving the $G$-invariance condition. However, the dynamical equations (that is, both, the instanton equation and the Yang-Mills equation with torsion) force the two fields to be proportional to one another or one field to be constant. In effect, also this case reduces to the well-known single-field scenario. The instanton equation on $\RR \times SU(3)/U(1)_{k,l}$ has already been analyzed in~\cite{Haupt:2011mg} and was reviewed in \secref{AW} for completeness. There we also studied the Yang-Mills equation with torsion on $\RR \times SU(3)/U(1)_{k,l}$. After solving the $G$-invariance condition, we showed that it can be reduced to a gradient system in three complex and two real scalar fields. The gradient system is determined by a quartic potential and subject to the Gauss-law constraint. Beyond the instanton constructions, we obtained two further analytical solutions in other special corners of configuration and parameter space. The first one corresponds to a single varying field which however diverges at $\tau\to\pm\infty$ either exponentially or quadratically depending on the values of the parameters. The second one describes a rescaled $\phi^4$ kink/anti-kink.

The second family of manifolds considered in this paper comprises seven-dimensional coset spaces with $SU(3)$-structure. Since $SU(3)$-structures on \emph{seven}-dimensional manifolds are less common in the string theory literature than $SU(3)$-structures on \emph{six}-dimensional manifolds, we reviewed key properties and differences between the two cases in \secref{SU3_struct}. We analyzed in detail four such coset spaces, namely $SO(5)/SO(3)_{A+B}$, $N^{pqr} = (SU(3)\times U(1)) / (U(1) \times U(1))$, $M^{pqr} = (SU(3)\times SU(2)\times U(1)) / (SU(2)\times U(1) \times U(1))$ and $Q^{pqr} = (SU(2)\times SU(2)\times SU(2)) / (U(1) \times U(1))$, where $(p,q,r)$ is a triple of mutually co-prime integers. In the latter two cases, an $SU(3)$-structure only exists for particular choices of $(p,q,r)$ and hence, we restricted to $M^{110}$ and $Q^{111}$, respectively. 

Then, in each of the four cases $SO(5)/SO(3)_{A+B}$, $N^{pqr}$, $M^{110}$ and $Q^{111}$, the $SU(3)$-structure induces two cocalibrated (or semi-parallel) $G_2$-structures on the coset space and two Spin$(7)$-structures on the cylinder thereover. After solving the $G$-invariance condition, we are left with five, seven, five and seven real $\tau$-dependent scalar fields, respectively. The instanton equation~\eqref*{insteq} reduces to a system of coupled non-linear first-order ordinary differential equations in those fields, supplemented by a number of (typically quadratic) algebraic constraints known as quiver relations. The full system of equations can generally not be solved analytically. We did present however several solutions in special corners of field- and parameter-space. These include the trivial case of constant (that is, $\tau$-independent) solutions, unphysical configurations exhibiting exponential growth, but also solutions that reduce to the well-known tanh-kink-type instanton~\eqref*{summary_insteq}.

The Yang-Mills equation with torsion for the four cases $SO(5)/SO(3)_{A+B}$, $N^{pqr}$, $M^{110}$ and $Q^{111}$ reduces \emph{inter alia} to a system of coupled non-linear second-order ordinary differential equations. This system of equations is always expressible (possibly, after performing some suitable field redefinitions) as a gradient system,
\begin{equation}\eqlabel{summary_gradsys}
 \ddot{\phi}_\alpha = \eta_{\alpha\beta} \frac{\partial V}{\partial\phi_\beta}  \; ,
\end{equation}
with a quartic potential $V = V(\{ \phi_\alpha \})$ and a constant metric $\eta_{\alpha\beta}$. In addition, we found a set of quadratic quiver relations and first-order constraints. The latter originate from the Gauss-law constraint~\eqref*{gauss_law} and are generally of the form $f_1(\{\phi_\alpha \dot{\phi}_\beta \}) = 0$, $f_2(\{\phi_\alpha \dot{\phi}_\beta \}) = 0$, etc., where $f_1$, $f_2$, etc. are some linear functions. 

In total, we faced an overdetermined system of coupled non-linear algebraic as well as differential equations. Any attempt at solving such a system is a formidable task. It is however possible to consistently restrict one's attention to a reduced set of scalar degrees of freedom such that the algebraic and first-order constraints are trivially satisfied, while retaining the non-trivial multi-dimensional nature of the motion. This is achieved by setting some of the scalar fields equal to one another or to constants and checking that this is compatible with all equations. The remaining equations still form a gradient system as in~\eqref*{summary_gradsys}, but for a subset of $\{ \phi_\alpha \}$.

We applied this procedure to each of the four cases $SO(5)/SO(3)_{A+B}$, $N^{pqr}$, $M^{110}$ and $Q^{111}$. The resulting gradient systems can be analyzed by standard methods. First, we computed the critical points of the respective potential $V$. We are interested in non-constant finite-energy solutions flowing from a critical point at $\tau\to -\infty$ to a critical point at $\tau\to +\infty$. Such solutions necessarily interpolate between zero-potential critical points.

In this way, we obtained a multitude of solutions on coset spaces that have not been studied before in this context. Besides constant solutions, linearly diverging solutions and single-field tanh-kink-type solutions, we found a couple of additional solutions which are new and interesting. For example in the case $G/H = SO(5)/SO(3)_{A+B}$, there exists a set-up where the dynamics is described by a single field $\phi$ governed by an equation of motion of the form,
\begin{equation}
 \ddot{\phi} = \phi^3 \mp \sfrac{1}{2 \sqrt{3}} \phi^2 - \sfrac{1}{2} \phi \; .
\end{equation}
Intriguingly, this equation can in general not be integrated to a first-order equation. However, we have not been able to find a non-constant finite-energy solution. For $G/H = M^{110}$, we showed that in a special corner of field-space the gradient system~\eqref*{summary_gradsys} de-couples, thus yielding two independent copies of~\eqref*{summary_eom} with different values of the parameters. We constructed several different types of analytical solutions for this case and showed their plots in \figref{M110_plot}. Most of them correspond to finite-energy configurations. Finally, for $G/H = Q^{111}$, we set one field to a constant and obtained a three-component $\phi^4$ theory invariant under $S_3 \times (Z_2)^3$. Non-trivial finite-energy configurations are given by a 3-vector with independent (rescaled) $\phi^4$ kinks/anti-kinks in each entry of the vector.

We end this paper by pointing out some open problems and possible future directions. Besides the obvious possibility of extending our methods to even more coset space constructions, it may also be fruitful to analyze in more detail the systems of coupled non-linear algebraic and differential equations obtained in this work. In particular, a more systematic and exhaustive exploration of the various parameter spaces may lead to new analytical or numerical solutions.

Another direction in which this work can be continued is to alter the properties of the cylinder direction. For example, one may consider the instanton and Yang-Mills equations on a Lorentzian cylinder $i\RR\times G/H$, that is sending $\tau \to i t$, or on $S^1 \times G/H$, that is identifying $\tau \simeq \tau + L$. The corresponding finite-energy solutions offer an interpretation as dyons (for $i\RR\times G/H$) and sphalerons (for $S^1 \times G/H$), respectively.

Finally, it would be interesting to examine possible embeddings into string theory or supergravity. A promising candidate for the embedding is provided by heterotic supergravity. One may attempt a construction analogous to~\cite{Chatzistavrakidis:2009mh,Klaput:2011mz,Gray:2012md,Haupt:2014ufa}, where ten-dimensional space-time is taken to be a (possibly) warped product of a four-dimensional external part and a compact six-dimensional $SU(3)$-structure manifold. With the set-up of this paper, one may instead consider a product of a three-dimensional external part and a compact seven-dimensional $G_2$-structure coset space. Some explicit constructions of this type have already been realized, for example, in~\cite{Harland:2011zs,Gemmer:2012pp}. However, these constructions are based on specific assumptions, which may not exhaust the possible types of embeddings. It is left for future work to determine whether other types of embeddings are also viable.

\section*{Acknowledgments}

The author is very grateful to Maike~Torm\"ahlen for collaboration in the early stages of this work and for constructive comments on the first version of the manuscript. The author is indebted to Olaf~Lechtenfeld and Felix~Lubbe for helpful discussions. The author thanks the Max-Planck-Institute for Physics in Munich, in particular its director Dieter~L\"ust, and the CERN Theory Division for warm hospitality and generous financial support during part of the work. This work was supported by the German Science Foundation (DFG) under the Collaborative Research Center (SFB) 676 ``Particles, Strings and the Early Universe''.

\appendix
\section*{Appendix}

\section{Structure Constants}\applabel{struct_consts}

In this appendix, we list the structure constants of the coset spaces appearing in this paper. Note that $f^A_{[BC]} = f^A_{BC}$, by definition. For most coset spaces, we list below only $f^A_{BC}$ for $B<C$. It is understood that $f^A_{BC} = - f^A_{CB}$, if $B>C$.

\stoptocwriting
\subsection{\texorpdfstring{$SO(5)/SO(3)_{\text{max}}$}{Berger space}}\applabel{struct_consts_SO5_SO3_max}

Here, we list the structure constants $f^A_{BC}$ of $\mathfrak{g}=\mathfrak{so}(5)$ used in \secref{SO5_SO3_max}. They were extracted from eq.~(B.2) in~\cite{Castellani:1983yg} and rescaled, $f^C_{AB} \rightarrow \sfrac{1}{\sqrt{6}} f^C_{AB}$. The only non-vanishing components are given by
\begin{align}
f^{1}_{23}&=m_1^+   , & f^{1}_{56}&=m_2^-   , & f^{1}_{59}&=m_3^-   , & f^{1}_{68}&=m_3^-   , & f^{1}_{7,10}&=m_4^+   , & f^{1}_{89}&=m_5^+   , & \\
f^{2}_{13}&=m_1^-   , & f^{2}_{46}&=m_2^+   , & f^{2}_{49}&=m_3^-   , & f^{2}_{67}&=m_3^-   , & f^{2}_{79}&=m_5^-   , & f^{2}_{8,10}&=m_4^+   , & \\
f^{3}_{12}&=m_1^+   , & f^{3}_{45}&=m_2^-   , & f^{3}_{48}&=m_3^-   , & f^{3}_{57}&=m_3^-   , & f^{3}_{78}&=m_5^+   , & f^{3}_{9,10}&=m_4^+   , & \\
f^{4}_{26}&=m_2^-   , & f^{4}_{29}&=m_3^+   , & f^{4}_{35}&=m_2^+   , & f^{4}_{38}&=m_3^+   , & f^{4}_{56}&=m_1^+   , & f^{4}_{7,10}&=m_1^+   , & \\
f^{5}_{16}&=m_2^+   , & f^{5}_{19}&=m_3^+   , & f^{5}_{34}&=m_2^-   , & f^{5}_{37}&=m_3^+   , & f^{5}_{46}&=m_1^-   , & f^{5}_{79}&=m_1^+   , & \\
f^{6}_{15}&=m_2^-   , & f^{6}_{18}&=m_3^+   , & f^{6}_{24}&=m_2^+   , & f^{6}_{27}&=m_3^+   , & f^{6}_{45}&=m_1^+   , & f^{6}_{78}&=m_1^-   , & \\
f^{4}_{89}&=m_1^-   , & f^{5}_{8,10}&=m_1^+   , & f^{6}_{9,10}&=m_1^+   , & & & & \\
f^{7}_{1,10}&=m_4^-   , & f^{7}_{26}&=m_3^-   , & f^{7}_{29}&=m_5^+   , & f^{7}_{35}&=m_3^-   , & f^{7}_{38}&=m_5^-   , & f^{7}_{4,10}&=m_1^-   , & \\
f^{8}_{16}&=m_3^-   , & f^{8}_{19}&=m_5^-   , & f^{8}_{2,10}&=m_4^-   , & f^{8}_{34}&=m_3^-   , & f^{8}_{37}&=m_5^+   , & f^{8}_{49}&=m_1^+   , & \\
f^{9}_{15}&=m_3^-   , & f^{9}_{18}&=m_5^+   , & f^{9}_{24}&=m_3^-   , & f^{9}_{27}&=m_5^-   , & f^{9}_{3,10}&=m_4^-   , & f^{9}_{48}&=m_1^-   , & \\
f^{7}_{59}&=m_1^-   , & f^{7}_{68}&=m_1^+   , & f^{8}_{5,10}&=m_1^-   , & f^{8}_{67}&=m_1^-   , & f^{9}_{57}&=m_1^+   , & f^{9}_{6,10}&=m_1^-   , & \\
f^{10}_{17}&=m_4^+   , & f^{10}_{28}&=m_4^+   , & f^{10}_{39}&=m_4^+   , & f^{10}_{47}&=m_1^+   , & f^{10}_{58}&=m_1^+   , & f^{10}_{69}&=m_1^+   , &
\end{align}
where $m_1^\pm = \pm 1/\sqrt{30}$, $m_2^\pm = \pm \sqrt{3/40}$, $m_3^\pm = \pm 1/(2 \sqrt{2})$, $m_4^\pm = \pm \sqrt{2/15}$ and $m_5^\pm = \pm 1/(2 \sqrt{30})$. The two sets of indices $\{1,2,3\}$ and $\{4,\ldots,10\}$ correspond to the Lie subalgebra $\mathfrak{h}=\mathfrak{so}(3)$ and the coset space directions $\mathfrak{m}$, respectively.

\subsection{\texorpdfstring{$Sp(2)\times Sp(1)/Sp(1)^2$}{Squashed seven-sphere}}\applabel{struct_consts_SquashedS7}

Here, we list the structure constants $f^A_{BC}$ of $\mathfrak{g}=\mathfrak{sp}(2)\oplus\mathfrak{sp}(1)$ used in \secref{SquashedS7}. They were extracted from eq.~(3.8) in~\cite{Bais:1983wc} and rescaled according to~\eqref*{SquashedS7_rescalings}. The only non-vanishing components are given by
\begin{align}
f^{1}_{25}&=\ell_1^+ \; , & f^{1}_{28}&=\ell_2^- \; , & f^{1}_{2,11}&=\ell_3^+ \; , & f^{1}_{36}&=\ell_1^+ \; , & f^{1}_{39}&=\ell_2^- \; , & \\ 
    f^{1}_{3,12}&=\ell_3^+ \; , & f^{1}_{47}&=\ell_1^+ \; , & f^{1}_{4,10}&=\ell_2^- \; , & f^{1}_{4,13}&=\ell_3^+ \; , & \\
f^{2}_{15}&=\ell_1^- \; , & f^{2}_{18}&=\ell_2^+ \; , & f^{2}_{1,11}&=\ell_3^- \; , & f^{2}_{37}&=\ell_1^+ \; , & f^{2}_{3,10}&=\ell_2^+ \; , & \\ 
    f^{2}_{3,13}&=\ell_3^+ \; , & f^{2}_{46}&=\ell_1^- \; , & f^{2}_{49}&=\ell_2^- \; , & f^{2}_{4,12}&=\ell_3^- \; , & \\
f^{3}_{16}&=\ell_1^- \; , & f^{3}_{19}&=\ell_2^+ \; , & f^{3}_{1,12}&=\ell_3^- \; , & f^{3}_{27}&=\ell_1^- \; , & f^{3}_{2,10}&=\ell_2^- \; , & \\ 
    f^{3}_{2,13}&=\ell_3^- \; , & f^{3}_{45}&=\ell_1^+ \; , & f^{3}_{48}&=\ell_2^+ \; , & f^{3}_{4,11}&=\ell_3^+ \; , & \\
f^{4}_{17}&=\ell_1^- \; , & f^{4}_{1,10}&=\ell_2^+ \; , & f^{4}_{1,13}&=\ell_3^- \; , & f^{4}_{26}&=\ell_1^+ \; , & f^{4}_{29}&=\ell_2^+ \; , & \\ 
    f^{4}_{2,12}&=\ell_3^+ \; , & f^{4}_{35}&=\ell_1^- \; , & f^{4}_{38}&=\ell_2^- \; , & f^{4}_{3,11}&=\ell_3^- \; , & \\
f^{5}_{12}&=\ell_1^+ \; , & f^{5}_{34}&=\ell_1^+ \; , & f^{5}_{67}&=\ell_1^- \; , & f^{5}_{6,13}&=\ell_4^+ \; , & f^{5}_{7,12}&=\ell_4^- \; , & \\
f^{6}_{13}&=\ell_1^+ \; , & f^{6}_{24}&=\ell_1^- \; , & f^{6}_{57}&=\ell_1^+ \; , & f^{6}_{5,13}&=\ell_4^- \; , & f^{6}_{7,11}&=\ell_4^+ \; , & \\
f^{7}_{14}&=\ell_1^+ \; , & f^{7}_{23}&=\ell_1^+ \; , & f^{7}_{56}&=\ell_1^- \; , & f^{7}_{5,12}&=\ell_4^+ \; , & f^{7}_{6,11}&=\ell_4^- \; , & \\
f^{8}_{12}&=\ell_2^- \; , & f^{8}_{34}&=\ell_2^+ \; , & f^{8}_{9,10}&=\ell_5^+ \; , & \\
f^{9}_{13}&=\ell_2^- \; , & f^{9}_{24}&=\ell_2^- \; , & f^{9}_{8,10}&=\ell_5^- \; , & \\
f^{10}_{14}&=\ell_2^- \; , & f^{10}_{23}&=\ell_2^+ \; , & f^{10}_{89}&=\ell_5^+ \; , & \\
f^{11}_{12}&=\ell_3^+ \; , & f^{11}_{34}&=\ell_3^+ \; , & f^{11}_{67}&=\ell_4^+ \; , & f^{11}_{12,13}&=\ell_4^+ \; , & \\
f^{12}_{13}&=\ell_3^+ \; , & f^{12}_{24}&=\ell_3^- \; , & f^{12}_{57}&=\ell_4^- \; , & f^{12}_{11,13}&=\ell_4^- \; , & \\
f^{13}_{14}&=\ell_3^+ \; , & f^{13}_{23}&=\ell_3^+ \; , & f^{13}_{56}&=\ell_4^+ \; , & f^{13}_{11,12}&=\ell_4^+ \; ,
\end{align}
where $\ell_1^\pm = \pm 1/\sqrt{30}$, $\ell_2^\pm = \pm 1/(2 \sqrt{3})$, $\ell_3^\pm = \pm 1/(2 \sqrt{5})$, $\ell_4^\pm = \pm 1/\sqrt{5}$ and $\ell_5^\pm = \pm 1/\sqrt{3}$. The two sets of indices $\{1,\ldots,7\}$ and $\{8,\ldots,13\}$ correspond to the coset space directions $\mathfrak{m}$ and the Lie subalgebra $\mathfrak{h}=\mathfrak{sp}(1)\oplus\mathfrak{sp}(1)$, respectively.

\subsection{\texorpdfstring{$SU(3)/U(1)_{k,l}$}{Aloff-Wallach spaces}}\applabel{struct_consts_AW}

Here, we list the structure constants $f^A_{BC}$ of $\mathfrak{g}=\mathfrak{su}(3)$ used in \secref{AW}. They were extracted from section 2 of~\cite{Haupt:2011mg}. The only non-vanishing components are given by
\begin{align}
f^{1}_{27}&=\sfrac{4 k-2 l}{\Delta  \mu } \; , & f^{1}_{28}&=-\sfrac{2 (k+2 l)}{\sqrt{3} \Delta  \mu } \; , & f^{1}_{35}&=-\sfrac{\zeta _2 \zeta _3}{\zeta _1} \; , & f^{1}_{46}&=\sfrac{\zeta _2 \zeta _3}{\zeta _1} \; , \\
f^{2}_{17}&=\sfrac{2 l-4 k}{\Delta  \mu } \; , & f^{2}_{18}&=\sfrac{2 (k+2 l)}{\sqrt{3} \Delta  \mu } \; , & f^{2}_{36}&=\sfrac{\zeta _2 \zeta _3}{\zeta _1} \; , & f^{2}_{45}&=\sfrac{\zeta _2 \zeta _3}{\zeta _1} \; , \\
f^{3}_{15}&=\sfrac{\zeta _1 \zeta _3}{\zeta _2} \; , & f^{3}_{26}&=-\sfrac{\zeta _1 \zeta _3}{\zeta _2} \; , & f^{3}_{47}&=-\sfrac{2 (k-2 l)}{\Delta  \mu } \; , & f^{3}_{48}&=\sfrac{2 (2 k+l)}{\sqrt{3} \Delta  \mu } \; , \\
f^{4}_{16}&=-\sfrac{\zeta _1 \zeta _3}{\zeta _2} \; , & f^{4}_{25}&=-\sfrac{\zeta _1 \zeta _3}{\zeta _2} \; , & f^{4}_{37}&=\sfrac{2 (k-2 l)}{\Delta  \mu } \; , & f^{4}_{38}&=-\sfrac{2 (2 k+l)}{\sqrt{3} \Delta  \mu } \; , \\
f^{5}_{13}&=-\sfrac{\zeta _1 \zeta _2}{\zeta _3} \; , & f^{5}_{24}&=\sfrac{\zeta _1 \zeta _2}{\zeta _3} \; , & f^{5}_{67}&=-\sfrac{2 (k+l)}{\Delta  \mu } \; , & f^{5}_{68}&=\sfrac{2 (l-k)}{\sqrt{3} \Delta  \mu } \; , \\
f^{6}_{14}&=\sfrac{\zeta _1 \zeta _2}{\zeta _3} \; , & f^{6}_{23}&=\sfrac{\zeta _1 \zeta _2}{\zeta _3} \; , & f^{6}_{57}&=\sfrac{2 (k+l)}{\Delta  \mu } \; , & f^{6}_{58}&=\sfrac{2 (k-l)}{\sqrt{3} \Delta  \mu } \; , \\
f^{7}_{12}&=\sfrac{2 \zeta _1^2 k \mu }{\Delta } \; , & f^{7}_{34}&=\sfrac{2 \zeta _2^2 l \mu }{\Delta } \; , & f^{7}_{56}&=-\sfrac{2 \zeta _3^2 \mu  (k+l)}{\Delta } \; , \\
f^{8}_{12}&=-\sfrac{2 \sqrt{3} \zeta _1^2 l \mu }{\Delta } \; , & f^{8}_{34}&=\sfrac{2 \sqrt{3} \zeta _2^2 k \mu }{\Delta } \; , & f^{8}_{56}&=\sfrac{2 \sqrt{3} \zeta _3^2 \mu  (l-k)}{\Delta } \; ,
\end{align}
where $\De^2:=2(k^2+l^2)$ and $\s_1,\s_2,\s_3,\mu\in\RR$ are $(k,l)$-dependent rescaling parameters appropriately chosen, such that the coset space metric $g_{G/H} = \delta_{ab}\,\et^a \otimes \et^b$ is Einstein for a connection with a torsion 3-form $P$ given by~\eqref*{P_AW}.

It is useful to also consider the complexified Lie algebra $\mathfrak{su}(3)\otimes\mathbb{C}$. The corresponding structure constants can be read off from eq.~(2.29) in~\cite{Haupt:2011mg},
\begin{align}
&\Ct^{1}_{3\2}=-\sfrac{\s_2\,\s_3}{\s_1}=-\Ct^{\1}_{2\3}\ , &
&\Ct^{2}_{3\1}=\sfrac{\s_3\,\s_1}{\s_2}= -\Ct^{\2}_{1\3}\ , &
&\Ct^{3}_{12}=\sfrac{\s_1\,\s_2}{\s_3}= \Ct^{\3}_{\1\2}\ ,\\
&\Ct^1_{71}=\sfrac{2(2k-l)}{\mu\De}=-\Ct^{\1}_{7\1}\ , &
&\Ct^2_{72}=\sfrac{2(2l-k)}{\mu\De}=-\Ct^{\2}_{7\2}\ , &
&\Ct^3_{73}=\sfrac{2(k+l)}{\mu\De}=-\Ct^{\3}_{7\3}\ ,\\
&\Ct^1_{81}{=}{-}\sfrac{2(k{+}2l)}{\sqrt{3}\mu\De}{=}{-}\Ct^{\1}_{8\1}\ , &
&\Ct^2_{82}{=}\sfrac{2(2k{+}l)}{\sqrt{3}\mu\De}{=}{-}\Ct^{\2}_{8\2}\ , &
&\Ct^3_{83}{=}\sfrac{2(k{-}l)}{\sqrt{3}\mu\De}{=}{-}\Ct^{\3}_{8\3}\ ,\\
&\Ct^7_{1\1}=-\sfrac{\mu\s_1^2}{\De}k\ , &
&\Ct^7_{2\2}=-\sfrac{\mu\s_2^2}{\De}l \ , &
&\Ct^7_{3\3}=-\sfrac{\mu\s_3^2}{\De}(k+l) \ ,\\
&\Ct^8_{1\1}=\sfrac{\sqrt{3}\mu\s_1^2}{\De}l\ , &
&\Ct^8_{2\2}=-\sfrac{\sqrt{3}\mu\s_2^2}{\De}k \ , &
&\Ct^8_{3\3}=\sfrac{\sqrt{3}\mu\s_3^2}{\De}(l-k) \ .
\end{align}

\subsection{\texorpdfstring{$SO(5)/SO(3)_{A+B}$}{SO5/SO3 subscript (A+B)}}\applabel{struct_consts_SO5_SO3_AB}

Here, we list the structure constants $f^A_{BC}$ of $\mathfrak{g}=\mathfrak{so}(5)$ used in \secref{SO5_SO3_AB}. They were extracted from eq.~(D.12) in~\cite{Micu:2006ey} and rescaled, $f^C_{AB} \rightarrow \sfrac{1}{\sqrt{6}} f^C_{AB}$. The only non-vanishing components are given by
\begin{align}
f^{1}_{2,10}&=c_+ \; , & f^{1}_{39}&=c_- \; , & f^{1}_{45}&=c_- \; , &
f^{2}_{1,10}&=c_- \; , & f^{2}_{38}&=c_+ \; , & f^{2}_{46}&=c_- \; , \\
f^{3}_{19}&=c_+ \; , & f^{3}_{28}&=c_- \; , & f^{3}_{47}&=c_- \; , &
f^{4}_{15}&=c_+ \; , & f^{4}_{26}&=c_+ \; , & f^{4}_{37}&=c_+ \; , \\
f^{5}_{14}&=c_- \; , & f^{5}_{6,10}&=c_+ \; , & f^{5}_{79}&=c_- \; , &
f^{6}_{24}&=c_- \; , & f^{6}_{5,10}&=c_- \; , & f^{6}_{78}&=c_+ \; , \\
f^{7}_{34}&=c_- \; , & f^{7}_{59}&=c_+ \; , & f^{7}_{68}&=c_- \; , &
f^{8}_{23}&=c_+ \; , & f^{8}_{67}&=c_+ \; , & f^{8}_{9,10}&=c_+ \; , \\
f^{9}_{13}&=c_- \; , & f^{9}_{57}&=c_- \; , & f^{9}_{8,10}&=c_- \; , &
f^{10}_{12}&=c_+ \; , & f^{10}_{56}&=c_+ \; , & f^{10}_{89}&=c_+ \; ,
\end{align}
where $c_\pm = \pm 1/\sqrt{6}$. The two sets of indices $\{1,\ldots,7\}$ and $\{8,9,10\}$ correspond to the coset space directions $\mathfrak{m}$ and the Lie subalgebra $\mathfrak{h}=\mathfrak{so}(3)$, respectively.

\subsection{\texorpdfstring{$N^{pqr} = (SU(3)\times U(1)) / (U(1) \times U(1))$}{Npqr}}\applabel{struct_consts_Npqr}

Here, we list the structure constants $f^A_{BC}$ of $\mathfrak{g}=\mathfrak{su}(3)\oplus\mathfrak{u}(1)$ used in \secref{Npqr}. They were extracted from eq.~(2.6) in~\cite{Castellani:1983tc} and rescaled according to~\eqref*{Npqr_rescalings}. The only non-vanishing components are given by
\begin{align}
f^{1}_{27}&=\sfrac{q}{\sqrt{3} \eta} \; , & f^{1}_{28}&=\sqrt{\sfrac{2}{3}} \sfrac{q r}{\eta \zeta} \; , & f^{1}_{29}&=\sfrac{p}{\eta} \; , & f^{1}_{36}&=\sfrac{1}{2 \sqrt{3}} \; , & f^{1}_{45}&=-\sfrac{1}{2 \sqrt{3}} \; , & \\
f^{2}_{17}&=-\sfrac{q}{\sqrt{3} \eta} \; , & f^{2}_{18}&=-\sqrt{\sfrac{2}{3}} \sfrac{q r}{\eta \zeta} \; , & f^{2}_{19}&=-\sfrac{p}{\eta} \; , & f^{2}_{35}&=\sfrac{1}{2 \sqrt{3}} \; , & f^{2}_{46}&=\sfrac{1}{2 \sqrt{3}} \; , & \\
f^{3}_{16}&=-\sfrac{1}{2 \sqrt{3}} \; , & f^{3}_{25}&=-\sfrac{1}{2 \sqrt{3}} \; , & f^{3}_{47}&=\sfrac{3 p+q}{2 \sqrt{3} \eta} \; , & f^{3}_{48}&=\sfrac{r (3 p+q)}{\sqrt{6} \eta \zeta} \; , & f^{3}_{49}&=\sfrac{p-q}{2 \eta} \; , & \\
f^{4}_{15}&=\sfrac{1}{2 \sqrt{3}} \; , & f^{4}_{26}&=-\sfrac{1}{2 \sqrt{3}} \; , & f^{4}_{37}&=-\sfrac{3 p+q}{2 \sqrt{3} \eta} \; , & f^{4}_{38}&=-\sfrac{r (3 p+q)}{\sqrt{6} \eta \zeta} \; , & f^{4}_{39}&=-\sfrac{p-q}{2 \eta} \; , & \\
f^{5}_{14}&=-\sfrac{1}{2 \sqrt{3}} \; , & f^{5}_{23}&=\sfrac{1}{2 \sqrt{3}} \; , & f^{5}_{67}&=\sfrac{3 p-q}{2 \sqrt{3} \eta} \; , & f^{5}_{68}&=\sfrac{r (3 p-q)}{\sqrt{6} \eta \zeta} \; , & f^{5}_{69}&=-\sfrac{p+q}{2 \eta} \; , & \\
f^{6}_{13}&=\sfrac{1}{2 \sqrt{3}} \; , & f^{6}_{24}&=\sfrac{1}{2 \sqrt{3}} \; , & f^{6}_{57}&=-\sfrac{3 p-q}{2 \sqrt{3} \eta} \; , & f^{6}_{58}&=-\sfrac{r (3 p-q)}{\sqrt{6} \eta \zeta} \; , & f^{6}_{59}&=\sfrac{p+q}{2 \eta} \; , & \\
f^{7}_{12}&=\sfrac{q \eta}{\sqrt{3} \zeta^2} \; , & f^{7}_{34}&=\sfrac{\eta (3 p+q)}{2 \sqrt{3} \zeta^2} \; , & f^{7}_{56}&=\sfrac{\eta (3 p-q)}{2 \sqrt{3} \zeta^2} \; , & \\
f^{8}_{12}&=\sqrt{\sfrac{2}{3}} \sfrac{q r}{\eta \zeta} \; , & f^{8}_{34}&=\sfrac{r (3 p+q)}{\sqrt{6} \eta \zeta} \; , & f^{8}_{56}&=\sfrac{r (3 p-q)}{\sqrt{6} \eta \zeta} \; , & \\
f^{9}_{12}&=\sfrac{p}{\eta} \; , & f^{9}_{34}&=\sfrac{p-q}{2 \eta} \; , & f^{9}_{56}&=-\sfrac{p+q}{2 \eta} \; ,
\end{align}
where $\zeta := \sqrt{3 p^2+q^2+2 r^2}$ and $\eta := \sqrt{3 p^2+q^2}$. The two sets of indices $\{1,\ldots,7\}$ and $\{8,9\}$ correspond to the coset space directions $\mathfrak{m}$ and the Lie subalgebra $\mathfrak{h}=\mathfrak{u}(1)\oplus\mathfrak{u}(1)$, respectively.

\subsection{\texorpdfstring{$M^{pqr} = (SU(3)\times SU(2)\times U(1)) / (SU(2)\times U(1) \times U(1))$}{Mpqr}}\applabel{struct_consts_Mpqr}

Here, we list the structure constants $f^A_{BC}$ of $\mathfrak{g}=\mathfrak{su}(3)\oplus\mathfrak{su}(2)\oplus\mathfrak{u}(1)$ used in \secref{Mpqr}. They were extracted from appendix~F of~\cite{Karthauser:2006wb} and rescaled according to~\eqref*{Mpqr_rescalings}. The only non-vanishing components are given by
\begin{align}
f^{1}_{27}&=\ell_1^+ \; , & f^{1}_{2,10}&=\ell_2^+ \; , & f^{1}_{2,11}&=\ell_3^+ \; , & f^{1}_{2,12}&=\ell_4^- \; , & f^{1}_{39}&=\ell_2^+ \; , & f^{1}_{48}&=\ell_2^+ \; , \\
f^{2}_{17}&=\ell_1^- \; , & f^{2}_{1,10}&=\ell_2^- \; , & f^{2}_{1,11}&=\ell_3^- \; , & f^{2}_{1,12}&=\ell_4^+ \; , & f^{2}_{38}&=\ell_2^- \; , & f^{2}_{49}&=\ell_2^+ \; , \\
f^{3}_{19}&=\ell_2^- \; , & f^{3}_{28}&=\ell_2^+ \; , & f^{3}_{47}&=\ell_1^+ \; , & f^{3}_{4,10}&=\ell_2^- \; , & f^{3}_{4,11}&=\ell_3^+ \; , & f^{3}_{4,12}&=\ell_4^- \; , \\
f^{4}_{18}&=\ell_2^- \; , & f^{4}_{29}&=\ell_2^- \; , & f^{4}_{37}&=\ell_1^- \; , & f^{4}_{3,10}&=\ell_2^+ \; , & f^{4}_{3,11}&=\ell_3^- \; , & f^{4}_{3,12}&=\ell_4^+ \; , \\
f^{5}_{67}&=\sfrac{q}{\rho } \; , & f^{5}_{6,11}&=\ell_5^+ \; , & f^{5}_{6,12}&=\ell_6^+ \; , & f^{8}_{14}&=\ell_7^+ \; , & f^{8}_{23}&=\ell_7^- \; , & f^{8}_{9,10}&=1 \; , \\
f^{6}_{57}&=-\sfrac{q}{\rho } \; , & f^{6}_{5,11}&=\ell_5^- \; , & f^{6}_{5,12}&=\ell_6^- \; , & f^{9}_{13}&=\ell_7^+ \; , & f^{9}_{24}&=\ell_7^+ \; , & f^{9}_{8,10}&=-1 \; , \\
f^{7}_{12}&=\sfrac{p \rho }{2 \zeta ^2} \; , & f^{7}_{34}&=\sfrac{p \rho }{2 \zeta ^2} \; , & f^{7}_{56}&=\sfrac{q \rho }{2 \zeta ^2} \; , & f^{10}_{12}&=\ell_7^+ \; , & f^{10}_{34}&=\ell_7^- \; , & f^{10}_{89}&=1 \; , \\
f^{11}_{12}&=p \ell_8 \; , & f^{11}_{34}&=p \ell_8 \; , & f^{11}_{56}&=q \ell_8 \; , & f^{12}_{12}&=\ell_9 \; , & f^{12}_{34}&=\ell_9 \; , & f^{12}_{56}&=\sfrac{\sqrt{3} p}{2 \eta } \; , 
\end{align}
where $\ell_1^\pm = \pm 3 p/(2 \rho)$, $\ell_2^\pm = \pm 1/2$, $\ell_3^\pm = \pm 3 p r / (\sqrt{2} \zeta  \eta )$, $\ell_4^\pm = \pm \sqrt{3} q/ (2 \eta )$, $\ell_5^\pm = \pm \sqrt{2} q r / (\zeta  \eta )$, $\ell_6^\pm = \pm \sqrt{3} p / \eta$, $\ell_7^\pm = \pm 1/6$, $\ell_8 = r / (\sqrt{2} \zeta  \eta )$ and $\ell_9 = -q / (2 \sqrt{3} \eta )$. In addition, we define $\zeta := \sqrt{3 p^2+q^2+2 r^2}$, $\eta := \sqrt{3 p^2+q^2}$ and $\rho := \sqrt{9 p^2+2 q^2}$. The two sets of indices $\{1,\ldots,7\}$ and $\{8,\ldots,12\}$ correspond to the coset space directions $\mathfrak{m}$ and the Lie subalgebra $\mathfrak{h}=\mathfrak{su}(2)\oplus\mathfrak{u}(1)\oplus\mathfrak{u}(1)$, respectively.

\subsection{\texorpdfstring{$Q^{pqr} = (SU(2)\times SU(2)\times SU(2)) / (U(1) \times U(1))$}{Qpqr}}\applabel{struct_consts_Qpqr}

Here, we list the structure constants $f^A_{BC}$ of $\mathfrak{g}=\mathfrak{su}(2)\oplus\mathfrak{su}(2)\oplus\mathfrak{su}(2)$ used in \secref{Qpqr}. They were extracted from appendix~H of~\cite{Karthauser:2006wb} and rescaled, $f^C_{AB} \rightarrow \sfrac{1}{\sqrt{2}} f^C_{AB}$. The only non-vanishing components are given by
\begin{align}
f^{1}_{27}&=\sfrac{p}{\sqrt{2} \zeta }  , & f^{1}_{28}&=\sfrac{p r}{\sqrt{2} \zeta  \eta }  , & f^{1}_{29}&=\sfrac{q}{\sqrt{2} \eta }  , &
f^{2}_{17}&=\sfrac{-p}{\sqrt{2} \zeta }  , & f^{2}_{18}&=\sfrac{-p r}{\sqrt{2} \zeta  \eta }  , & f^{2}_{19}&=\sfrac{-q}{\sqrt{2} \eta }  , \\
f^{3}_{47}&=\sfrac{q}{\sqrt{2} \zeta }  , & f^{3}_{48}&=\sfrac{q r}{\sqrt{2} \zeta  \eta }  , & f^{3}_{49}&=\sfrac{-p}{\sqrt{2} \eta }  , &
f^{4}_{37}&=\sfrac{-q}{\sqrt{2} \zeta }  , & f^{4}_{38}&=\sfrac{-q r}{\sqrt{2} \zeta  \eta }  , & f^{4}_{39}&=\sfrac{p}{\sqrt{2} \eta }  , \\
f^{5}_{67}&=\sfrac{r}{\sqrt{2} \zeta }  , & f^{5}_{68}&=\sfrac{-\eta }{\sqrt{2} \zeta }  , &
f^{6}_{57}&=\sfrac{-r}{\sqrt{2} \zeta }  , & f^{6}_{58}&=\sfrac{\eta }{\sqrt{2} \zeta }  , &
f^{9}_{12}&=\sfrac{q}{\sqrt{2} \eta }  , & f^{9}_{34}&=\sfrac{-p}{\sqrt{2} \eta }  , \\
f^{7}_{12}&=\sfrac{p}{\sqrt{2} \zeta }  , & f^{7}_{34}&=\sfrac{q}{\sqrt{2} \zeta }  , & f^{7}_{56}&=\sfrac{r}{\sqrt{2} \zeta }  , &
f^{8}_{12}&=\sfrac{p r}{\sqrt{2} \zeta  \eta }  , & f^{8}_{34}&=\sfrac{q r}{\sqrt{2} \zeta  \eta }  , & f^{8}_{56}&=\sfrac{-\eta }{\sqrt{2} \zeta }  ,
\end{align}
where $\zeta := \sqrt{p^2+q^2+r^2}$ and $\eta := \sqrt{p^2+q^2}$. The two sets of indices $\{1,\ldots,7\}$ and $\{8,9\}$ correspond to the coset space directions $\mathfrak{m}$ and the Lie subalgebra $\mathfrak{h}=\mathfrak{u}(1)\oplus\mathfrak{u}(1)$, respectively.

\resumetocwriting


\end{document}